\documentclass{article}

\PassOptionsToPackage{numbers, compress}{natbib}

\usepackage[preprint]{neurips_2025}

\usepackage[utf8]{inputenc} 
\usepackage[T1]{fontenc}    
\usepackage{url}            
\usepackage{booktabs}       
\usepackage{amsfonts}       
\usepackage{nicefrac}       
\usepackage{microtype}      
\usepackage{xcolor}         
\usepackage{dblfloatfix}
\usepackage{placeins}
\usepackage{caption}
\usepackage{graphicx}
\usepackage{enumitem}
\usepackage{amsmath}
\usepackage[colorlinks]{hyperref}

\usepackage[normalem]{ulem} 
\usepackage{color}          

\title{MIRAGE: Robust multi-modal architectures translate fMRI-to-image models from vision to mental imagery}

\author{%
  Reese Kneeland$^{1,2,*}$\quad
  Cesar Kadir Torrico Villanueva$^{2,\dagger}$\quad
  Tong Chen$^{2,3,\dagger}$\\
  \textbf{Jordyn Ojeda}$^{1}$\quad
  \textbf{Shubh Khanna}$^{4}$\quad 
  \textbf{Jonathan Xu}$^{2,5}$\quad
  \textbf{Paul S.\ Scotti}$^{2,6,7}$\\
  \textbf{Thomas Naselaris}$^{1}$\\
  \\
  $^{1}$University of Minnesota, Minneapolis, Minnesota, United States of America\\
  $^{2}$Medical AI Research Center (MedARC)\\
  $^{3}$University of Sydney, Sydney, Australia\\
  $^{4}$Stanford University, Palo Alto, California, United States of America\\
  $^{5}$Alljoined, San Francisco, California, Unites States of America\\
  $^{6}$Sophont, San Francisco, California, Unites States of America \\
  $^{7}$Princeton Neuroscience Institute, Princeton, New Jersey, United States of America \\
  $^{*}$Corresponding author: reesekneeland@gmail.com \\
  $^{\dagger}$Core contribution.
}

\begin{document}

\maketitle
\vspace{-20pt}

\begin{abstract}

\vspace{-5pt}
To be useful for downstream applications, vision decoding models that are trained to reconstruct seen images from human brain activity must be able to generalize to internally generated visual representations, i.e., mental images. In an analysis of the recently released NSD-Imagery dataset, we demonstrated that while some modern vision decoders can perform quite well on mental image reconstruction, some fail, and that state-of-the-art (SOTA) performance on seen image reconstruction is no guarantee of SOTA performance on mental image reconstruction. Motivated by these findings, we developed \textbf{MIRAGE}, a method explicitly designed to train on vision datasets and cross-decode mental images from brain activity. \textbf{MIRAGE} employs a linear backbone and multi-modal text and image features as input to a diffusion model. Feature metrics and human raters establish MIRAGE as SOTA for mental image reconstruction on the NSD-Imagery benchmark.  With ablation analysis we show that mental image reconstruction works best when decoders use image features with relatively few dimensions and include guidance from text-based and both high- and low-level image-based features. Our work indicates that--given the right architecture--existing large-scale datasets using external stimuli are viable training data for decoding mental images, and warrant optimism about the future success and utility of mental image reconstruction. 

\vspace{-3pt}
\paragraph{Author Summary}
Recent research has focused on developing "vision decoding models" that reconstruct images a person is currently viewing. While scientifically impressive, we argue that the tremendous potential of this technology lies in externalizing mental images—the private, internal world of our thoughts—rather than simply decoding what is already visible to the naked eye. This capability is a necessary step towards future applications, such as diagnostic instruments and communication tools for patients with disorders of communication or consciousness. In this study, we demonstrate that state-of-the-art vision decoding models are often not state-of-the-art when it comes to decoding mental imagery, and that the field’s current focus on complex architectures optimized for vision has led to models that are often too fragile to capture the fainter, noisier signals of the imagination. We introduce MIRAGE, a method explicitly designed to bridge this gap by prioritizing robustness over complexity. Our findings indicate that making progress toward practical brain-computer interfaces requires a shift in focus: researchers must target mental imagery directly rather than assuming vision-based performance will translate in the service of impactful downstream use-cases.

\end{abstract}

  {
  \captionsetup[figure]{margin=0.5in}
  \begin{figure}[!htb]
    \centering
    \vspace{-5pt}
    \includegraphics[width=0.84\linewidth]{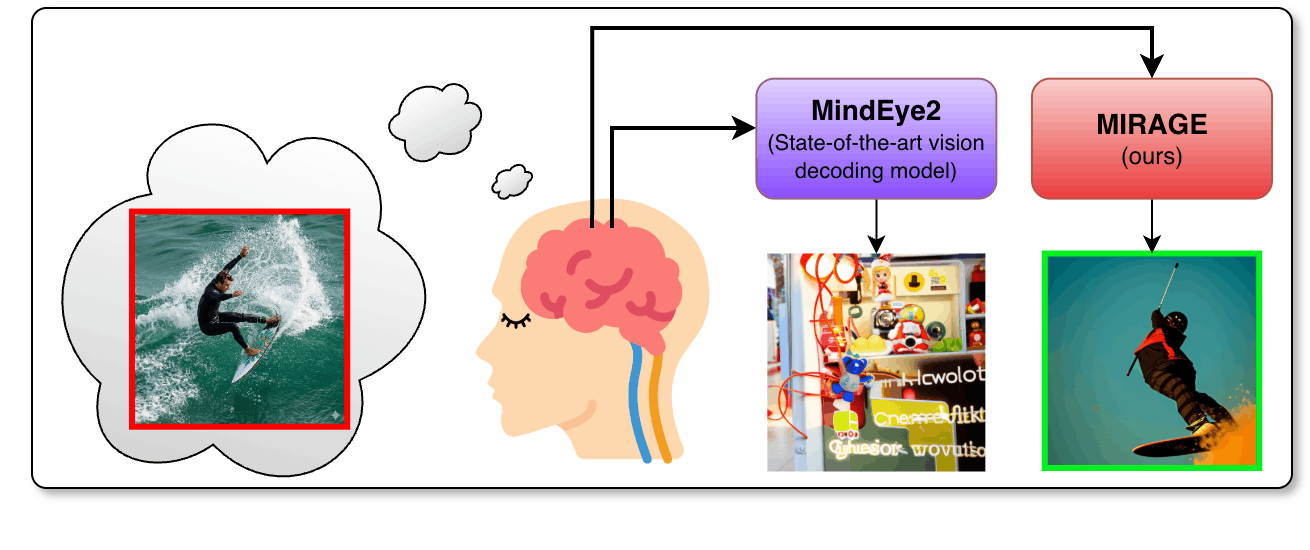}
    \vspace{-15pt}
    \caption{\textbf{MIRAGE} (ours) vs MindEye2 \cite{Scotti2024MindEye2} reconstructions of an imagined image from fMRI brain activity.}
    \vspace{-12pt}
    \label{figure:header}
  \end{figure}%
  }

\section{Introduction}
\label{intro}
The ability to decode and reconstruct mental images—internally generated visual representations not driven by sensory input—from brain activity has tremendous potential for downstream applications such as brain-computer interfaces and medical diagnostics for patients with disorders of communication or consciousness. Externalizing mental images also provides insights into cognitive processes that would be otherwise inaccessible.

Recent research on decoding has focused on developing ``vision decoding models" that are trained and tested on brain responses to seen images. To utilize such vision decoding models for the downstream applications we envision, it is important to demonstrate that these vision decoding models generalize well when tested on mental imagery. An evaluation of a suite of high-performing vision decoding models on the recently-released NSD-Imagery dataset \cite{NSDImagery}--which contains samples of brain activity patterns measured with 7T fMRI in human subjects as they generate mental images--has revealed that decoding performance on seen images is a poor predictor of decoding performance on mental images: while some of the vision decoders evaluated performed quite well on mental image reconstruction, some failed, and state-of-the-art (SOTA) performance on seen image reconstruction was no guarantee of SOTA performance on mental image reconstruction. This indicates that making progress toward a decoder with practical utility will require careful consideration of the factors that make mental image reconstruction challenging, and of the attributes that make vision decoders more or less likely to generalize. To that end, we present the following contributions:
\vspace{-5pt}
\begin{enumerate} 
\item We introduce \textbf{MIRAGE} (\textbf{M}ental \textbf{I}mage \textbf{R}econstruction using \textbf{A}dvanced \textbf{G}enerative Mod\textbf{E}ls), demonstrating that the principled integration of linear decoding backbones with low-dimensional multi-modal feature spaces is an effective technique for enabling cross-decoding of internally generated mental images from vision-only training data. 
\vspace{-3pt}
\item We establish \textbf{MIRAGE} as a SOTA mental imagery decoding method by comparing evaluations across a broad selection of image feature metrics and human preference ratings derived from large-scale behavioral experiments.
\vspace{-3pt}
\item We conduct a detailed ablation analysis to provide empirical evidence for the architectural choices that facilitate generalization from seen to mental image reconstruction, and identify the specific technical
reasons why MIRAGE successfully generalizes to NSD-Imagery where MindEye2 and other more complex vision decoding models fail (Fig \ref{figure:header}). 
\end{enumerate}
\vspace{-10pt}
\subsection{Related Work}
\label{previouswork}
\cite{clip_encoding} Basic neuroscience has demonstrated an extensive overlap in the representation of seen and mental images \cite{kosslyn2006case,stokes_top-down_2009,goebel_reading_2022,naselaris_voxel-wise_2015,reddy_reading_2010}. Nonetheless, differences between vision and mental imagery make cross-decoding from vision to imagery challenging \cite{lee2012disentangling, pearson2019human}. Compared to vision, brain activity during mental imagery has a lower signal-to-noise ratio (SNR) \cite{imagerysnr}, varies along fewer signal dimensions \cite{saharoycompressed}, and encodes imagined stimuli with less spatial resolution \cite{BREEDLOVE20202211, favila2019spatial}.

Previous studies have reported successful $n$-way classification of mental images \cite{cichy2012imagery, lee2012disentangling, albers2013shared}, retrieval of mental images of natural scenes using a visual encoding model \cite{naselaris_voxel-wise_2015,styvespredictmi}, and reconstruction of simple blobs,
letters, or singular natural objects \cite{thirion_inverse_2006,goebel_reading_2022,Senden_Emmerling_van,lee_reconstructing_2016,shen_deep_2019, KOIDEMAJIMA2024349}. With the open releases of CLIP \cite{radford2021learning}, Stable Diffusion \cite{stablediffusion}, and large-scale fMRI datasets like NSD \cite{allen_massive_2022}, newer vision decoding models now yield highly accurate reconstructions of natural scenes \cite{takagi2022_decoding,takagi2023improving,ozcelik2023braindiffuser,scotti_reconstructing_2023,kneeland2023reconstructing,kneeland_second_2023,kneeland_brain-optimized_2023,ferrante_through_2023,chen_seeing_2023,sun_contrast_2023,mai_unibrain_2023}. These methods map fMRI brain activity patterns to embeddings of pre-trained deep learning models that are used to drive a diffusion model \cite{vdvae,podell_sdxl_2023,xu_versatile_2023,Scotti2024MindEye2} to generate image reconstructions of the content present in visual cortex. 

Tests of open-source vision decoding models \cite{ozcelik2023braindiffuser, scotti_reconstructing_2023,Scotti2024MindEye2,shen_deep_2019,takagi2022_decoding,takagi2023improving} on mental imagery activity in the NSD-Imagery dataset \cite{NSDImagery} showed that improved performance on vision decoding does not necessarily translate to mental imagery decoding. For example, MindEye2 \cite{Scotti2024MindEye2} was the SOTA model on the test set of NSD, but not on the imagery trials of NSD-Imagery (Fig \ref{figure:header}). Furthermore, in that study it was observed that methods with linear backbones, compact representations, and multimodal guidance yielded the best performance.  We interpreted these observations as follows: First, \cite{wang_incorporating_2022} showed that linear mappings from CLIP to brain activity explains much of variance in activity in visual cortex. Thus, we should expect linear methods to work quite well when decoding CLIP. Second, as models increase in complexity, the risk of overfitting to noise in the input increases; this danger is especially acute in our case, given that imagery has very low SNR. Thus, representations with fewer dimensions are to be favored, all else being equal. Third, imagery and visual activity are most aligned in brain areas with language-like representations (\cite{lee2012disentangling, BREEDLOVE20202211, clip_encoding}). Thus, it would make sense to include high-quality language-like representations in the decoding pipeline for mental imagery. These ideas informed the design of MIRAGE. Our results, including detailed ablation analyses, support our reasoning, and explain why MIRAGE is SOTA for mental image reconstruction.

\section{Results}

\begin{table*}[!b]
    \vspace{-10pt}
    \centering
    \setlength{\tabcolsep}{2pt}
    \small
    \resizebox{\textwidth}{!}{
    \begin{tabular}{lcccccccccccccccc}
        \toprule
        Method & \multicolumn{4}{c}{Low‑Level} & \multicolumn{4}{c}{High‑Level} & \multicolumn{3}{c}{Brain Correlation} & \multicolumn{4}{c}{Captions}\\
        \cmidrule(lr){2-5}\cmidrule(lr){6-9}\cmidrule(lr){10-12}\cmidrule(l){13-16}
        & PixCorr $\uparrow$ & SSIM $\uparrow$ & Alex(2) $\uparrow$ & Alex(5) $\uparrow$ 
        & Incep $\uparrow$ & CLIP $\uparrow$ & Eff $\downarrow$ & SwAV $\downarrow$ 
        & Early Vis. $\uparrow$ & Higher Vis. $\uparrow$ & Visual Cortex $\uparrow$ 
        & ROUGE‑L $\uparrow$ & METEOR $\uparrow$ & Sentence $\uparrow$ & CLIP‑L $\uparrow$\\
        \midrule
        \multicolumn{16}{c}{\textbf{Mental Imagery Reconstructions}}\\
        \midrule
        \textbf{MIRAGE (ours)}        & \underline{0.104} & 0.398 & \textbf{63.92\%} & \textbf{62.46\%} & 52.25\% & \textbf{57.46\%} & \textbf{0.914} & 0.575 & \textbf{0.204} & \underline{0.142} & \textbf{0.168} & \textbf{0.154} & \textbf{0.095} & \textbf{0.176} & \underline{0.469}\\
        
        MindEye1 \cite{scotti_reconstructing_2023} 
                                       & 0.086 & 0.349 & \underline{59.56\%} & \underline{61.00\%} & 52.03\% & \underline{54.72\%} & 0.948 & \underline{0.564} & \underline{0.180} & 0.135 & \underline{0.155} & – & – & – & –\\
        Brain Diffuser \cite{ozcelik2023braindiffuser} 
                                       & 0.064 & \underline{0.401} & 52.14\% & 58.35\% & \textbf{52.73\%} & 54.07\% & 0.935 & 0.585 & 0.133 & 0.127 & 0.141 & – & – & – & –\\
        iCNN \cite{shen_deep_2019}     & \textbf{0.108} & 0.340 & 50.57\% & 55.25\% & 49.39\% & 41.72\% & 0.994 & \textbf{0.560} & 0.113 & 0.062 & 0.081 & – & – & – & –\\
        MindEye2 \cite{Scotti2024MindEye2} 
                                       & 0.036 & \textbf{0.414} & 47.60\% & 55.38\% & 46.02\% & 50.78\% & 0.966 & 0.591 & 0.069 & 0.055 & 0.061 & \underline{0.143} & \underline{0.080} & \underline{0.162} & \textbf{0.484}\\
        MindBridge \cite{mindbridge}                     & 0.030 & 0.200 & 44.13\% & 52.46\% & 45.55\% & 49.70\% & 0.980 & 0.627 & 0.079 & 0.079 & 0.075 & – & – & – & –\\
        NeuroPictor \cite{neuropictor}                    & 0.022 & 0.305 & 43.18\% & 44.85\% & 44.15\% & 46.40\% & 0.994 & 0.612 & 0.084 & 0.054 & 0.140 & – & – & – & –\\
        BrainRAM \cite{brainram}                       & 0.056 & 0.372 & 51.78\% & 56.29\% & \underline{52.52\%} & 53.73\% & \underline{0.927} & 0.577 & 0.139 & \textbf{0.145} & 0.112 & – & – & – & –\\
        \midrule
        \multicolumn{16}{c}{\textbf{Vision Reconstructions}}\\
        \midrule
        \textbf{MIRAGE (ours)}        & \underline{0.221} & 0.442 & \textbf{79.03\%} & 76.57\% & \textbf{69.75\%} & \textbf{66.69\%} & \textbf{0.879} & 0.546 & 0.363 & \textbf{0.262} & 0.316 & \underline{0.157} & \underline{0.105} & \underline{0.216} & \underline{0.469}\\
        MindEye1 \cite{scotti_reconstructing_2023} 
                                       & 0.218 & 0.412 & \underline{73.56\%} & \underline{80.81\%} & 62.44\% & 65.34\% & \underline{0.881} & \textbf{0.510} & \underline{0.374} & \underline{0.253} & 0.311 & – & – & – & –\\
        Brain Diffuser \cite{ozcelik2023braindiffuser} 
                                       & 0.107 & \underline{0.455} & 60.34\% & 72.84\% & 60.95\% & 58.31\% & 0.908 & 0.555 & 0.247 & 0.229 & 0.255 & – & – & – & –\\
        iCNN \cite{shen_deep_2019}     & \textbf{0.224} & 0.385 & 71.67\% & \textbf{81.35\%} & 61.16\% & 49.03\% & 0.926 & 0.524 & \textbf{0.442} & 0.246 & \underline{0.338} & – & – & – & –\\
        MindEye2 \cite{Scotti2024MindEye2} 
                                       & 0.161 & \textbf{0.480} & 70.10\% & 77.52\% & \underline{62.69\%} & \underline{65.93\%} & 0.886 & \underline{0.512} & 0.352 & 0.237 & 0.290 & \textbf{0.171} & \textbf{0.118} & \textbf{0.249} & \textbf{0.515}\\
        MindBridge \cite{mindbridge}                     & 0.117 & 0.352 & 58.47\% & 70.76\% & 58.83\% & 64.49\% & 0.915 & 0.565 & 0.245 & 0.227 & 0.232 & – & – & – & –\\
        NeuroPictor \cite{neuropictor}                    & 0.055 & 0.364 & 62.27\% & 66.42\% & 49.92\% & 53.49\% & 0.949 & 0.571 & 0.272 & 0.192 & \textbf{0.363} & – & – & – & –\\
        BrainRAM \cite{brainram}                       & 0.097 & 0.409 & 63.11\% & 67.99\% & 58.50\% & 59.79\% & 0.894 & 0.530 & 0.215 & 0.226 & 0.175 & – & – & – & –\\
        \bottomrule
\end{tabular}
    }
    \vspace{-5pt}
    \caption{Quantitative comparison between reconstruction methods for both imagery and vision trials on simple and complex stimuli (conceptual stimuli have no ground truth images). PixCorr is the pixel-level correlation score. SSIM is the structural similarity index metric \cite{wang_image_2004}. AlexNet($2$) and AlexNet($5$) are the 2-way comparisons (2WC) of layers 2 and 5 of AlexNet \cite{alexnet}. CLIP is the 2WC of the output layer of the CLIP ViT-L/14 Vision model \cite{radford2021learning}. Inception is the 2WC of the last pooling layer of InceptionV3 \cite{inceptionv3}. EffNet-B and SwAV are distance metrics gathered from EfficientNet-B13 \cite{tan_efficientnet_2020} and SwAV-ResNet50 \cite{caron_unsupervised_2021} models. Each brain correlation score is calculated using voxels from within the respective regions of the visual cortex. For EffNet-B and SwAV distances, lower is better. For all other metrics, higher is better. Bold indicates best performance, and underlines second-best performance. Additional details on the metrics used, including explanations of 2-way comparisons and brain correlation scores, are in Appendix \ref{app:metrics} in S1 Text. A breakdown of model performance across the different types of stimuli is in Appendix \ref{app:stimtypes} in S1 Text. Details for our implementation of iCNN are provided in Appendix \ref{app:icnnchanges} in S1 Text.} 
    \label{table:combined} %
\end{table*}

\begin{figure}[ht]
\centering
\includegraphics[height=0.87\textheight]{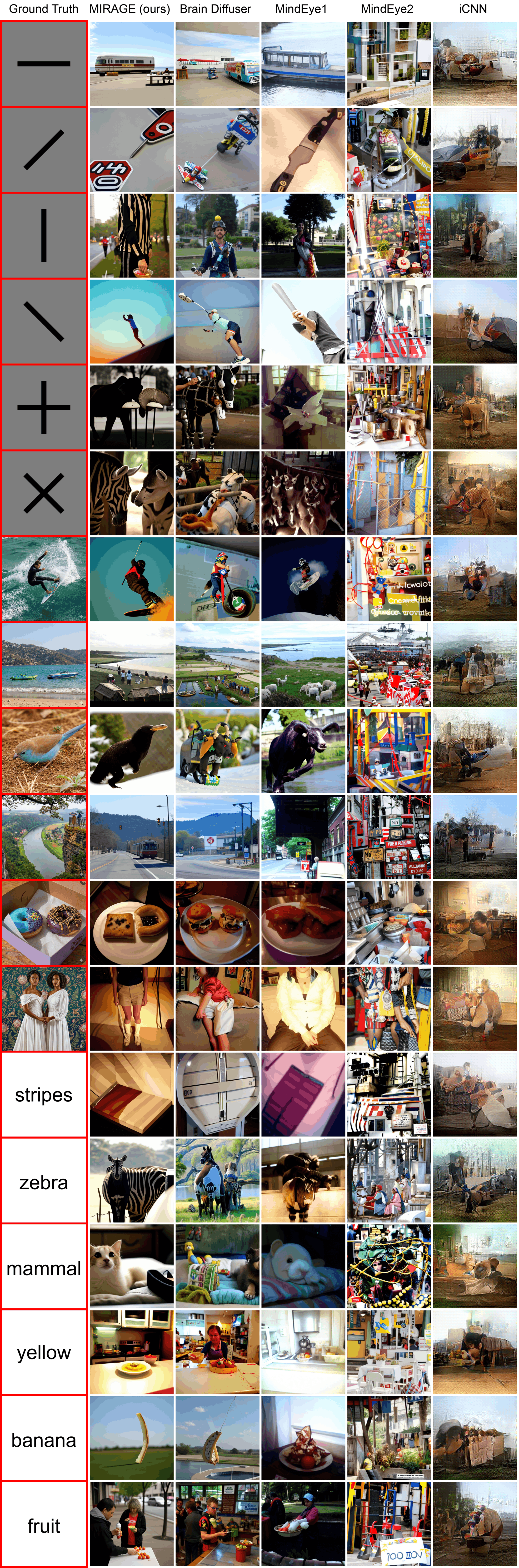}
\vspace{-5pt}
\caption{Qualitative comparison of reconstruction methods on imagined stimuli from NSD-Imagery. Reconstructions selected are the outputs sampled from each method and stimulus with the highest scores on all of the image feature metrics in Table \ref{table:combined}. Examples from more methods can be seen in Appendix \ref{app:moremethods} in S1 Text.}
\label{figure:imagery}
\end{figure}

\vspace{-5pt}
To generate baseline reconstructions for the NSD-Imagery benchmark, we utilized the official open-source implementations provided by the authors of each method, and sampled 10 reconstructions from the posterior distribution of each model for our analysis. The mental image reconstructions produced by \textbf{MIRAGE} (Fig \ref{figure:imagery} and \ref{figure:imagery_best_rebuttal}) are noticeably more faithful to the ground truth images the subjects were instructed to imagine than Brain Diffuser (the previous SOTA model applied to NSD-Imagery), and the other methods we evaluate against. For simple stimuli, the overall structure and orientation of the ground truth images are noticeably improved. For complex stimuli, we successfully reconstruct images that contain similar categories and themes—such as donuts, a man riding a surfboard, an animal with a beak—and with notably improved structural accuracy for complex objects such as the pose of the surfer and the position of the donuts. For conceptual stimuli, the images reconstructed clearly reflect content that is related or identical to the corresponding concept word, with a zebra for ``zebra", a cat for ``mammal", and a recognizable banana for the ``banana" prompt. 

\twocolumn

Vision reconstructions for simple and complex stimuli (Fig \ref{figure:vision} and \ref{figure:vision_best_rebuttal} in S1 Text) also perform well, despite vision decoding performance not being our primary target. The simple stimuli reconstructions are less visually distorted as a result of our low-level guidance, and the complex stimuli hold up well to previous methods. Median and worst-case reconstructions are in Fig \ref{figure:vision_median}, \ref{figure:imagery_median}, \ref{figure:vision_worst}, \ref{figure:imagery_worst}, \ref{figure:vision_median_rebuttal}, \ref{figure:imagery_median_rebuttal}, \ref{figure:vision_worst_rebuttal}, and \ref{figure:imagery_worst_rebuttal} in S1 Text.

\vspace{-5pt}
\subsection{Feature metric evaluations}
\vspace{-5pt}
\label{featuremetrics}
We provide performance benchmarks against existing methods evaluated on NSD-Imagery \cite{NSDImagery} using the metrics in Table \ref{table:combined}. For all methods, we output 10 reconstructions per test sample from each method and report averaged metrics across them. Metrics can fluctuate due to the stochastic nature of fMRI-to-image models, and this averaging step increases the reliability of results. Statistical significance measures can be found in Table \ref{table:combined_stats} in S1 Text. Across the majority of metrics, \textbf{MIRAGE} shows SOTA performance on mental image reconstructions. We note that although these metrics are often used as a proxy for human judgment, many research efforts have established that these metrics do not closely approximate or align with human assessments of content \cite{peceptualsimilarity} or quality \cite{pickapic}. We observe that they are particularly volatile with a dataset as small as NSD-Imagery. For this reason, we provide extensive behavioral evaluations of our results by human raters in Section \ref{humanratings}. Benchmarks separated by stimulus type are provided in Tables \ref{table:simple_stimuli} and \ref{table:complex_stimuli} in S1 Text. Results of our method on the NSD shared1000 test set are also provided in Table \ref{table:shared1000} in S1 Text.

\subsection{Human ratings of reconstruction quality}
\label{humanratings}
To ensure downstream applicability, reconstruction methods must produce outputs that are meaningful to human observers. Although we report standard image feature metrics, prior work indicates that these automated scores often dissociate from human perceptual judgments of quality and semantic content \cite{peceptualsimilarity,pickapic,Scotti2024MindEye2}. These factors, in addition to the high volatility of automated metrics on small datasets such as NSD-Imagery, lead us to treat human evaluation as the definitive standard for assessing model performance. We thus conducted several large-scale online behavioral experiments in which human raters (n=$500$) assessed the quality of the reconstructions (See Appendix \ref{app:behavioral} in S1 Text).

\textbf{Experiment 1} 
To perform a systematic controlled experiment analyzing our results, we had human raters perform a 2-alternative forced choice (2AFC) judgment about whether a reconstruction was more similar to the ground truth image than 

\begin{figure}[!htb]
\vspace{-10pt}
  \centering
  \begin{minipage}[t]{0.58\textwidth}
    \centering
    \textbf{A}\\[4pt]
    \captionsetup{font=small}%
    \setlength{\tabcolsep}{3pt}%
    \small
    \resizebox{\columnwidth}{!}{%
      \begin{tabular}{lcccc}
        \\
        \multicolumn{5}{c}{\textbf{Human Identification Accuracy}} \\
        \midrule
        Method & All Stimuli ↑ & Simple ↑ & Complex ↑ & Conceptual ↑ \\
        \midrule
        \multicolumn{5}{c}{\textbf{Mental Imagery Reconstructions}} \\
        \midrule
        \textbf{MIRAGE (ours)} & \textbf{78.30\%}  & \textbf{73.93\%}  & \textbf{83.19\%}  & \textbf{77.68\%}  \\
        \textbf{MindEye1} & 73.00\% & \underline{71.01\%} & 82.28\% & 65.68\% \\
        \textbf{Brain Diffuser} & \underline{73.95\%} & 68.20\% & \underline{82.70\%} & \underline{71.01\%} \\
        \textbf{iCNN } & 66.15\% & 66.81\% & 70.04\% & 61.60\% \\
        \textbf{MindEye2} & 56.96\% & 50.21\% & 64.83\% & 55.74\% \\
        \midrule
        \multicolumn{5}{c}{\textbf{Vision Reconstructions}} \\
        \midrule
        \textbf{MIRAGE (ours)} & \underline{84.22\%}  & \textbf{80.77\%}  & 87.66\%  & \textit{N/A} \\
        \textbf{MindEye1 } & \textbf{84.29\%} & \underline{79.32\%} & \underline{89.32\%} & \textit{N/A} \\
        \textbf{Brain Diffuser} & 80.13\% & 71.79\% & 88.46\% & \textit{N/A} \\
        \textbf{iCNN } & 74.52\% & 75.85\% & 73.19\% & \textit{N/A} \\
        \textbf{MindEye2} & 83.05\% & 75.54\% & \textbf{90.56\%} & \textit{N/A} \\
        \bottomrule
      \end{tabular}%
    }
  \end{minipage}%
  \hfill
  \begin{minipage}[t]{0.41\textwidth}
    \centering
    \textbf{B}\\[4pt]
    \includegraphics[width=\columnwidth]{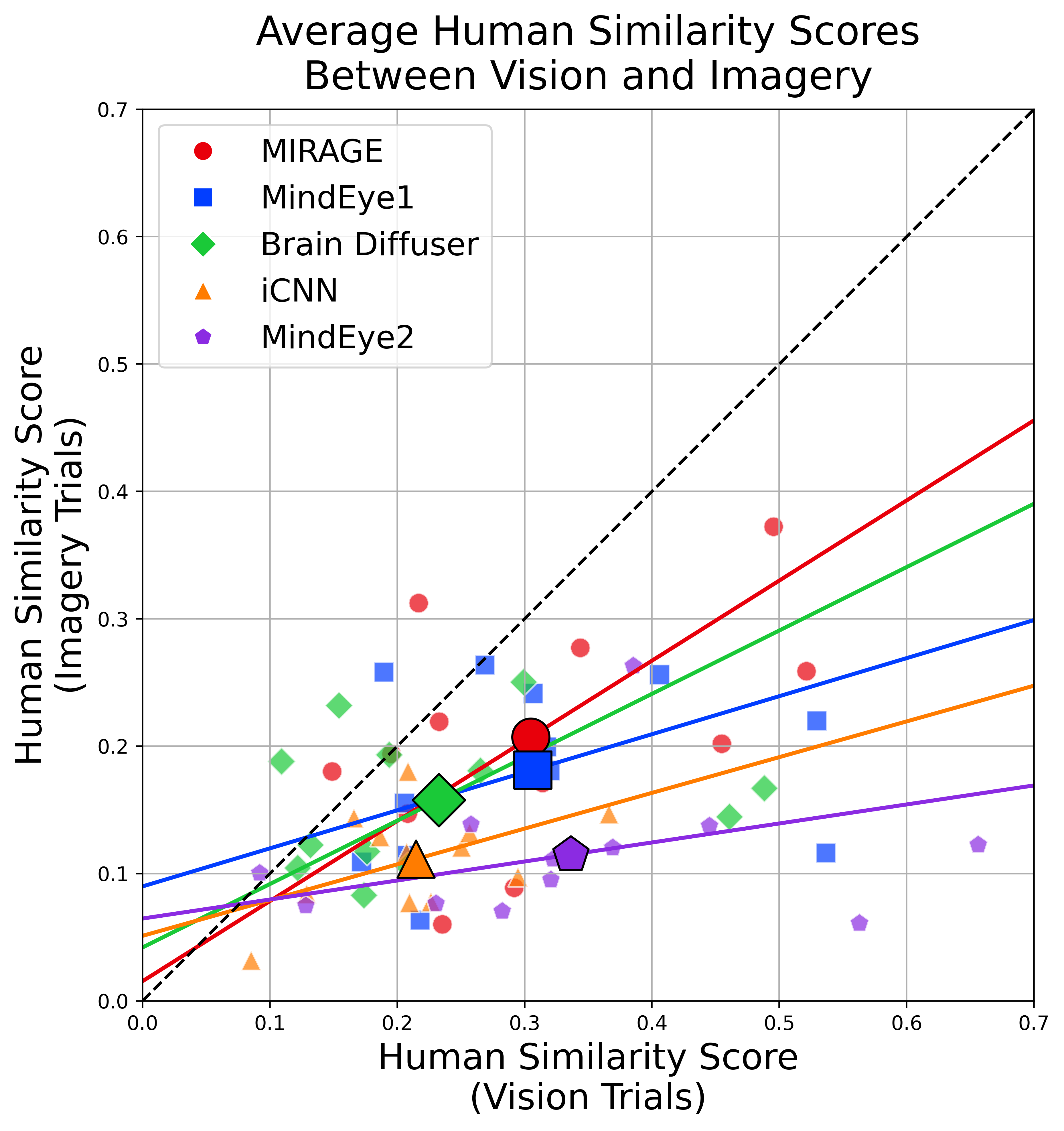}
  \end{minipage}

  \vspace{-4pt}
  \captionof{table}{\textbf{(A)} Human identification accuracy scores for the vision and mental imagery trials of the NSD-Imagery benchmark. Scores are provided for each method and stimulus type; best values are bolded and second best underlined (chance = 50\%, all $p<0.001$).}
  \captionof{figure}{%
    \textbf{(B)} Human similarity scores for simple and complex stimuli: X-axis = vision, Y-axis = imagery; each point is the mean over 12 samples (larger bold points are the overall means), colored/shaped by method. PCA-fit slopes closer to unity indicate tighter imagery–vision correspondence; dashed unity line shown.
  }
  \vspace{-10pt}
  \label{fig:combined_2afc_exp3}
\end{figure}

\onecolumn

a randomly selected "distractor" reconstruction of a brain activity pattern originating from a different stimulus that was sampled from the same stimulus type, method, and NSD subject. Results (Table \ref{fig:combined_2afc_exp3}A) confirm \textbf{MIRAGE} as SOTA for every stimulus type ($p<0.001$).

\textbf{Experiment 2}
Human raters viewed a ground truth image, a reconstruction of that stimulus from a vision trial, and a reconstruction of the same stimulus from an imagery trial. Raters provided continuous measures of similarity between each reconstruction and the ground truth image. The rating provided a direct comparison of the similarity between vision and imagery reconstructions. \textbf{MIRAGE} shows SOTA generalization from vison to imagery (\cite{wang_incorporating_2022}).

\begin{figure}[!htb]
  \centering
  \begin{minipage}[t]{0.52\textwidth}
    \centering
    \textbf{A}\\[4pt]
    \includegraphics[width=\linewidth]{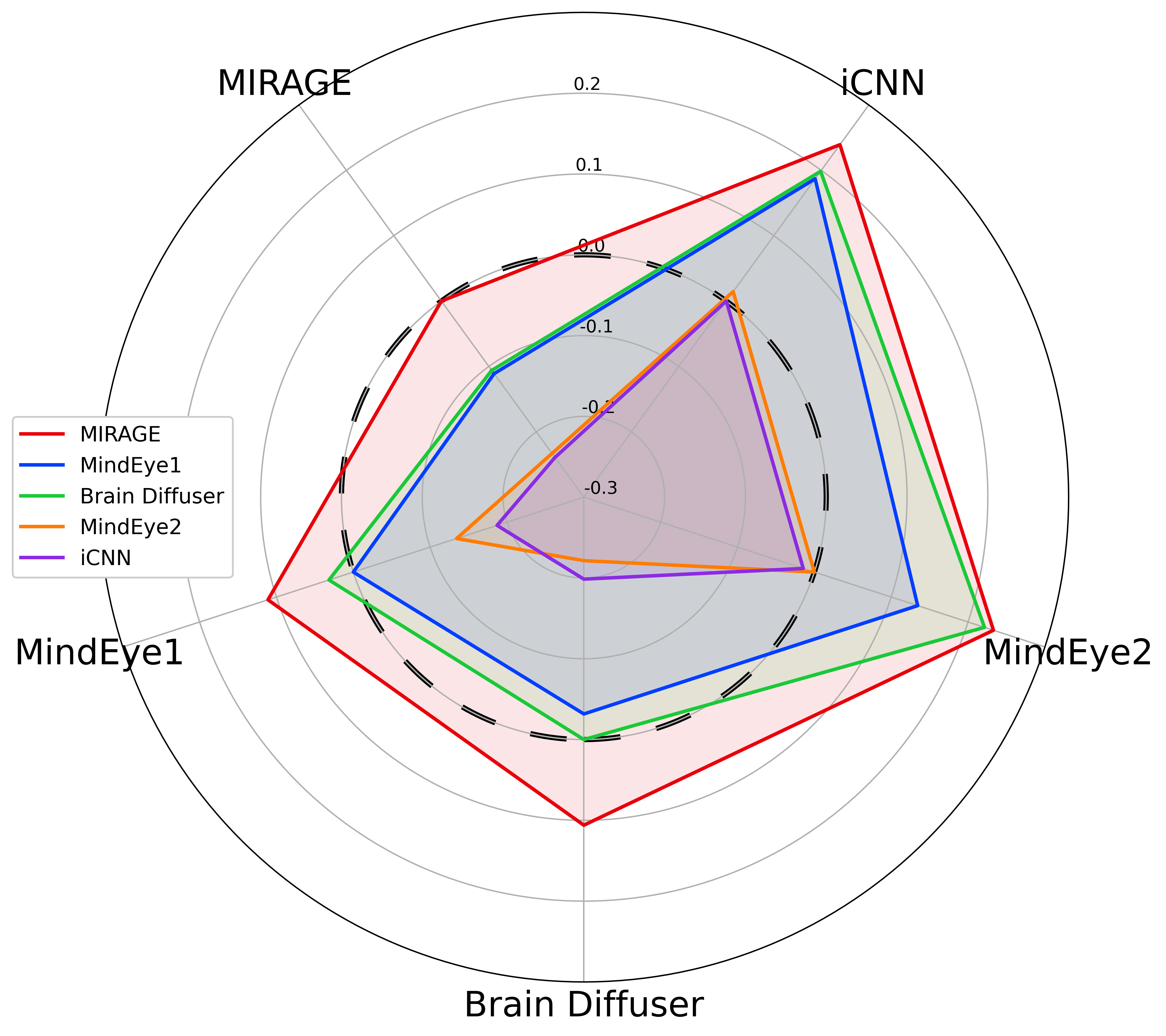}
  \end{minipage}\hfill
  \begin{minipage}[t]{0.47\textwidth}
    \centering
    \textbf{B}\\[4pt]
    \includegraphics[width=\linewidth]{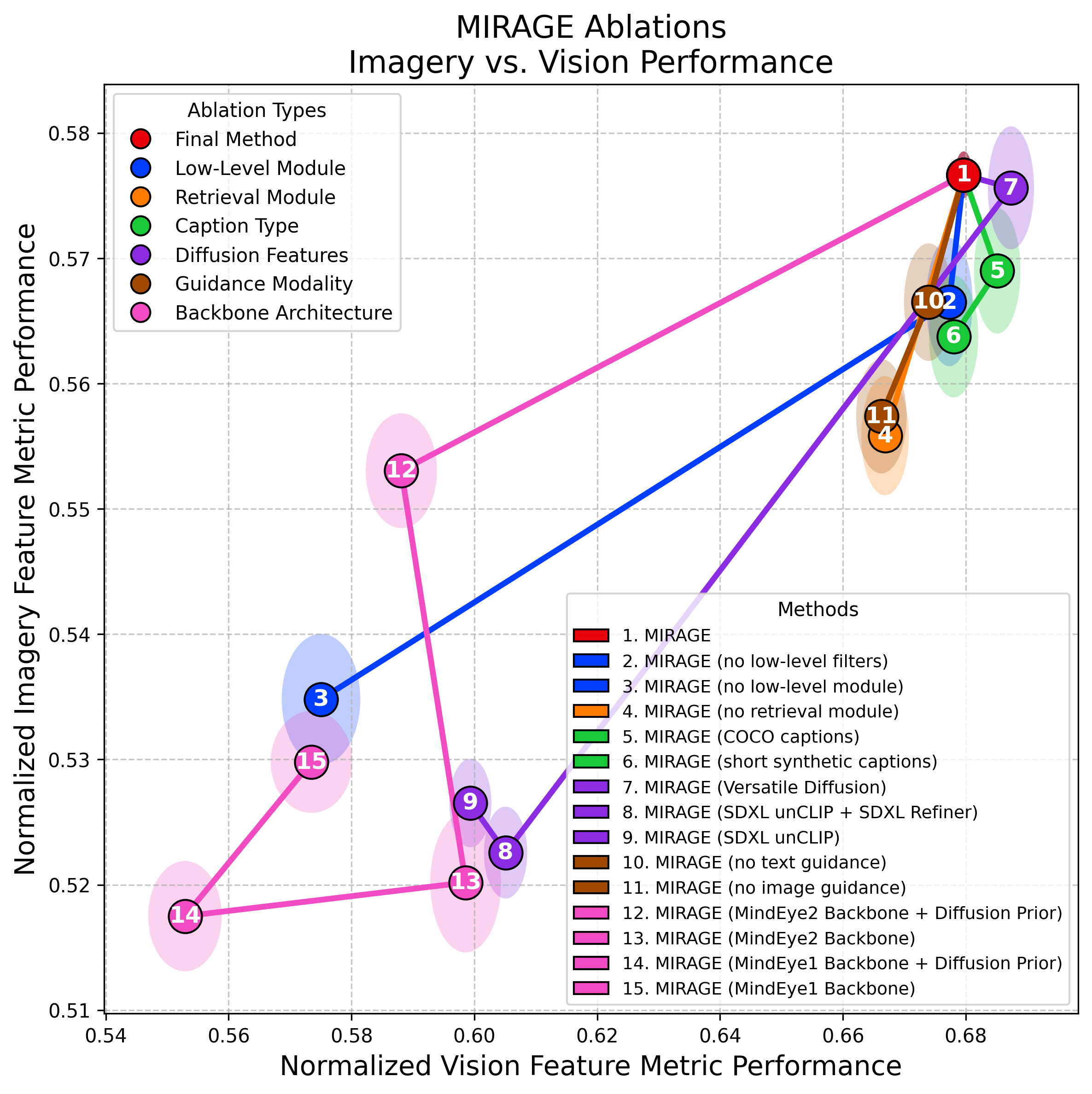}
  \end{minipage}

  \vspace{-6pt}
  \caption{%
    \textbf{(A)} Head-to-head human similarity score results for the conceptual stimuli. The Y-axis represents the similarity score advantage (difference between target method’s score and the alternative, on the radial X-axis); a larger colored polygon area indicates a stronger advantage, and the dashed circle at unity denotes equal performance. MIRAGE outperforms all other methods ($p<0.001$).
    \textbf{(B)} Ablation analyses: model variants (numbered circles) under each ablation type (color) are assessed via the normalized average of all feature metrics (Table \ref{table:combined}), with vision on the x-axis and imagery on the y-axis. The standard error of each point along each axis is visualized with a shaded ellipse of the same color. For details on how we compute the normalized average metric scores, see Appendix \ref{app:featureaverage} in S1 Text.
  }
  \vspace{-23pt}
  \label{fig:combined_exp4_ablation}
\end{figure}

\textbf{Experiment 3}
Mental image reconstructions of the conceptual stimuli are particularly difficult to evaluate, as they do not have associated ground truth images or a meaningful match to vision trials. To compare the performance of our method on these stimuli relative to other models trained on NSD, we conducted a third behavioral experiment that pitted reconstruction methods head-to-head by presenting human raters with a ground truth conceptual stimulus and two reconstructions of that stimulus sampled at random from the collection of methods we evaluate against in this work. Every trial in this experiment is a head-to-head comparison between two random reconstruction methods, and so over the course of the whole experiment, we gather lots of trials between all combinations of two methods. The “similarity score advantage” is the average difference between one method and another across the trials where both methods were presented. Fig \ref{fig:combined_exp4_ablation}A plots the average ``similarity score advantage" for all combinations of methods. Reconstructions from our method are the most strongly preferred in head-to-head comparisons ($p < 0.001$) and have the largest advantage in all comparison cases evaluated.

\subsection{Ablation Study}
\label{ablations}
\definecolor{final}{RGB}{232, 0, 11}
\definecolor{lowlevel}{RGB}{2, 62, 255}
\definecolor{retrieval}{RGB}{255, 124, 0}
\definecolor{caption}{RGB}{26, 201, 56}
\definecolor{diffusion}{RGB}{139, 43, 226}
\definecolor{guidance}{RGB}{159, 72, 0}
\definecolor{backbone}{RGB}{241, 76, 193}

We systematically ablated model components to identify which were most important for mental imagery reconstruction. Colored numeric identifiers refer to Fig \ref{fig:combined_exp4_ablation}B. \textbf{MIRAGE} is identified as \textcolor{final}{(1)}. We also provide a hyperparameter search over the ridge parameter $\lambda$ in Appendix \ref{app:ridge_parameter} in S1 Text.


\textbf{Low-Level Module}
We found that despite the relatively low spatial resolution of mental imagery, the low-level decoding module provides much of the needed structure to accurately reconstruct the target stimulus \textcolor{lowlevel}{(3)}. We also observe that our image filtering technique induces a small performance increase for mental imagery reconstructions \textcolor{lowlevel}{(2)}, suggesting that this technique partially mitigates the loss of structural detail in mental vs. seen image reconstructions.  

\textbf{Retrieval Module}
We see a notable boost from our retrieval procedure \textcolor{retrieval}{(4)}, demonstrating that high-dimensional image embeddings can be decoded and used to guide a retrieval module to improve performance, even if those same embeddings are not suitable to drive an image generator.

\textbf{Caption Types}
We examine the results of using COCO captions \textcolor{caption}{(5)}, as well as a set of synthetic ``short" captions \textcolor{caption}{(6)} that aim to replicate the length of the COCO captions (the median word count is $8$ vs $10$ for COCO), and evaluate the quality of captions provided by LLaVA v1.5-13B. These short synthetic captions did not provide a boost over COCO, but the longer synthetic captions (average word count $=34$) utilized in our final method \textcolor{final}{(1)} provided an improvement.

\textbf{Diffusion Features}
We replaced the $1\times768$ image embeddings and $77\times1280$ text embeddings used to drive Stable Cascade with the features used to drive Versatile Diffusion in MindEye1 and Brain Diffuser \textcolor{diffusion}{(7)} \cite{xu_versatile_2023} ($257\times768$ image embedding, $77\times768$ text embedding), and SDXL unCLIP \cite{Scotti2024MindEye2}, the embedding used to drive MindEye2 \textcolor{diffusion}{(8,9)} ($257\times1664$ image embedding). Our smaller image embedding and larger text embedding provided the most robust guidance for reconstructing mental images \textcolor{final}{(1)}.

\textbf{Guidance Modality}
For both vision and mental imagery, image-only guidance \textcolor{guidance}{(10)} to Stable Cascade afforded better performance than text-only guidance \textcolor{guidance}{(11)}. In both cases, but especially for mental imagery, the combination of text and image guidance \textcolor{final}{(1)} provided a large performance boost, suggesting that multimodal guidance plays an important role for mental imagery reconstruction. 

\textbf{Backbone Architecture}
The MLP backbone and diffusion prior architectures of MindEye1 \textcolor{backbone}{(14,15)} and MindEye2 \textcolor{backbone}{(12,13)} perform worse than our ridge regression backbone \textcolor{final}{(1)} for both vision and mental imagery, suggesting that these architectures tend to overfit to the training data in the core NSD experiment. Evidently, the potential for increased expressivity afforded by an MLP is a disadvantage when attempting to generalize vision reconstruction performance to new sessions and stimulus types, or to mental imagery.   

\textbf{Scaling Behavior} We assess the scaling behavior of \textbf{MIRAGE} and other decoding pipelines with respect to the number of trial repetitions folded into the brain activity averages (Appendix \ref{app:trial_reps} in S1 Text), and the number of training data hours (Appendix \ref{app:scaling} in S1 Text). In both cases, the scaling behavior of \textbf{MIRAGE} outperforms all other decoding pipelines. %

\textbf{Contribution of the Natural Image Prior} To assess the relative contributions of the natural image prior on reconstruction performance, we implemented two systematic controls. First, in Experiment 1, distractor images were generated by decoding randomly selected brain activity patterns. Thus, both target and distractor outputs were equally shaped by the diffusion prior. The high identification accuracy observed ($p < 0.001$; see Fig \ref{fig:combined_2afc_exp3}A) therefore reflects genuine signal decoding rather than the prior's semantic bias. Second, we assessed reconstruction performance across varying strengths of diffusion guidance. We found that reconstructions with minimal guidance (with nearly pure VDVAE decoding) remained highly identifiable, while increasing guidance strength primarily enhanced image quality rather than semantic content (Fig \ref{fig:priorstrength} in S1 Text), mitigating recent concerns about the authenticity of fMRI-to-image 
reconstructions \cite{Shirakawa2024SpuriousRF, brainbits}. 

\section{Discussion}
In this work, we addressed the challenge of generalizing fMRI decoding models from visual perception to mental imagery, investigating the specific architectural constraints required to robustly reconstruct internal mental states using models trained exclusively on seen images. Specifically, we sought to identify why state-of-the-art vision decoders fail on mental imagery, and how to bridge the signal-to-noise gap between these two modalities. \textbf{MIRAGE} improves mental imagery reconstruction over vision reconstruction pipelines shown to be SOTA on NSD shared1000 test samples (i.e., MindEye2 \cite{Scotti2024MindEye2}) and on the NSD-imagery dataset (i.e., BrainDiffuser  \cite{ozcelik2023braindiffuser}). Ablation studies revealed that the keys to the success of MIRAGE are (1) a simple linear decoding backbone, (2) a reduction in the dimension of latent image representations, and (3) the inclusion of high-quality multi-modal guidance to the diffusion model.  

\subsection{Neuroscientific Interpretations}
Our ablation studies (Section \ref{ablations}) identify two critical factors for generalization: reduced feature dimensionality and the inclusion of text-based guidance. These empirical results align with known properties of the visual cortex. First, the success of lower-dimensional image embeddings likely reflects the reduced signal-to-noise ratio (SNR) and coarser spatial resolution of mental imagery compared to vision \cite{saharoycompressed, BREEDLOVE20202211}. High-dimensional embeddings, while expressive for high-SNR vision trials, are empirically prone to overfitting and harming generalization to the noisier imagery signal. 

Second, the efficacy of text guidance supports our hypothesis that mental imagery relies heavily on semantic representations. Previous work has demonstrated that natural language supervision improves encoding models of higher-level visual cortex \cite{clip_encoding}, suggesting that these regions encode visual information in a format that is semantically aligned with language. Higher-level visual cortex is also the area of the brain that most heavily overlaps with activations created by mental imagery \cite{BREEDLOVE20202211}, and so by incorporating text embeddings, we speculate that MIRAGE taps into this semantic overlap, allowing the model to stabilize reconstructions even when fine-grained visual details in the brain signal are degraded.

\subsection{Societal Impact}
Our work is a necessary step towards applications of mental image reconstruction, including diagnostic instruments for psychiatric conditions \cite{holmes2010mental} and disorders of consciousness \cite{Giacino_Kalmar_1997,consciousdetection,tbimortality}, as well as expressive alternative communication methods for patients with traumatic brain injuries \cite{bci_tbi}, amyotrophic lateral sclerosis (ALS) \cite{bci_communication}, and locked-in syndrome \cite{bci_lockedin}. While a significant fraction of this communicative value could be provided by a “concept decoder” whose output is linguistic or otherwise compressed, however, visual representations of internal states could complement linguistic representations, for example by revealing how a specific concept or text prompt is visually interpreted. As the saying goes, 'a picture is worth a thousand words'. While our work makes no claims about being able to decode what is unique and idiosyncratic about a specific individual’s mental imagery, it does represent a first step towards customized bespoke representations of internal states, and is a very literal attempt at this goal motivated by the overlap between representations of vision and mental imagery, and measured against a known ground truth visual stimulus image. 

Of course, the development of this technology raises concerns about the potential for misuse \cite{sethics}. We propose that when deployed in a clinical setting, brain decoding should be defined as a medical procedure that yields private health information and should therefore be subject to all relevant laws pertaining to patient consent, risk / benefit assessment, and the protection of privacy. In all other settings, it seems obvious that laws governing brain decoding should require informed consent and, where necessary, parental guidance. 

\subsection{Current Limitations}

The \textbf{MIRAGE} method does exhibit some notable weaknesses and biases when applied to NSD-Imagery. The median and worst case reconstructions from \textbf{MIRAGE} and other methods can be seen in Appendix \ref{app:median_worst} in S1 Text, and in particular we note how the method is heavily influenced by the prior of the NSD training dataset. Objects that are more well represented in NSD, such as surfers, tend to be reconstructed much more successfully than objects more sparsely observed, such as birds. This is seen even more clearly on the simple stimuli, which are not represented at all in NSD (which contains only natural scenes), and so MIRAGE and other methods tend to produce reconstructions with consistent low-level structural features, but seemingly random semantic content reconstructed in the details of the image.

The NSD-Imagery dataset utilized as a validation set for this work also presents a limitation, as it contains only $18$ stimuli, and thus does not allow for large-scale model training or fine-tuning of existing vision decoding models, necessitating a cross-decoding approach. Future models for downstream applications would surely be improved by training or fine-tuning on mental imagery datasets. Such datasets must be a priority for research in this space.

In addition to dataset availability, practical clinical implementation faces the challenge of data acquisition time per subject. Collecting the volume of fMRI data typically required to train robust decoding models is often infeasible for patient populations, however, recent advancements in the ``novel subject'' problem, such as MindEye2 \cite{Scotti2024MindEye2}, have demonstrated the potential to fine-tune models on as little as one hour of data by leveraging pre-training on large public datasets. While current data-efficient methods struggle to generalize to mental imagery, we anticipate that future research will bridge this gap, and our results suggest a linear backbone with multimodal guidance to be a promising direction for future subject-adaptive mental imagery decoding.

Currently, the computational requirements to run these models are also an obstacle to realizing medical applications. We trained our ridge regression modules on computing hardware with 512GB of RAM and performed inference for our models using an NVIDIA A100 with 40GB of VRAM. The requirement of such hardware limits the use of our method to researchers with considerable compute resources.

All of these limitations we believe to be rich ground for future work, and we look forward to follow-up research addressing them in more detail.

\vspace{-5pt}

\section{Methods}
\subsection{NSD-Imagery Dataset}
\label{sec:dataset_details}

To evaluate model performance on internally generated visual representations, we utilize the NSD-Imagery dataset, an extension of the Natural Scenes Dataset (NSD) \cite{allen_massive_2022} where the same eight subjects completed a session of additional mental imagery tasks. For a full description of the dataset, please see \citet{NSDImagery}. This dataset consists of high-resolution 7T fMRI responses collected using identical acquisition protocols to NSD.

\begin{figure}[!htb]
\vspace{-8pt}
\setlength{\abovecaptionskip}{2pt}    
\setlength{\belowcaptionskip}{-15pt}    
\includegraphics[width=\columnwidth]{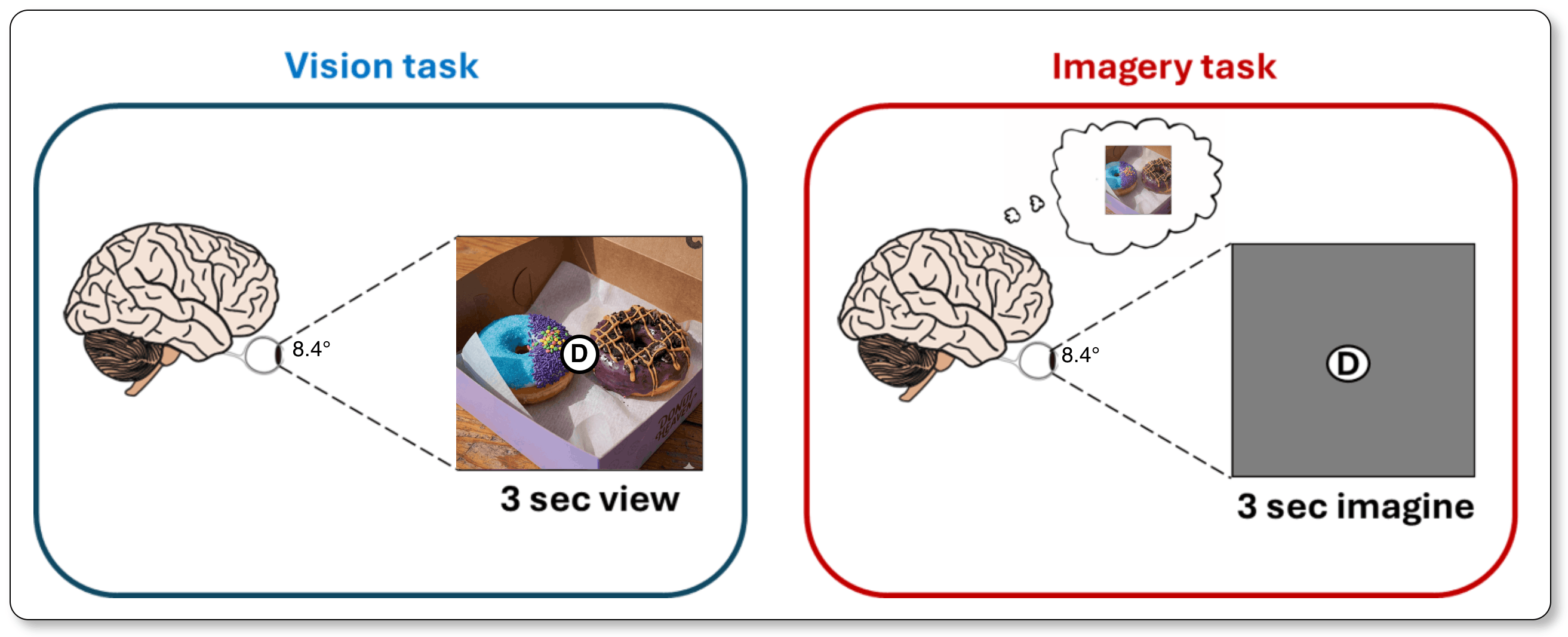}
\caption{Overview of the tasks utilized for the NSD-Imagery benchmark.} 
\label{figure:dataset}
\end{figure}

\paragraph{Stimuli}
Prior to scanning, participants in the NSD-Imagery experiment memorized a unique single-letter cue associated with each of the 18 stimuli (Fig \ref{figure:dataset}), which span three distinct categories to assess reconstruction across varying levels of visual and semantic complexity:
\begin{enumerate}[label=(\Alph*)]
    \item \textit{Simple Stimuli:} Six geometric shapes constructed from black bars on a gray background, including four oriented bars ($0^\circ, 45^\circ, 90^\circ, 135^\circ$) and two crosses (``+'' and ``$\times$'').
    \item \textit{Complex Stimuli:} Five natural scenes selected from the NSD \texttt{shared1000} set and one artwork (``The Two Sisters'' by Kehinde Wiley), chosen based on recognizability scores from the original NSD sessions.
    \item \textit{Conceptual Stimuli:} Six abstract single-word concepts (e.g., ``stripes'', ``mammal'', ``banana'') rather than specific fixed images. As the vision trials for these concepts involved viewing multiple variant images, we exclude the vision trials for this condition and evaluate only on the imagery trials.
\end{enumerate}

\paragraph{Task Protocol}
The experiment consisted of alternating run types. In \textit{Vision Runs}, participants viewed the target image and its corresponding letter cue for 3 seconds, followed by a 1-second rest. Participants performed a one-back task indicating via button press if the image matched the cue. In \textit{Imagery Runs}, participants were presented only with the letter cue and an empty frame (spanning $8.4^\circ \times 8.4^\circ$ visual angle) for 3 seconds. They were instructed to vividly visualize the corresponding stimulus projected into the frame. This was followed by a 1-second rest and a button-press vividness rating. We note that the NSD-Imagery data were collected in a dedicated scanning session separate from the main NSD experiment. Consequently, \textbf{none of the trials in this dataset were used to train MIRAGE} or any other vision decoding models evaluated in this work, which were trained exclusively on the separate visual presentation trials from the NSD core dataset. Furthermore, within the NSD-Imagery session, vision and imagery trials were separated into distinct runs. Vision runs were presented prior to imagery runs to ensure participants correctly recalled the stimuli, but the temporal separation ensures that the decoding of mental imagery is not confounded by hemodynamic responses from preceding visual stimuli. Because there is no session overlaps between the training set (NSD) and the test set (NSD-Imagery), there are also no temporal relationships or block paradigm artifacts for the model to exploit during inference.

\subsection{MIRAGE}

\subsubsection{Datasets}
Our method is trained exclusively on the Natural Scenes Dataset (NSD) \cite{allen_massive_2022} which consists of between $22$k and $30$k fMRI-image pairs per subject ($8$ subjects). NSD stimuli are sourced from the Common Objects in Context dataset (COCO) \cite{microsoftcoco}. We train models for subjects $1, 2, 5,$ and $7$, as only these subjects completed the full $30$k trials of the NSD experiment. We applied the provided nsdgeneral voxel mask at a 1.8 mm resolution to the preprocessed fMRI signals, encompassing numerous visual areas ranging from the early visual cortex to higher-level visual regions. Data from the other $4$ NSD subjects are used as a hyperparameter tuning set, as discussed in Section \ref{regression}.

To evaluate performance on mental images—our primary decoding target—we apply our method to the NSD-Imagery benchmark discussed in Section \ref{sec:dataset_details}, a small extension of NSD where the same subjects completed mental imagery trials. 

\begin{figure}[!htb]
\vspace{-5pt}
\begin{center}
\includegraphics[width=\columnwidth]{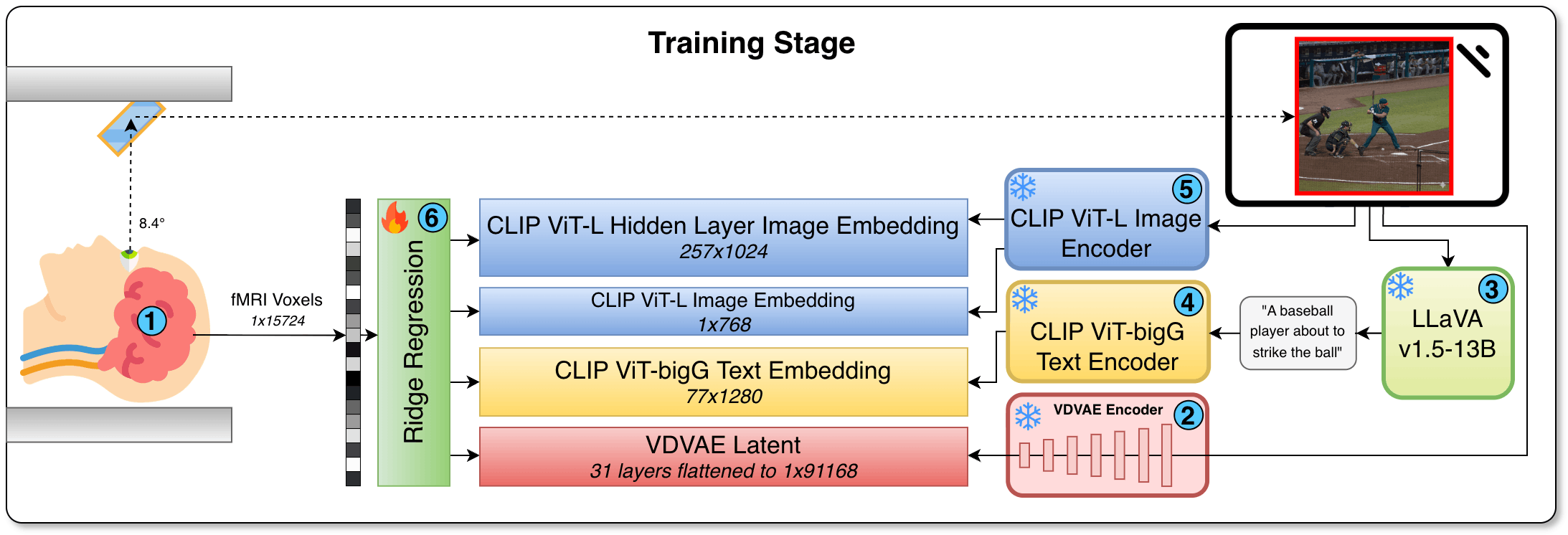}
\end{center}
\vspace{-10pt}
\caption{\textbf{MIRAGE} training pipeline. $(1)$ Brain activity (7T fMRI) acquired as NSD subjects view $>10K$ stimuli. $(2)$ Stimuli are passed to VDVAE encoder \cite{vdvae} yielding ($1\times91168$) latents $(3)$ LLaVA v1.5-13B \cite{llava1,llava2} generates synthetic captions. $(4)$ Captions are encoded into CLIP ViT-bigG/14 text embeddings ($77\times1280$) \cite{openclip2}. $(5)$ Stimuli are also passed through the CLIP ViT-L/14 image encoder \cite{radford2021learning} to generate both CLS token ($1\times768$) and hidden layer ($257\times1024$) image embeddings. $(6)$ Parallel ridge regression modules are trained from the measured fMRI brain activity to the various feature spaces.} 
\vspace{-10pt}
\label{figure:train_pipeline}
\end{figure}

\subsubsection{Methodology}
We propose a model, (\textbf{MIRAGE}), that respects two important desiderata for decoders that generalize well from vision to mental imagery:

\textbf{Reduced model complexity to accommodate the relatively low SNR of mental imagery activity}. Complex, expressive models trained in a high SNR regime may exhibit unacceptably high levels of error variance when tested in a lower SNR setting \cite{Geman_Bienenstock_Doursat_1992}. Brain activity is known to have much lower SNR during mental imagery than it does during vision, and recent work \cite{NSDImagery} has found that relatively complex SOTA vision decoding pipelines generalize poorly when applied to mental images. Our decoding pipeline therefore prioritizes robustness over expressivity. In particular, we implemented a linear ridge regression backbone (Fig \ref{figure:train_pipeline})—in contrast to the non-linear MLP backbone use more commonly in vision decoding. Ridge regression is effective in high-dimensional, low-SNR settings \cite{ridgereg}, and is singularly effective for aligning noisy brain activity patterns across individual brains \cite{ferrante_through_2023}. 

\begin{figure}[!htb]
\begin{center}
\includegraphics[width=\textwidth]{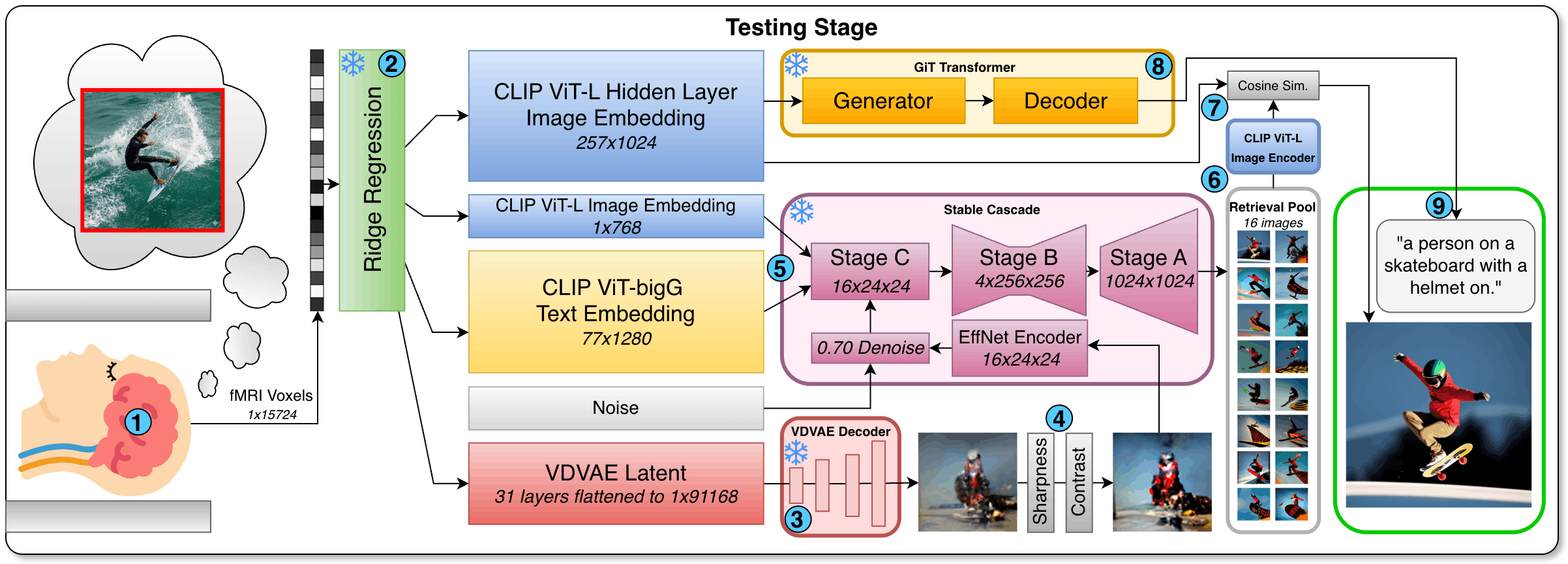}
\end{center}
\vspace{-10pt}
\caption{\textbf{MIRAGE} inference pipeline. $(1)$ The NSD subjects imagine stimuli from letter cues under 7T fMRI. $(2)$ A set of feature embeddings is predicted by passing the measured fMRI brain activity through our frozen ridge regression models. $(3)$ The VDVAE \cite{vdvae} latents are reconstructed into a low-level image. $(4)$ The image is filtered to boost its structure. $(5)$ The filtered low-level image, decoded image embedding, and decoded text embedding are used as input to Stable Cascade \cite{pernias2024wrstchen} to generate a retrieval pool of 16 candidate reconstructions. $(6)$ Each of the candidate reconstructions is encoded into a CLIP ViT-L/14 hidden layer image embedding \cite{radford2021learning}. $(7)$ The final reconstruction is automatically selected as the image with the highest cosine similarity to the decoded CLIP ViT-L/14 hidden layer image embedding. $(8)$ The same hidden layer image embedding is passed through the GiT \cite{wang_git_2022} captioning model. $(9)$ A decoded caption is generated along with our final reconstruction.} 
\vspace{-15pt}
\label{figure:inference_pipeline}
\end{figure}

\textbf{Representational alignment to mental images}.  Brain activity patterns that represent seen and mental images, respectively, overlap in early visual cortex \cite{naselaris_voxel-wise_2015} where structural details of stimuli are encoded, but are most closely aligned in higher-level visual brain areas \cite{BREEDLOVE20202211} that are known to represent semantic and/or linguistic aspects of stimuli \cite{bettermodels}. Accordingly, our method drives an image generator with the CLS token of a CLIP ViT-L/14 image embedding, and incorporates multi-modal guidance from decoded CLIP ViT-bigG/14 text features (Fig \ref{figure:inference_pipeline}).

\vspace{-5pt}
\subsubsection{Ridge regression backbone}
\label{regression}
To map preprocessed fMRI data to feature representations in our decoding pipeline, we employ parallel $L_2$ regularized ridge regression models trained on each individual feature set. In keeping with our first constraint, ridge regression is chosen because it is known to be effective in low signal-to-noise regimes, in contrast to MLPs that offer advantages for capturing nonlinear relationships, but can also be fragile in brain decoding contexts when the input data are of low SNR. For each set of features, we train a parallel ridge regression model to predict the feature value from our fMRI responses, minimizing the loss function in:
\begin{equation}
    \mathcal{L}(\mathbf{w}, \mathbf{b}) = \| \mathbf{X} \mathbf{w} + \mathbf{b} - \mathbf{y} \|_2^2 + \lambda \| \mathbf{w} \|_2^2
\end{equation}

where $\mathbf{X}\in{\mathbb{R}}^{n \times v}$ is the fMRI data matrix of $n$ fMRI trials by $v$ voxels, $\mathbf{y}\in{\mathbb{R}}^{n \times d}$ is the matrix of target features $d$ for each trial $n$, $\mathbf{w} \in \mathbb{R}^{v \times d}$ is the weight vector mapping fMRI voxels to the feature dimension $d$, $\mathbf{b} \in \mathbb{R}^{d}$ is the bias vector added to each trial $n$, and $\lambda$ is the ridge parameter controlling the strength of the $L_2$ penalty. To select an optimal value for $\lambda$, we treat the four NSD subjects who are not selected for reconstruction as a hyperparameter tuning set, and perform a grid search over possible values of $\lambda$ on this set to pick the optimal value for generalizing to fMRI responses of mental images. Using this process, we select $\lambda=100,000$ for all modules of the regression backbone. Details on our hyperparameter search are in Appendix \ref{app:ridge_parameter} in S1 Text.

\subsubsection{Low-Level Module}
\label{low-level}
To begin the reconstruction process, we decode and reconstruct a ``low-level" image that captures the structural layout of the target image. Inspired by the Brain Diffuser method \cite{ozcelik2023braindiffuser}, our approach utilizes a Very Deep Variational Autoencoder (VDVAE) \cite{vdvae} for this task. During training, images are fed into the encoder of the VDVAE to extract latent variables from the first $31$ layers of the VDVAE model, which are concatenated into a $1\times91168$ dimensional feature vector. In the testing phase, we predict these features using our ridge regression backbone and pass them through the VDVAE decoder to generate reconstructed images at $64\times64$ pixel resolution. Details about our implementation of the VDVAE are in Appendix \ref{app:vdvae} in S1 Text.

We observed that when decoding low-level images from fMRI responses to mental images, often the low-level reconstructions have significantly deteriorated features and reduced contrast. To counteract this and boost the influence of our structural layout on subsequent stages of the reconstruction process, we apply a set of image filters to boost the sharpness and contrast of the low-level images. These reconstructions serve as an initial estimate for the diffusion model employed in the subsequent stage of our pipeline.

\subsubsection{Guidance Features}

\textbf{Image Features:}
Noting that the large high-dimensional hidden layer CLIP ViT-bigG/14 image embeddings used in the MindEye2 architecture ($257\times1664$) fail to robustly drive image generators when decoded from mental images, we use only the CLS token of the CLIP ViT-L/14 model ($1\times768$) for our image embeddings. We hypothesize that the removal of the spatial patch tokens used in the hidden layer of ViT embeddings increases the alignment of our features with the encoding of mental images in the brain. 

\textbf{Text Features:}
Motivated by recent findings suggesting that representations in higher-level visual cortex are well-modeled by language-supervised models \cite{clip_encoding}, we incorporate multi-modal guidance into our method through the addition of an unpooled CLIP ViT-bigG/14 text embedding ($77 \times 1280$). We hypothesize that this semantic guidance helps bridge the gap between the visual specificity of seen images and the more abstract, concept-driven nature of mental imagery. In practice, we do not observe the same performance dropoff with high-dimensional text features as we do with high-dimensional image features for driving the diffusion model.

\textbf{Recaptioning:}
To increase the quality of our decoded text features, we replace the COCO captions for the training set with synthetic captions generated by LLaVA v1.5-13B. Each of the original training captions from the COCO dataset is short (average length of only 10 words). We improve the downstream training performance of our model by generating our own captions that are longer and more descriptive.

\subsubsection{Reconstruction}
\label{sec:reconstruction}
\textbf{Diffusion Architecture:}
To generate the final reconstructions, we utilize Stable Cascade, one of the latest multi-modal diffusion models by Stability AI. Stable Cascade uses Würstchen \cite{pernias2024wrstchen}, a text-to-image architecture that also has support for CLIP Image guidance. This architecture consists of three stages, which separately train an image reconstruction module as a VQ-GAN, a latent image diffusion decoder, and a text/image-conditional latent image diffusion generator. By decoupling the text/image condition from the rest of the generation and decoding, the model learns robust representations and how to translate them into images. This architecture accepts both our small $1\times768$ CLIP ViT-L/14 image embeddings, $77\times1280$ CLIP ViT-bigG/14 text embeddings, and a partially diffused input image for img2img mode \cite{sdedit}. The model's multi-modal guidance capabilities, native img2img support, and robust multi-stage architecture make it well-suited for our mental imagery reconstruction task.

\textbf{Retrieval Pooling:}
Because diffusion models are stochastic generators, the quality of the reconstructions they yield will vary from sample to sample. Our pipeline therefore includes a retrieval step to select the best reconstruction from among a small pool samples output by the diffusion model. We decode a CLIP ViT-L/14 hidden layer image embedding ($257\times1024$) that has been unit normalized to approximate a cosine similarity loss in the regression training stage. We then generate $16$ candidate reconstructions to form a retrieval pool, pass each candidate through the CLIP ViT-L/14 image encoder to get hidden layer image embeddings, and select only the highest-scoring reconstruction via cosine similarity with our decoded retrieval embedding as our output. We note that using a high-dimensional image embedding for this retrieval operation does not come with the same drawbacks as using a high-dimensional image embedding to drive the image generator, and we observe our process of using a cosine similarity comparison to differentiate between image embeddings to be robust even when the embeddings are decoded from brain activity responses to mental images.

\textbf{Caption Decoding:}
We implement a caption decoding module that provides a text description of the visual content being decoded. To decode captions, we reuse the CLIP ViT-L/14 hidden layer image features decoded in the retrieval module and pass them through a frozen GiT~\cite{wang_git_2022} transformer (which contains a generator and a decoder) to generate a predicted caption, an approach originally pioneered by \citet{ferrante_brain_2023} and MindEye2 \cite{Scotti2024MindEye2}.

\section*{Acknowledgments}
We would like to thank the many collaborators in the Medical AI Research Center (MedARC) who supported and contributed to the project.

\bibliographystyle{unsrtnat}
\bibliography{main}

@String(CVPR= {IEEE Conf. Comput. Vis. Pattern Recog.})

@String(ICCV= {Int. Conf. Comput. Vis.})

@String(ECCV= {Eur. Conf. Comput. Vis.})

@String(CVPR  = {CVPR})

@String(ICCV  = {ICCV})

@String(ECCV  = {ECCV})

@inproceedings{takagi2022_decoding,
  title={High-resolution image reconstruction with latent diffusion models from human brain activity},
  author={Takagi, Yu and Nishimoto, Shinji},
  booktitle={Proceedings of the IEEE/CVF Conference on Computer Vision and Pattern Recognition},
  pages={14453--14463},
  year={2023}
}

@misc{takagi2023improving,
      title={Improving visual image reconstruction from human brain activity using latent diffusion models via multiple decoded inputs}, 
      author={Takagi, Yu and Nishimoto, Shinji},
      year={2023},
      eprint={2306.11536},
      archivePrefix={arXiv},
      primaryClass={q-bio.NC}
}

@article {clip_encoding,
author = {Wang, Aria and Kay, Kendrick and Naselaris, Thomas and Tarr, Michael and Wehbe, Leila},
year = {2023},
month = {11},
pages = {1-12},
title = {Better models of human high-level visual cortex emerge from natural language supervision with a large and diverse dataset},
volume = {5},
journal = {Nature Machine Intelligence},
doi = {10.1038/s42256-023-00753-y}
}

@article{ozcelik2023braindiffuser,
  title={Natural scene reconstruction from fMRI signals using generative latent diffusion},
  author={Furkan Ozcelik and Rufin VanRullen},
  journal={Scientific Reports},
  year={2023},
  volume={13},
  url={https://api.semanticscholar.org/CorpusID:260439960}
}

@InProceedings{radford2021learning,
  title = 	 {Learning Transferable Visual Models From Natural Language Supervision},
  author =       {Radford, Alec and Kim, Jong Wook and Hallacy, Chris and Ramesh, Aditya and Goh, Gabriel and Agarwal, Sandhini and Sastry, Girish and Askell, Amanda and Mishkin, Pamela and Clark, Jack and Krueger, Gretchen and Sutskever, Ilya},
  booktitle = 	 {Proceedings of the 38th International Conference on Machine Learning},
  pages = 	 {8748--8763},
  year = 	 {2021},
  editor = 	 {Meila, Marina and Zhang, Tong},
  volume = 	 {139},
  month = 	 {18--24 Jul},
  publisher =    {PMLR},
  url = 	 {https://proceedings.mlr.press/v139/radford21a.html}
}

@article{stablediffusion,
  author       = {Robin Rombach and
                  Andreas Blattmann and
                  Dominik Lorenz and
                  Patrick Esser and
                  Bj{\"{o}}rn Ommer},
  title        = {High-Resolution Image Synthesis with Latent Diffusion Models},
  journal      = {CoRR},
  volume       = {abs/2112.10752},
  year         = {2021},
  url          = {https://arxiv.org/abs/2112.10752},
  eprinttype    = {arXiv},
  eprint       = {2112.10752},
  bibsource    = {dblp computer science bibliography, https://dblp.org}
}

@InProceedings{microsoftcoco,
author="Lin, Tsung-Yi
and Maire, Michael
and Belongie, Serge
and Hays, James
and Perona, Pietro
and Ramanan, Deva
and Doll{\'a}r, Piotr
and Zitnick, C. Lawrence",
editor="Fleet, David
and Pajdla, Tomas
and Schiele, Bernt
and Tuytelaars, Tinne",
title="Microsoft COCO: Common Objects in Context",
booktitle="Computer Vision -- ECCV 2014",
year="2014",
publisher="Springer International Publishing",
address="Cham",
pages="740--755",
isbn="978-3-319-10602-1"
}

@article{BREEDLOVE20202211,
title = {Generative Feedback Explains Distinct Brain Activity Codes for Seen and Mental Images},
journal = {Current Biology},
volume = {30},
number = {12},
pages = {2211-2224.e6},
year = {2020},
issn = {0960-9822},
doi = {https://doi.org/10.1016/j.cub.2020.04.014},
url = {https://www.sciencedirect.com/science/article/pii/S0960982220304942},
author = {Jesse L. Breedlove and Ghislain St-Yves and Cheryl A. Olman and Thomas Naselaris}
}

@article{St-Yves_heirarchy,
  author    = {Ghislain St-Yves and Emily J. Allen and Yihan Wu and Kendrick Kay and Thomas Naselaris},
  title     = {Brain-optimized deep neural network models of human visual areas learn non-hierarchical representations},
  journal   = {Nature Communications},
  volume    = {14},
  number    = {1},
  pages     = {3329},
  year      = {2023},
  doi       = {10.1038/s41467-023-38674-4},
  url       = {https://doi.org/10.1038/s41467-023-38674-4},
  issn      = {2041-1723},
  date      = {2023/06/07},
}

@inproceedings{openclip2,
  title={Reproducible scaling laws for contrastive language-image learning},
  author={Cherti, Mehdi and Beaumont, Romain and Wightman, Ross and Wortsman, Mitchell and Ilharco, Gabriel and Gordon, Cade and Schuhmann, Christoph and Schmidt, Ludwig and Jitsev, Jenia},
  booktitle={Proceedings of the IEEE/CVF Conference on Computer Vision and Pattern Recognition},
  pages={2818--2829},
  year={2023}
}

@inproceedings{kneeland2023reconstructing,
      title={Reconstructing seen images from human brain activity via guided stochastic search}, 
      author={Reese Kneeland and Jordyn Ojeda and Ghislain St-Yves and Thomas Naselaris},
      doi = {10.32470/CCN.2023.1672-0},
      url = {https://2023.ccneuro.org/view_paper1337.html?PaperNum=1672},
      booktitle = {Conference on Cognitive Computational Neuroscience},
      year={2023},
      eprint={2305.00556},
      archivePrefix={arXiv},
      primaryClass={q-bio.NC}
}

@misc{mai_unibrain_2023,
	title = {{UniBrain}: {Unify} {Image} {Reconstruction} and {Captioning} {All} in {One} {Diffusion} {Model} from {Human} {Brain} {Activity}},
	shorttitle = {{UniBrain}},
	url = {http://arxiv.org/abs/2308.07428},
	doi = {10.48550/arXiv.2308.07428},
	urldate = {2024-01-28},
	publisher = {arXiv},
	author = {Mai, Weijian and Zhang, Zhijun},
	month = aug,
	year = {2023},
	note = {arXiv:2308.07428 [cs]},
}

@misc{ferrante_brain_2023,
	title = {Brain {Captioning}: {Decoding} human brain activity into images and text},
	shorttitle = {Brain {Captioning}},
	url = {http://arxiv.org/abs/2305.11560},
	doi = {10.48550/arXiv.2305.11560},
	urldate = {2024-01-24},
	publisher = {arXiv},
	author = {Ferrante, Matteo and Ozcelik, Furkan and Boccato, Tommaso and VanRullen, Rufin and Toschi, Nicola},
	month = may,
	year = {2023},
	note = {arXiv:2305.11560 [cs]},
}

@article{
wang_git_2022,
title={{GIT}: A Generative Image-to-text Transformer for Vision and Language},
author={Jianfeng Wang and Zhengyuan Yang and Xiaowei Hu and Linjie Li and Kevin Lin and Zhe Gan and Zicheng Liu and Ce Liu and Lijuan Wang},
journal={Transactions on Machine Learning Research},
issn={2835-8856},
year={2022},
url={https://openreview.net/forum?id=b4tMhpN0JC},
note={}
}

@misc{kneeland_second_2023,
	title = {Second {Sight}: {Using} brain-optimized encoding models to align image distributions with human brain activity},
	shorttitle = {Second {Sight}},
	url = {http://arxiv.org/abs/2306.00927},
	doi = {10.48550/arXiv.2306.00927},
	urldate = {2024-01-16},
	publisher = {arXiv},
	author = {Kneeland, Reese and Ojeda, Jordyn and St-Yves, Ghislain and Naselaris, Thomas},
	month = jun,
	year = {2023},
	note = {arXiv:2306.00927 [cs, q-bio]},
}

@inproceedings{
podell_sdxl_2023,
title={{SDXL}: Improving Latent Diffusion Models for High-Resolution Image Synthesis},
author={Dustin Podell and Zion English and Kyle Lacey and Andreas Blattmann and Tim Dockhorn and Jonas M{\"u}ller and Joe Penna and Robin Rombach},
booktitle={The Twelfth International Conference on Learning Representations},
year={2024},
url={https://openreview.net/forum?id=di52zR8xgf}
}

@misc{sun_contrast_2023,
	title = {Contrast, {Attend} and {Diffuse} to {Decode} {High}-{Resolution} {Images} from {Brain} {Activities}},
	url = {http://arxiv.org/abs/2305.17214},
	doi = {10.48550/arXiv.2305.17214},
	urldate = {2024-01-13},
	publisher = {arXiv},
	author = {Sun, Jingyuan and Li, Mingxiao and Chen, Zijiao and Zhang, Yunhao and Wang, Shaonan and Moens, Marie-Francine},
	month = dec,
	year = {2023},
	note = {arXiv:2305.17214 [cs]},
}

@article{ferrante_through_2023,
author = {Ferrante, Matteo and Boccato, Tommaso and Ozcelik, Furkan and VanRullen, Rufin and Toschi, Nicola},
year = {2024},
month = {04},
pages = {},
title = {Through their eyes: multi-subject Brain Decoding with simple alignment techniques},
volume = {2},
journal = {Imaging Neuroscience},
doi = {10.1162/imag_a_00170}
}

@misc{kneeland_brain-optimized_2023,
	title = {Brain-optimized inference improves reconstructions of {fMRI} brain activity},
	url = {http://arxiv.org/abs/2312.07705},
	doi = {10.48550/arXiv.2312.07705},
	urldate = {2024-01-13},
	publisher = {arXiv},
	author = {Kneeland, Reese and Ojeda, Jordyn and St-Yves, Ghislain and Naselaris, Thomas},
	month = dec,
	year = {2023},
	note = {arXiv:2312.07705 [cs, q-bio]},
}

@inproceedings{
scotti_reconstructing_2023,
title={Reconstructing the Mind's Eye: f{MRI}-to-Image with Contrastive Learning and Diffusion Priors},
author={Paul Steven Scotti and Atmadeep Banerjee and Jimmie Goode and Stepan Shabalin and Alex Nguyen and Cohen Ethan and Aidan James Dempster and Nathalie Verlinde and Elad Yundler and David Weisberg and Kenneth Norman and Tanishq Mathew Abraham},
booktitle={Thirty-seventh Conference on Neural Information Processing Systems},
year={2023},
url={https://openreview.net/forum?id=rwrblCYb2A}
}

@inproceedings{tan_efficientnet_2020,
  author       = {Mingxing Tan and
                  Quoc V. Le},
  editor       = {Kamalika Chaudhuri and
                  Ruslan Salakhutdinov},
  title        = {EfficientNet: Rethinking Model Scaling for Convolutional Neural Networks},
  booktitle    = {Proceedings of the 36th International Conference on Machine Learning,
                  {ICML} 2019, 9-15 June 2019, Long Beach, California, {USA}},
  series       = {Proceedings of Machine Learning Research},
  volume       = {97},
  pages        = {6105--6114},
  publisher    = {{PMLR}},
  year         = {2019},
  url          = {http://proceedings.mlr.press/v97/tan19a.html},
  bibsource    = {dblp computer science bibliography, https://dblp.org}
}

@article{caron_unsupervised_2021,
  author       = {Mathilde Caron and
                  Ishan Misra and
                  Julien Mairal and
                  Priya Goyal and
                  Piotr Bojanowski and
                  Armand Joulin},
  title        = {Unsupervised Learning of Visual Features by Contrasting Cluster Assignments},
  journal      = {CoRR},
  volume       = {abs/2006.09882},
  year         = {2020},
  url          = {https://arxiv.org/abs/2006.09882},
  eprinttype    = {arXiv},
  eprint       = {2006.09882},
  bibsource    = {dblp computer science bibliography, https://dblp.org}
}

@misc{wang_incorporating_2022,
	title = {Incorporating natural language into vision models improves prediction and understanding of higher visual cortex},
	copyright = {© 2022, Posted by Cold Spring Harbor Laboratory. The copyright holder for this pre-print is the author. All rights reserved. The material may not be redistributed, re-used or adapted without the author's permission.},
	url = {https://www.biorxiv.org/content/10.1101/2022.09.27.508760v1},
	doi = {10.1101/2022.09.27.508760},
	language = {en},
	urldate = {2023-05-13},
	publisher = {bioRxiv},
	author = {Wang, Aria Y. and Kay, Kendrick and Naselaris, Thomas and Tarr, Michael J. and Wehbe, Leila},
	month = sep,
	year = {2022},
	note = {Pages: 2022.09.27.508760
Section: New Results},
}

@article{stokes_top-down_2009,
	title = {Top-{Down} {Activation} of {Shape}-{Specific} {Population} {Codes} in {Visual} {Cortex} during {Mental} {Imagery}},
	volume = {29},
	copyright = {Copyright © 2009 Society for Neuroscience 0270-6474/09/291565-08\$15.00/0},
	issn = {0270-6474, 1529-2401},
	url = {https://www.jneurosci.org/content/29/5/1565},
	doi = {10.1523/JNEUROSCI.4657-08.2009},
	language = {en},
	number = {5},
	urldate = {2023-05-09},
	journal = {Journal of Neuroscience},
	author = {Stokes, Mark and Thompson, Russell and Cusack, Rhodri and Duncan, John},
	month = feb,
	year = {2009},
	pmid = {19193903},
	note = {Publisher: Society for Neuroscience
Section: Articles},
	pages = {1565--1572},
}

@inproceedings{goebel_reading_2022,
	title = {Reading {Imagined} {Letter} {Shapes} from the {Mind}’s {Eye} {Using} {Real}-time 7 {Tesla} {fMRI}},
	doi = {10.1109/BCI53720.2022.9735031},
	booktitle = {2022 10th {International} {Winter} {Conference} on {Brain}-{Computer} {Interface} ({BCI})},
	author = {Goebel, Rainer and van Hoof, Rick and Bhat, Salil and Lührs, Michael and Senden, Mario},
	month = feb,
	year = {2022},
	note = {ISSN: 2572-7672},
	pages = {1--3},
}

@article{naselaris_voxel-wise_2015,
	title = {A voxel-wise encoding model for early visual areas decodes mental images of remembered scenes},
	volume = {105},
	issn = {1053-8119},
	url = {https://www.sciencedirect.com/science/article/pii/S1053811914008428},
	doi = {10.1016/j.neuroimage.2014.10.018},
	language = {en},
	urldate = {2023-05-09},
	journal = {NeuroImage},
	author = {Naselaris, Thomas and Olman, Cheryl A. and Stansbury, Dustin E. and Ugurbil, Kamil and Gallant, Jack L.},
	month = jan,
	year = {2015},
	pages = {215--228},
}

@article{reddy_reading_2010,
	title = {Reading the mind's eye: {Decoding} category information during mental imagery},
	volume = {50},
	issn = {1053-8119},
	shorttitle = {Reading the mind's eye},
	url = {https://www.sciencedirect.com/science/article/pii/S1053811909012701},
	doi = {10.1016/j.neuroimage.2009.11.084},
	language = {en},
	number = {2},
	urldate = {2023-05-09},
	journal = {NeuroImage},
	author = {Reddy, Leila and Tsuchiya, Naotsugu and Serre, Thomas},
	month = apr,
	year = {2010},
	pages = {818--825},
}

@article{wang_image_2004,
	title = {Image quality assessment: from error visibility to structural similarity},
	volume = {13},
	issn = {1941-0042},
	shorttitle = {Image quality assessment},
	doi = {10.1109/TIP.2003.819861},
	number = {4},
	journal = {IEEE Transactions on Image Processing},
	author = {Wang, Zhou and Bovik, A.C. and Sheikh, H.R. and Simoncelli, E.P.},
	month = apr,
	year = {2004},
	note = {Conference Name: IEEE Transactions on Image Processing},
	pages = {600--612},
}

@article{chen_seeing_2023,
  title={Seeing Beyond the Brain: Conditional Diffusion Model with Sparse Masked Modeling for Vision Decoding},
  author={Zijiao Chen and Jiaxin Qing and Tiange Xiang and Wan Lin Yue and Juan Helen Zhou},
  journal={2023 IEEE/CVF Conference on Computer Vision and Pattern Recognition (CVPR)},
  year={2022},
  pages={22710-22720},
  url={https://api.semanticscholar.org/CorpusID:253510456}
}

@inproceedings{
vdvae,
title={Very Deep {\{}VAE{\}}s Generalize Autoregressive Models and Can Outperform Them on Images},
author={Rewon Child},
booktitle={International Conference on Learning Representations},
year={2021},
url={https://openreview.net/forum?id=RLRXCV6DbEJ}
}

@article{xu_versatile_2023,
  title={Versatile Diffusion: Text, Images and Variations All in One Diffusion Model},
  author={Xingqian Xu and Zhangyang Wang and Eric Zhang and Kai Wang and Humphrey Shi},
  journal={2023 IEEE/CVF International Conference on Computer Vision (ICCV)},
  year={2022},
  pages={7720-7731},
  url={https://api.semanticscholar.org/CorpusID:253523371}
}

@article{thirion_inverse_2006,
	title = {Inverse retinotopy: {Inferring} the visual content of images from brain activation patterns},
	volume = {33},
	issn = {10538119},
	shorttitle = {Inverse retinotopy},
	url = {https://linkinghub.elsevier.com/retrieve/pii/S1053811906007373},
	doi = {10.1016/j.neuroimage.2006.06.062},
	language = {en},
	number = {4},
	urldate = {2022-11-01},
	journal = {NeuroImage},
	author = {Thirion, Bertrand and Duchesnay, Edouard and Hubbard, Edward and Dubois, Jessica and Poline, Jean-Baptiste and Lebihan, Denis and Dehaene, Stanislas},
	month = dec,
	year = {2006},
	pages = {1104--1116},
}

@article{allen_massive_2022,
	title = {A massive {7T} {fMRI} dataset to bridge cognitive neuroscience and artificial intelligence},
	volume = {25},
	issn = {1097-6256, 1546-1726},
	url = {https://www.nature.com/articles/s41593-021-00962-x},
	doi = {10.1038/s41593-021-00962-x},
	language = {en},
	number = {1},
	urldate = {2022-11-01},
	journal = {Nature Neuroscience},
	author = {Allen, Emily J. and St-Yves, Ghislain and Wu, Yihan and Breedlove, Jesse L. and Prince, Jacob S. and Dowdle, Logan T. and Nau, Matthias and Caron, Brad and Pestilli, Franco and Charest, Ian and Hutchinson, J. Benjamin and Naselaris, Thomas and Kay, Kendrick},
	month = jan,
	year = {2022},
	pages = {116--126},
}

@article{shen_deep_2019,
	title = {Deep image reconstruction from human brain activity},
	volume = {15},
	issn = {1553-7358},
	url = {https://dx.plos.org/10.1371/journal.pcbi.1006633},
	doi = {10.1371/journal.pcbi.1006633},
	language = {en},
	number = {1},
	urldate = {2022-11-01},
	journal = {PLOS Computational Biology},
	author = {Shen, Guohua and Horikawa, Tomoyasu and Majima, Kei and Kamitani, Yukiyasu},
	editor = {O'Reilly, Jill},
	month = jan,
	year = {2019},
	pages = {e1006633},
}

@article{lee_reconstructing_2016,
	title = {Reconstructing perceived and retrieved faces from activity patterns in lateral parietal cortex},
	volume = {36},
	number = {22},
	journal = {Journal of Neuroscience},
	author = {Lee, Hongmi and Kuhl, Brice A.},
	year = {2016},
	note = {Publisher: Soc Neuroscience},
	pages = {6069--6082},
}

@inproceedings{alexnet,
 author = {Krizhevsky, Alex and Sutskever, Ilya and Hinton, Geoffrey E},
 booktitle = {Advances in Neural Information Processing Systems},
 editor = {F. Pereira and C.J. Burges and L. Bottou and K.Q. Weinberger},
 pages = {},
 publisher = {Curran Associates, Inc.},
 title = {ImageNet Classification with Deep Convolutional Neural Networks},
 url = {https://proceedings.neurips.cc/paper_files/paper/2012/file/c399862d3b9d6b76c8436e924a68c45b-Paper.pdf},
 volume = {25},
 year = {2012}
}

@article{inceptionv3,
  author       = {Christian Szegedy and
                  Vincent Vanhoucke and
                  Sergey Ioffe and
                  Jonathon Shlens and
                  Zbigniew Wojna},
  title        = {Rethinking the Inception Architecture for Computer Vision},
  journal      = {CoRR},
  volume       = {abs/1512.00567},
  year         = {2015},
  url          = {http://arxiv.org/abs/1512.00567},
  eprinttype    = {arXiv},
  eprint       = {1512.00567},
  bibsource    = {dblp computer science bibliography, https://dblp.org}
}

@inproceedings{saharoycompressed,
      title={Mental Imagery: Weak Vision or Compressed Vision?}, 
      author={Tiasha Saha Roy and Jesse Breedlove and Ghislain St-Yves and Kendrick Kay and Thomas Naselaris},
      doi = {10.32470/CCN.2023.1693-0},
      booktitle = {Conference on Cognitive Computational Neuroscience},
      year={2023},
      url= {https://2023.ccneuro.org/view_paper4eea.html?PaperNum=1693}
}

@inproceedings{styvespredictmi,
      title={Do better models of fMRI visual response better predict mental imagery responses?}, 
      author={Ghislain St-Yves and Jesse Breedlove and Kendrick Kay and Thomas Naselaris},
      doi = {10.32470/CCN.2023.1644-0},
      booktitle = {Conference on Cognitive Computational Neuroscience},
      year={2023},
      url={https://2023.ccneuro.org/view_paper37c6.html?PaperNum=1644}
}

@article{peceptualsimilarity,
author = {Pawan Sinha and Richard Russell},
title ={A Perceptually Based Comparison of Image Similarity Metrics},

journal = {Perception},
volume = {40},
number = {11},
pages = {1269-1281},
year = {2011},
doi = {10.1068/p7063},
    note ={PMID: 22416586},

URL = { 
    
        https://doi.org/10.1068/p7063
    
    

},
eprint = { 
    
        https://doi.org/10.1068/p7063
    
    

}
}

@inproceedings{pickapic,
title={Pick-a-Pic: An Open Dataset of User Preferences for Text-to-Image Generation},
author={Yuval Kirstain and Adam Polyak and Uriel Singer and Shahbuland Matiana and Joe Penna and Omer Levy},
booktitle={Thirty-seventh Conference on Neural Information Processing Systems},
year={2023},
url={https://openreview.net/forum?id=G5RwHpBUv0}
}

@article{imagerysnr,
  author    = {Tiasha Saha Roy and Jesse Breedlove and Ghislain St-Yves and Kendrick Kay and Thomas Naselaris},
  title     = {Comparison of Signal to Noise in Vision and Imagery for qualitatively different kinds of stimuli},
  journal   = {Journal of Vision},
  volume    = {23},
  number    = {9},
  pages     = {5961},
  year      = {2023},
  doi       = {10.1167/jov.23.9.5961},
  url       = {https://doi.org/10.1167/jov.23.9.5961},
  issn      = {1534-7362},
  date      = {2023/08/01},
  urldate   = {2024/03/02}
}

@article{consciousdetection, title={Early detection of consciousness in patients with acute severe traumatic brain injury}, volume={140}, DOI={10.1093/brain/awx176}, number={9}, journal={Brain}, author={Edlow, Brian L and Chatelle, Camille and Spencer, Camille A. and Chu, Catherine J. and Bodien, Yelena G. and O’Connor, Kathryn L. and Hirschberg, Ronald E. and Hochberg, Leigh R. and Giacino, Joseph T. and Rosenthal, Eric S. and et al.}, year={2017}, pages={2399–2414}}

@article{tbimortality, title={Mortality associated with withdrawal of life-sustaining therapy for patients with severe traumatic brain injury: A Canadian multicentre cohort study}, volume={183}, DOI={10.1503/cmaj.101786}, number={14}, journal={Canadian Medical Association Journal}, author={Turgeon, Alexis F. and Lauzier, François and Simard, Jean-François and Scales, Damon C. and Burns, Karen E.A. and Moore, Lynne and Zygun, David A. and Bernard, Francis and Meade, Maureen O. and Dung, Tran Cong and et al.}, year={2011}, pages={1581–1588}}

@article{Giacino_Kalmar_1997, 
title={The vegetative and minimally conscious states: A comparison of clinical features and functional outcome}, 
volume={12}, 
DOI={10.1097/00001199-199708000-00005}, 
number={4}, 
journal={Journal of Head Trauma Rehabilitation}, 
author={Giacino, Joseph T. and Kalmar, Kathleen}, 
year={1997}, 
pages={36–51}}

@inproceedings{Scotti2024MindEye2,
author = {Scotti, Paul S. and Tripathy, Mihir and Villanueva, Cesare Kadir Torrico and Kneeland, Reese and Chen, Tong and Narang, Ashutosh and Santhirasegaran, Charan and Xu, Jonathan and Naselaris, Thomas and Norman, Kenneth A. and Abraham, Tanishq Mathew},
title = {MindEye2: shared-subject models enable fMRI-to-image with 1 hour of data},
year = {2024},
booktitle = {Proceedings of the 41st International Conference on Machine Learning},
articleno = {1794},
numpages = {22},
location = {Vienna, Austria},
}

@article{KOIDEMAJIMA2024349,
title = {Mental image reconstruction from human brain activity: Neural decoding of mental imagery via deep neural network-based Bayesian estimation},
journal = {Neural Networks},
volume = {170},
pages = {349-363},
year = {2024},
issn = {0893-6080},
doi = {https://doi.org/10.1016/j.neunet.2023.11.024},
url = {https://www.sciencedirect.com/science/article/pii/S0893608023006470},
author = {Naoko Koide-Majima and Shinji Nishimoto and Kei Majima}
}

@article{Senden_Emmerling_van, title={Reconstructing imagined letters from early visual cortex reveals tight topographic correspondence between visual mental imagery and perception}, volume={224}, DOI={10.1007/s00429-019-01828-6}, number={3}, journal={Brain Structure and Function}, author={Senden, Mario and Emmerling, Thomas C. and van Hoof, Rick and Frost, Martin A. and Goebel, Rainer}, year={2019}, month={Jan}, pages={1167–1183}}

@book{kosslyn2006case,
  title={The case for mental imagery},
  author={Kosslyn, Stephen M and Thompson, William L and Ganis, Giorgio},
  year={2006},
  publisher={Oxford University Press}
}

@article{cichy2012imagery,
  title={Imagery and perception share cortical representations of content and location},
  author={Cichy, Radoslaw M and Heinzle, Jakob and Haynes, John-Dylan},
  journal={Cerebral cortex},
  volume={22},
  number={2},
  pages={372--380},
  year={2012},
  publisher={Oxford University Press}
}

@article{lee2012disentangling,
  title={Disentangling visual imagery and perception of real-world objects},
  author={Lee, Sue-Hyun and Kravitz, Dwight J and Baker, Chris I},
  journal={Neuroimage},
  volume={59},
  number={4},
  pages={4064--4073},
  year={2012},
  publisher={Elsevier}
}

@article{albers2013shared,
  title={Shared representations for working memory and mental imagery in early visual cortex},
  author={Albers, Anke Marit and Kok, Peter and Toni, Ivan and Dijkerman, H Chris and De Lange, Floris P},
  journal={Current Biology},
  volume={23},
  number={15},
  pages={1427--1431},
  year={2013},
  publisher={Elsevier}
}

@article{holmes2010mental,
  title={Mental imagery in emotion and emotional disorders},
  author={Holmes, Emily A and Mathews, Andrew},
  journal={Clinical psychology review},
  volume={30},
  number={3},
  pages={349--362},
  year={2010},
  publisher={Elsevier}
}

@article{pearson2019human,
  title={The human imagination: the cognitive neuroscience of visual mental imagery},
  author={Pearson, Joel},
  journal={Nature reviews neuroscience},
  volume={20},
  number={10},
  pages={624--634},
  year={2019},
  publisher={Nature Publishing Group UK London}
}

@article{favila2019spatial,
  title={Spatial perception and memory have distinct activation profiles in human visual cortex},
  author={Favila, Serra E and Kuhl, Brice A and Winawer, Jonathan},
  journal={BioRxiv},
  pages={811331},
  year={2019},
  publisher={Cold Spring Harbor Laboratory}
}

@inproceedings{
pernias2024wrstchen,
title={W\"urstchen: An Efficient Architecture for Large-Scale Text-to-Image Diffusion Models},
author={Pablo Pernias and Dominic Rampas and Mats Leon Richter and Christopher Pal and Marc Aubreville},
booktitle={The Twelfth International Conference on Learning Representations},
year={2024},
url={https://openreview.net/forum?id=gU58d5QeGv}
}

@InProceedings{NSDImagery,
    author    = {Kneeland, Reese and Scotti, Paul S. and St-Yves, Ghislain and Breedlove, Jesse and Kay, Kendrick and Naselaris, Thomas},
    title     = {NSD-Imagery: A Benchmark Dataset for Extending fMRI Vision Decoding Methods to Mental Imagery},
    booktitle = {Proceedings of the IEEE/CVF Conference on Computer Vision and Pattern Recognition (CVPR)},
    month     = {June},
    year      = {2025},
    pages     = {28852-28862}
}

@inproceedings{llava1,
    author      = {Liu, Haotian and Li, Chunyuan and Wu, Qingyang and Lee, Yong Jae},
    title       = {Visual Instruction Tuning},
    booktitle   = {NeurIPS},
    year        = {2023}
  }

@InProceedings{llava2,
    author    = {Liu, Haotian and Li, Chunyuan and Li, Yuheng and Lee, Yong Jae},
    title     = {Improved Baselines with Visual Instruction Tuning},
    booktitle = {Proceedings of the IEEE/CVF Conference on Computer Vision and Pattern Recognition (CVPR)},
    month     = {June},
    year      = {2024},
    pages     = {26296-26306}
}

@article{bci_lockedin, 
title={Boosting brain–computer interfaces with functional electrical stimulation: Potential applications in people with locked-in syndrome}, 
volume={20}, 
DOI={10.1186/s12984-023-01272-y}, 
number={1}, 
journal={Journal of NeuroEngineering and Rehabilitation}, 
author={Canny, Evan and Vansteensel, Mariska J. and van der Salm, Sandra M. and Müller-Putz, Gernot R. and Berezutskaya, Julia}, 
year={2023}, 
month={Nov}}

@article{bci_communication, 
title={Brain-computer interface: Applications to speech decoding and synthesis to augment communication}, 
volume={19}, 
DOI={10.1007/s13311-022-01190-2}, 
number={1}, 
journal={Neurotherapeutics}, 
author={Luo, Shiyu and Rabbani, Qinwan and Crone, Nathan E.}, 
year={2022}, 
month={Jan}, 
pages={263–273}}

@Article{bci_tbi,
AUTHOR = {Popa, Livia Livinț and Chira, Diana and Strilciuc, Ștefan and Mureșanu, Dafin F.},
TITLE = {Non-Invasive Systems Application in Traumatic Brain Injury Rehabilitation},
JOURNAL = {Brain Sciences},
VOLUME = {13},
YEAR = {2023},
NUMBER = {11},
ARTICLE-NUMBER = {1594},
URL = {https://www.mdpi.com/2076-3425/13/11/1594},
PubMedID = {38002552},
ISSN = {2076-3425},
DOI = {10.3390/brainsci13111594}
}

@article{sdedit,
  author       = {Chenlin Meng and
                  Yang Song and
                  Jiaming Song and
                  Jiajun Wu and
                  Jun{-}Yan Zhu and
                  Stefano Ermon},
  title        = {{SDEdit}: Image Synthesis and Editing with Stochastic Differential Equations},
  journal      = {CoRR},
  volume       = {abs/2108.01073},
  year         = {2021},
  url          = {https://arxiv.org/abs/2108.01073},
  eprinttype    = {arXiv},
  eprint       = {2108.01073},
  bibsource    = {dblp computer science bibliography, https://dblp.org}
}

@article{Shirakawa2024SpuriousRF,
  title={Spurious reconstruction from brain activity: The thin line between reconstruction, classification, and hallucination},
  author={Ken Shirakawa and Yoshihiro Nagano and Misato Tanaka and Shuntaro C. Aoki and Kei Majima and Yusuke Muraki and Yukiyasu Kamitani},
  journal={Journal of Vision},
  year={2024},
  url={https://api.semanticscholar.org/CorpusID:269791182}
}

@inproceedings{
brainbits,
title={BrainBits: How Much of the Brain are Generative Reconstruction Methods Using?},
author={David Mayo and Christopher Wang and Asa Harbin and Abdulrahman Alabdulkareem and Albert Eaton Shaw and Boris Katz and Andrei Barbu},
booktitle={The Thirty-eighth Annual Conference on Neural Information Processing Systems},
year={2024},
url={https://openreview.net/forum?id=KAAUvi4kpb}
}

@article{Geman_Bienenstock_Doursat_1992, title={Neural networks and the bias/variance dilemma}, volume={4}, DOI={10.1162/neco.1992.4.1.1}, number={1}, journal={Neural Computation}, author={Geman, Stuart and Bienenstock, Elie and Doursat, René}, year={1992}, month={Jan}, pages={1–58}}

@article{ridgereg,
 ISSN = {00401706},
 URL = {http://www.jstor.org/stable/1267351},
 author = {Arthur E. Hoerl and Robert W. Kennard},
 journal = {Technometrics},
 number = {1},
 pages = {55--67},
 publisher = {[Taylor & Francis, Ltd., American Statistical Association, American Society for Quality]},
 title = {Ridge Regression: Biased Estimation for Nonorthogonal Problems},
 urldate = {2025-01-30},
 volume = {12},
 year = {1970}
}

@article{bettermodels,
title = "Better models of human high-level visual cortex emerge from natural language supervision with a large and diverse dataset",
author = "Wang, {Aria Y.} and Kendrick Kay and Thomas Naselaris and Tarr, {Michael J.} and Leila Wehbe",
note = "Publisher Copyright: 2023, The Author(s), under exclusive licence to Springer Nature Limited.",
year = "2023",
month = dec,
doi = "10.1038/s42256-023-00753-y",
language = "English (US)",
volume = "5",
pages = "1415--1426",
journal = "Nature Machine Intelligence",
issn = "2522-5839",
publisher = "Springer International Publishing",
number = "12",
}

@inproceedings{brainram,
author = {Xie, Dian and Zhao, Peiang and Zhang, Jiarui and Wei, Kangqi and Ni, Xiaobao and Xia, Jiong},
title = {BrainRAM: Cross-Modality Retrieval-Augmented Image Reconstruction from Human Brain Activity},
year = {2024},
isbn = {9798400706868},
publisher = {Association for Computing Machinery},
address = {New York, NY, USA},
url = {https://doi.org/10.1145/3664647.3681296},
doi = {10.1145/3664647.3681296},
booktitle = {Proceedings of the 32nd ACM International Conference on Multimedia},
pages = {3994–4003},
numpages = {10},
location = {Melbourne VIC, Australia},
series = {MM '24}
}

@inproceedings{mindbridge,
  title={Mindbridge: A cross-subject brain decoding framework},
  author={Wang, Shizun and Liu, Songhua and Tan, Zhenxiong and Wang, Xinchao},
  booktitle={Proceedings of the IEEE/CVF Conference on Computer Vision and Pattern Recognition},
  pages={11333--11342},
  year={2024}
}

@misc{neuropictor,
        title={NeuroPictor: Refining fMRI-to-Image Reconstruction via Multi-individual Pretraining and Multi-level Modulation}, 
        author={Jingyang Huo and Yikai Wang and Xuelin Qian and Yun Wang and Chong Li and Jianfeng Feng and Yanwei Fu},
        year={2024},
        eprint={2403.18211},
        archivePrefix={arXiv},
        primaryClass={cs.CV}
  }

@article{sethics,
    doi = {10.1371/journal.pbio.3002899},
    author = {Gordon, Emma C. AND Seth, Anil K.},
    journal = {PLOS Biology},
    publisher = {Public Library of Science},
    title = {Ethical considerations for the use of brain–computer interfaces for cognitive enhancement},
    year = {2024},
    month = {10},
    volume = {22},
    url = {https://doi.org/10.1371/journal.pbio.3002899},
    pages = {1-15},
    number = {10},
}

\section*{Supporting information}

\textbf{S1 Text}

\textbf{Fig A}: Hyperparameter logarithmic grid search over possible values of $\lambda$ for use in Equation 1 (Section \ref{regression}). Metrics are the normalized average of all metrics in Table 1 of the manuscript, with imagery performance on the Y axis and vision on the X axis.

\textbf{Fig B}: Qualitative comparison of reconstruction methods on stimuli seen during the vision trials of NSD-Imagery. Samples selected are the best scoring according to the reconstruction metrics in Table \ref{table:combined} of the manuscript.

\textbf{Fig C}: Median-case vision reconstructions from the vision trials of NSD-Imagery. Samples selected as median scoring based on metrics in Table \ref{table:combined} of the manuscript.

\textbf{Fig D}: Median-case imagery reconstructions
from the imagery trials of NSD-Imagery. Samples
selected as in Fig C.

\textbf{Fig E}: Worst-case vision reconstructions from the vision trials of NSD-Imagery. Samples selected as lowest scoring based on metrics in Table
\ref{table:combined} of the manuscript.

\textbf{Fig F}: Worst-case imagery reconstructions from the imagery trials of NSD-Imagery. Samples selected as in Fig E.

\textbf{Fig G}: Best-case vision reconstructions (additional methods) from vision trials of NSD-Imagery. Samples selected as highest scoring based on metrics in Table \ref{table:combined} of the manuscript.

\textbf{Fig H}: Best-case imagery reconstructions (additional methods) from imagery trials of NSD-Imagery. Samples selected as in Fig G.

\textbf{Fig I}: Median-case vision reconstructions (additional methods) from vision trials of NSD-Imagery. Samples selected as median scoring based on metrics in Table \ref{table:combined} of the manuscript.

\textbf{Fig J}: Median-case imagery reconstructions (additional methods) from imagery trials of NSD-Imagery. Samples selected as in Fig I.

\textbf{Fig K}: Worst-case vision reconstructions (additional methods) from vision trials of NSD-Imagery. Samples selected as lowest scoring based on metrics in Table \ref{table:combined} of the manuscript.

\textbf{Fig L}: Worst-case imagery reconstructions (additional methods) from imagery trials of NSD-Imagery. Samples selected as in Fig K.

\textbf{Table A}: Standard error measurements for evaluation metrics of fMRI-to-Image reconstruction models evaluated on both the vision and mental imagery trials of NSD-Imagery. Values correspond to the standard error spread of values in Table \ref{table:combined} in the manuscript.

\textbf{Table B}: Quantitative comparison between reconstruction methods on the NSD Shared1000 Test Set. Metrics are the same as Table \ref{table:combined} of the manuscript.

\textbf{Table C}: Quantitative comparison between reconstruction methods for both imagery and vision trials on simple stimuli. Metrics are the same as Table \ref{table:combined} of the manuscript.

\textbf{Table D}: Quantitative comparison between reconstruction methods for both imagery and vision trials on complex stimuli. Metrics are the same as Table \ref{table:combined} of the manuscript.

\textbf{Fig M}: Performance of \textbf{MIRAGE} and other methods when averaging across brain activity responses to multiple trial repetitions of the same stimulus. Y-axis is the normalized average of all metrics in Table \ref{table:combined} of the manuscript, X-axis is the number of averaged trial repetitions.

\textbf{Fig N}: Performance of \textbf{MIRAGE} and other methods on NSD-Imagery for Subject 1 when trained on different numbers of fMRI sessions present in NSD. Each session includes approximately one hour of fMRI data. Metrics are the normalized average of all metrics in Table \ref{table:combined} of the manuscript, with imagery performance on the Y axis and vision on the X axis. Methods are indicated by color, with the number of training sessions indicated by the numbers in each dot.

\textbf{Fig O}: Examples of reconstructions provided at different diffusion strength parameters, images are the ground truth (outlined in red) and reconstructions provided at 0.4, 0.6, 0.8, and 1.0 diffusion strength respectively.

\textbf{Fig P}: Human identification accuracy of \textbf{MIRAGE} (with no CLIP-Image guidance) as a function of diffusion model strength for imagery trials (orange line), vision trials (green line), and a control experiment that used the features directly from the ground truth image and caption (blue line). A dashed line is placed at the 50\% chance threshold. Results are from a behavioral experiment that is identical to Experiment 1 (Fig \ref{fig:combined_2afc_exp3}), but varied across strength parameters.

\textbf{Table E}: Quantitative comparison between \textbf{MIRAGE} and two Top-1 Retrieval baselines (pooled and hidden layer CLIP ViT-L/14 embeddings), separated by simple and complex stimuli and averaged across all subjects. Metrics are the same as Table 1. Bold indicates the best performance between \textbf{MIRAGE} and the retrieval baselines within each stimulus category, and underlines indicate second-best.

\textbf{Fig Q}: \textbf{Top-K retrieval performance vs. pool size for Subjects 1, 2, 5, and 7.} Accuracy (y-axis) is evaluated across varying distractor pool sizes (x-axis) for both mental imagery (left: \textbf{A, C}) and vision trials (right: \textbf{B, D}). The top row (\textbf{A, B}) evaluates retrieval in the pooled ViT-L/14 image embedding space used to drive the MIRAGE generative model, while the bottom row (\textbf{C, D}) uses the hidden layer ViT-L/14 space utilized in the retrieval pooling step (Section \ref{sec:reconstruction}). Curves denote top-1, top-5, and top-10 performance for simple (light lines) and complex (dark lines) stimuli, with chance levels indicated by corresponding dotted lines. To calculate accuracy, the ground-truth NSD-Imagery stimulus is shuffled with $N$ random distractor images from the NSD shared1000 pool; a success is recorded when the target image ranks within the top $K$ closest matches to the subject's brain-predicted embedding. All curves are bootstrapped across 100 randomly sampled distractor pools for each value of $N$.

\textbf{Fig R}: An example of the 2 alternative forced choice task used in the first behavioral experiment performed by human raters.

\textbf{Fig S}: An example of similarity score task used in experiments 2 and 3 of the behavioral experiment performed by human raters.

\textbf{Appendix A.17}: AI-Generated Images and Copyright Compliance



\newpage
\appendix
\setcounter{figure}{0} 
\setcounter{table}{0} 
\renewcommand{\thefigure}{\Alph{figure}}
\renewcommand{\thetable}{\Alph{table}}

\section{Appendix}
\subsection{Finding the optimal value of $\lambda$  for the regression backbone}
\label{app:ridge_parameter}

\begin{figure}[!htb]   
\centering
\includegraphics[width=0.6\linewidth, trim=0 0 0 0]{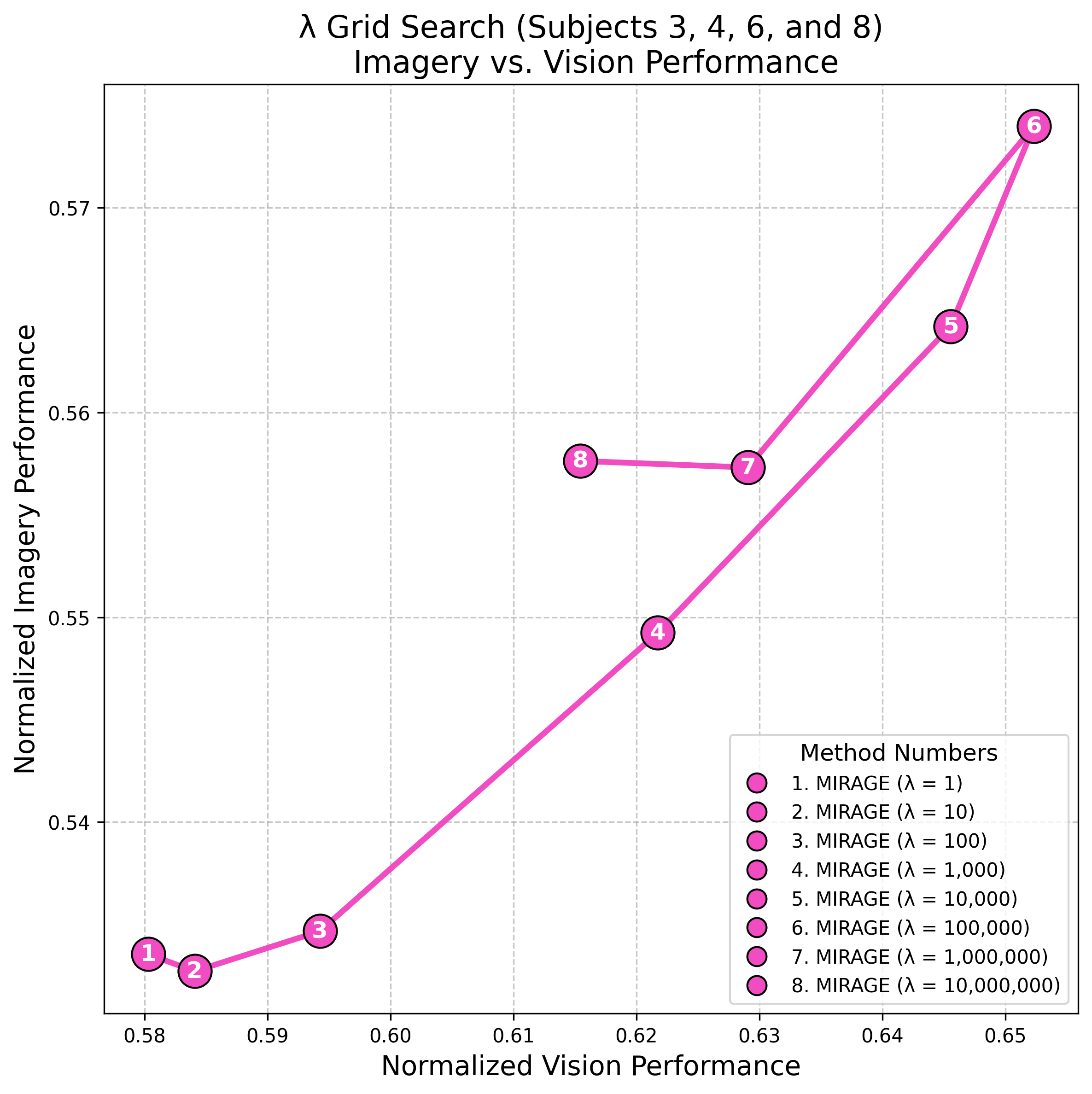}
\caption{Hyperparameter logarithmic grid search over possible values of $\lambda$ for use in Equation 1 (Section \ref{regression}). Metrics are the normalized average of all metrics in Table $1$ of the manuscript, with imagery performance on the Y axis and vision on the X axis.} 
\label{figure:weight_decay}
\end{figure}

To select the optimal $L_2$ weight decay parameter $\lambda$ for use in our regression backbone detailed in Equation $1$ of Section \ref{regression}, we performed a logarithmic grid search over possible values of $\lambda$ on subjects $3, 4, 6$, and $8$ of the NSD-Imagery dataset, which we use a hyperparameter tuning set. From this analysis (Fig \ref{figure:weight_decay}), we select $\lambda = 100,000$.

\subsection{VDVAE implementation}
\label{app:vdvae}
For decoding the low-level image in our pipeline, we utilize the Very Deep Variational Autoencoder (VDVAE) model introduced in \citet{vdvae}. VDVAEs are generative models that learn to represent an input distribution—such as an image dataset—through a low-dimensional latent space constrained by a predefined prior distribution, typically Gaussian. The VDVAE utilizes a hierarchical structure with multiple layers of conditionally dependent latent variables organized hierarchically, with each layer capturing different levels of detail from coarse to fine as one moves from the top to the bottom of the hierarchy. Appendix Eq. (2) shows the factorization of the variational posterior, where each latent $z_n$ corresponds to one layer of the VDVAE. Eq. (3) shows the factorization of the prior.

\begin{equation}
    q_{\phi} (z | x) = q_{\phi} (z_0 | x) q_{\phi} (z_1 | z_0 ,x)...q_{\phi} (z_N | z_{<N},x)
\end{equation}
\begin{equation}
    p_{\theta} (z) = p_{\theta} (z_0) p_{\theta} (z_1 | z_0)...p_{\theta} (z_N | z_{<N})
\end{equation}

In our approach, we utilize the VDVAE model \cite{vdvae} trained on the ImageNet dataset at a resolution of $64\times64$ pixels and consisting of $75$ hierarchical layers. We use the latent variables from the first $31$ layers, as including additional layers does not yield significant improvements in reconstruction quality. In the testing phase, our predicted latents for the first $31$ layers are concatenated with the remaining $44$ layers sampled from Eq. (2), and passed through the latent-to-pixel decoder module of the VDVAE to generate reconstructed images at $64\times64$ pixel resolution.

Following the generation of these initial reconstructions, we apply a post-processing step to enhance their visual clarity using the \texttt{PIL.ImageEnhance} module in Python. Specifically, we boost the sharpness and contrast of the low-level images using the following implementation:

\begin{verbatim}
blurred_image = ImageEnhance.Sharpness(blurred_image).enhance(20)
blurred_image = ImageEnhance.Contrast(blurred_image).enhance(1.5)
\end{verbatim}

We did not perform a formal hyperparameter optimization for these specific values; rather, they were selected heuristically. These parameters were chosen because they qualitatively produced the desired visual corrections to the low-level images when testing the pipeline on the held-out NSD subjects (Subjects 3, 4, 6, and 8) that were not included in the main evaluation.

\FloatBarrier
\clearpage

\subsection{Best case vision reconstructions}
\label{app:vision_best}
\begin{figure}[!htb]    
\centering
\includegraphics[width=0.6\textwidth]{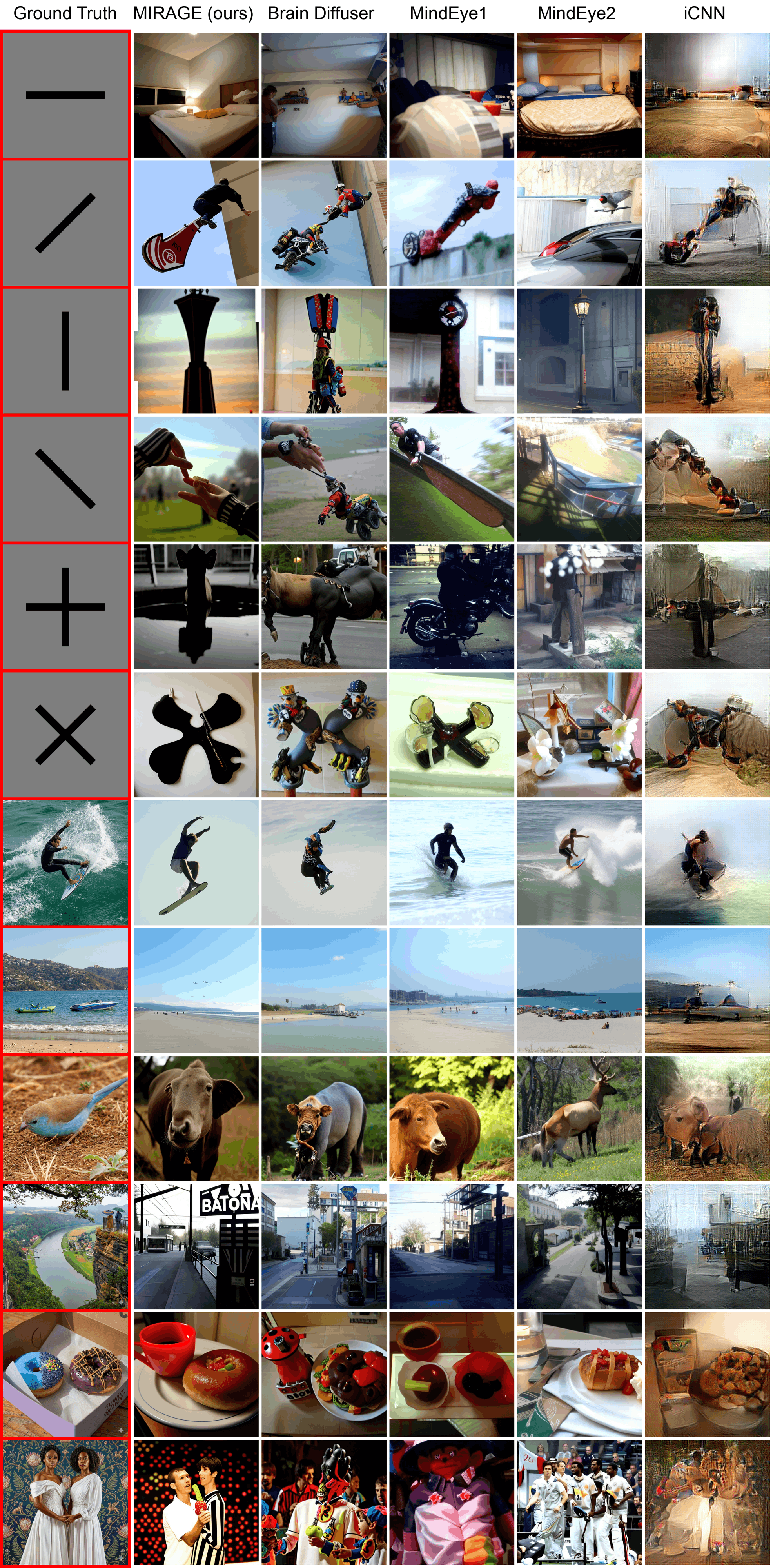}
\caption{Qualitative comparison of reconstruction methods on stimuli seen during the vision trials of NSD-Imagery. Samples selected are the best scoring according to the reconstruction metrics in Table $1$ of the manuscript.} 
\label{figure:vision}
\end{figure}

\FloatBarrier
\clearpage

\subsection{Median and worst-case reconstructions}
\label{app:median_worst}
\begin{figure*}[!htb]
\centering
\begin{minipage}[t]{0.49\textwidth}
    \vspace{0pt}
    \centering
    \includegraphics[width=0.95\textwidth]{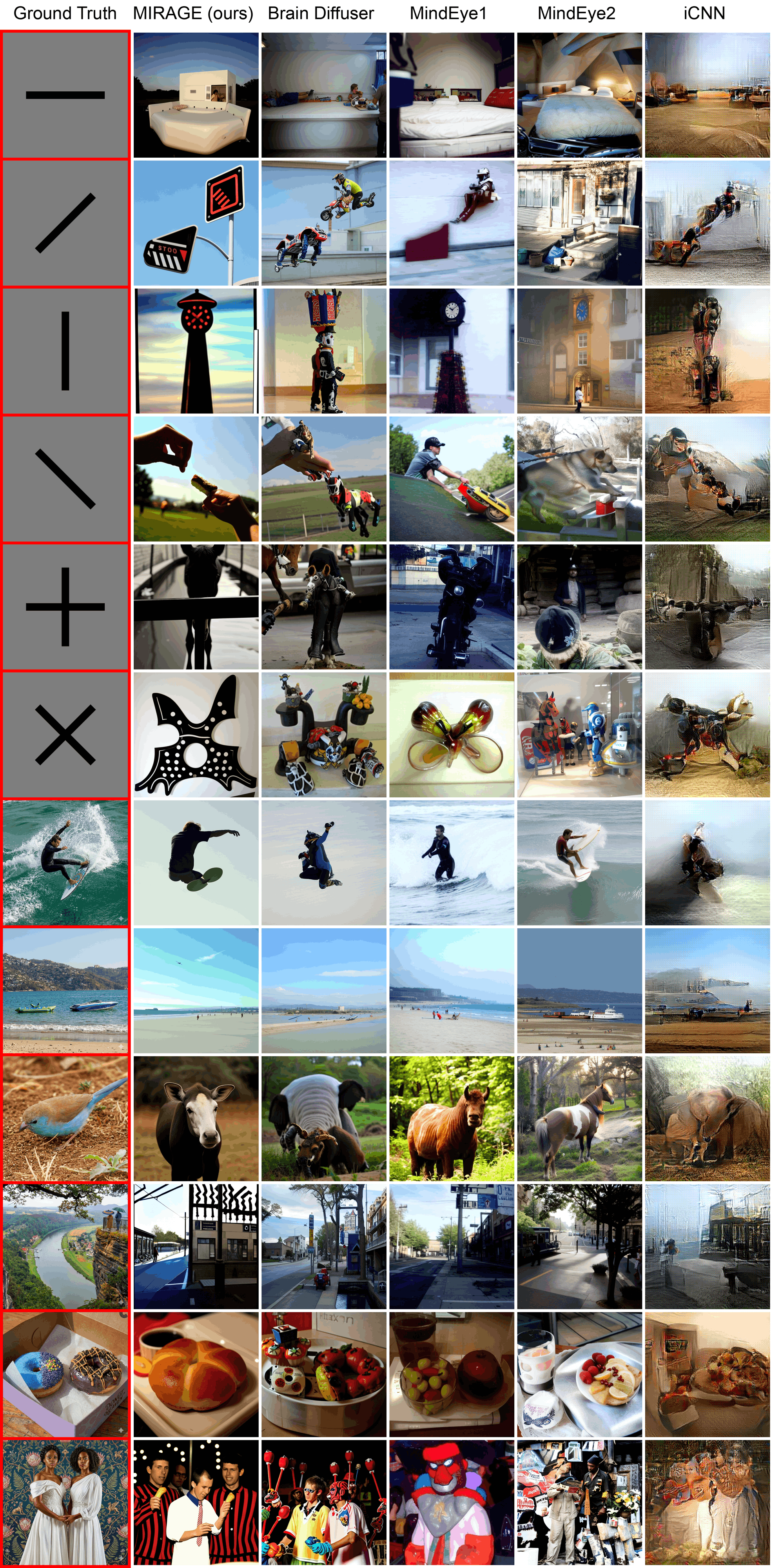}
    \caption{Median-case vision reconstructions from the vision trials of NSD-Imagery. Samples selected as median scoring based on metrics in Table 1 of the manuscript.}
    \label{figure:vision_median}
\end{minipage}\hfill
\begin{minipage}[t]{0.49\textwidth}
    \vspace{0pt}
    \centering
    \includegraphics[width=0.95\textwidth]{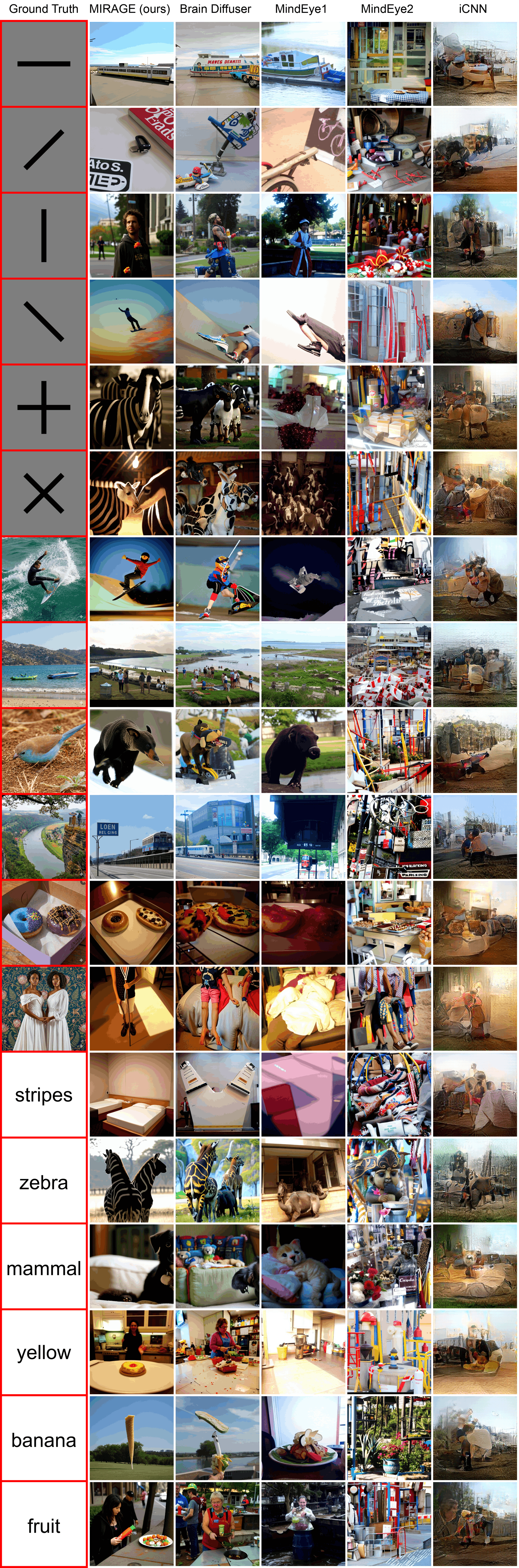}
    \caption{Median-case imagery reconstructions from the imagery trials of NSD-Imagery. Samples selected as in Fig \ref{figure:vision_median}.}
    \label{figure:imagery_median}
\end{minipage}
\end{figure*}
\FloatBarrier
\clearpage

\begin{figure*}[!htb]
\centering
\begin{minipage}[t]{0.49\textwidth}
    \vspace{0pt}
    \centering
    \includegraphics[width=0.95\textwidth]{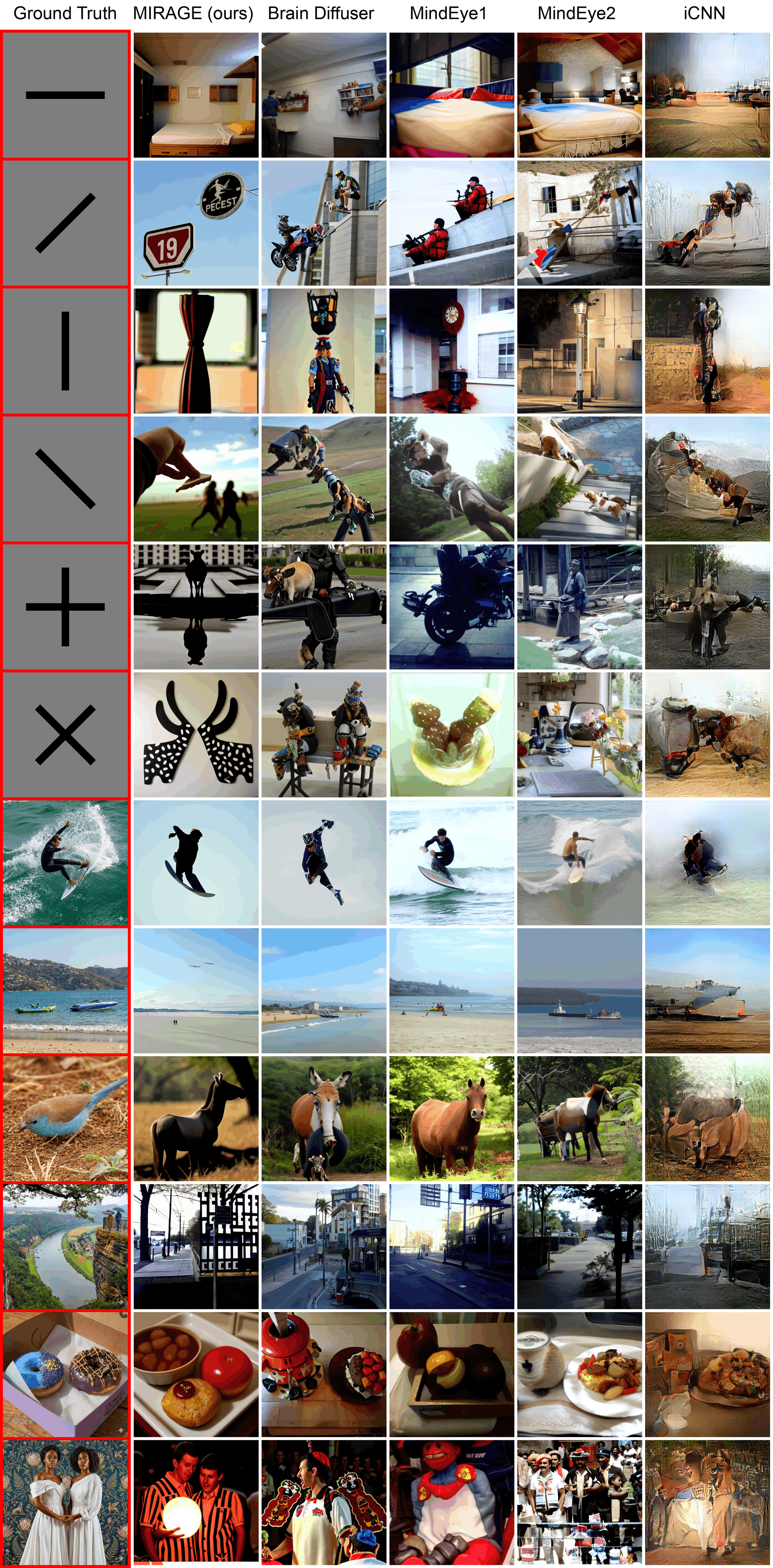}
    \caption{Worst-case vision reconstructions from the vision trials of NSD-Imagery. Samples selected as lowest scoring based on metrics in Table 1 of the manuscript.}
    \label{figure:vision_worst}
\end{minipage}\hfill
\begin{minipage}[t]{0.49\textwidth}
    \vspace{0pt}
    \centering
    \includegraphics[width=0.95\textwidth]{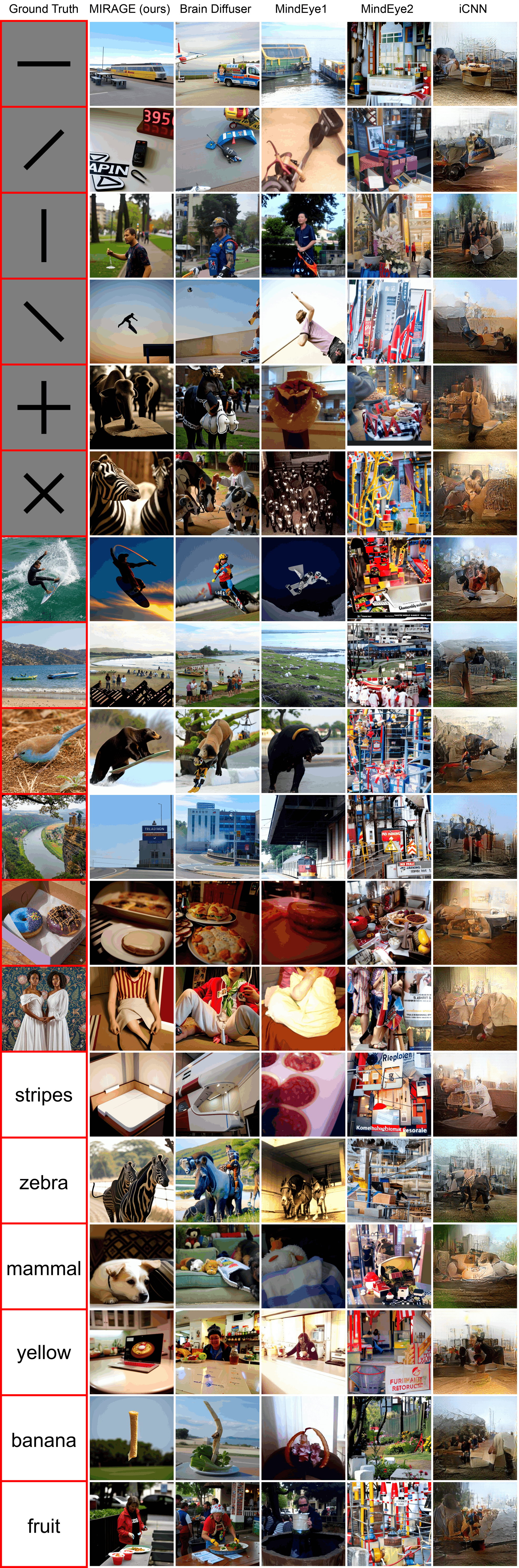}
    \caption{Worst-case imagery reconstructions from the imagery trials of NSD-Imagery. Samples selected as in Fig \ref{figure:vision_worst}.}
    \label{figure:imagery_worst}
\end{minipage}
\end{figure*}
\FloatBarrier
\clearpage

\subsection{Reconstructions from additional methods on NSD-Imagery}
\label{app:moremethods}
\begin{figure*}[!htb]
\centering
\begin{minipage}[t]{0.49\textwidth}
    \vspace{0pt}
    \centering
    \includegraphics[width=0.80\textwidth]{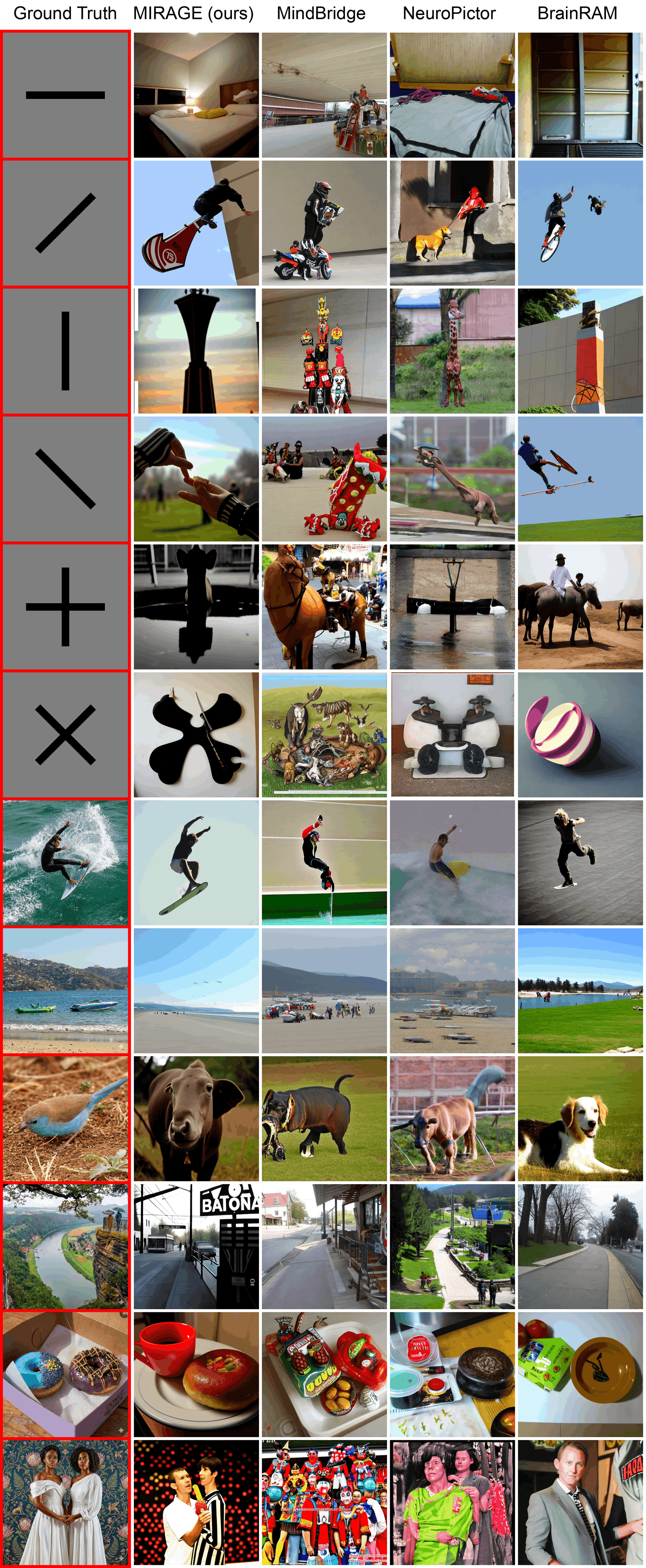}
    \caption{Best-case vision reconstructions (additional methods) from vision trials of NSD-Imagery. Samples selected as highest scoring based on metrics in Table 1 of the manuscript.}
    \label{figure:vision_best_rebuttal}
\end{minipage}\hfill
\begin{minipage}[t]{0.49\textwidth}
    \vspace{0pt}
    \centering
    \includegraphics[width=0.80\textwidth]{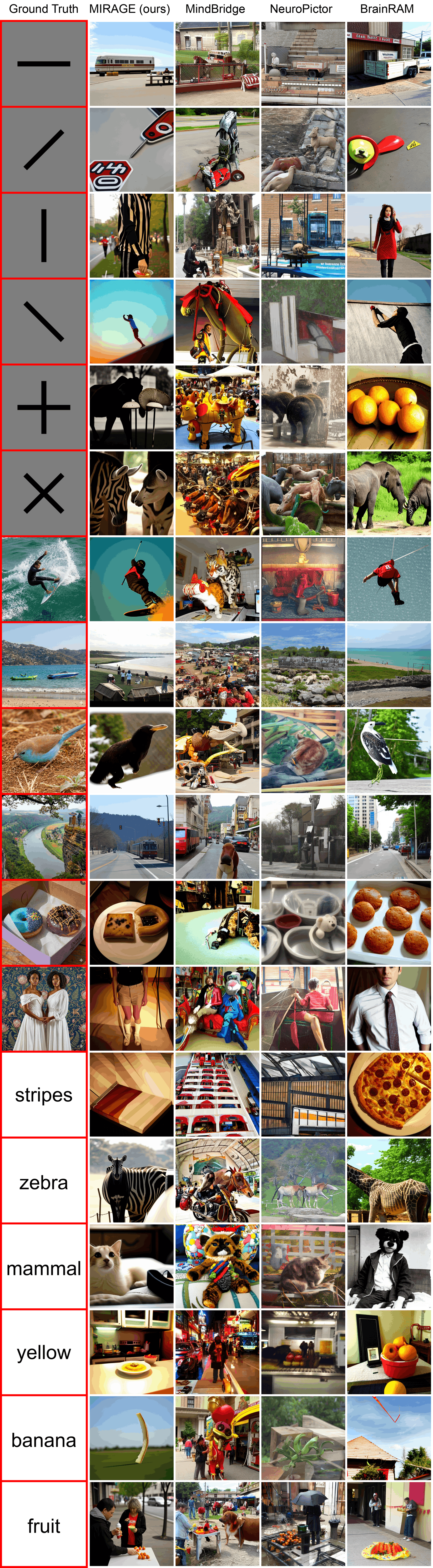}
    \caption{Best-case imagery reconstructions (additional methods) from imagery trials of NSD-Imagery. Samples selected as in Fig \ref{figure:vision_best_rebuttal}.}
    \label{figure:imagery_best_rebuttal}
\end{minipage}
\end{figure*}
\FloatBarrier
\clearpage

\begin{figure*}[!htb]
\centering
\begin{minipage}[t]{0.49\textwidth}
    \vspace{0pt}
    \centering
    \includegraphics[width=0.80\textwidth]{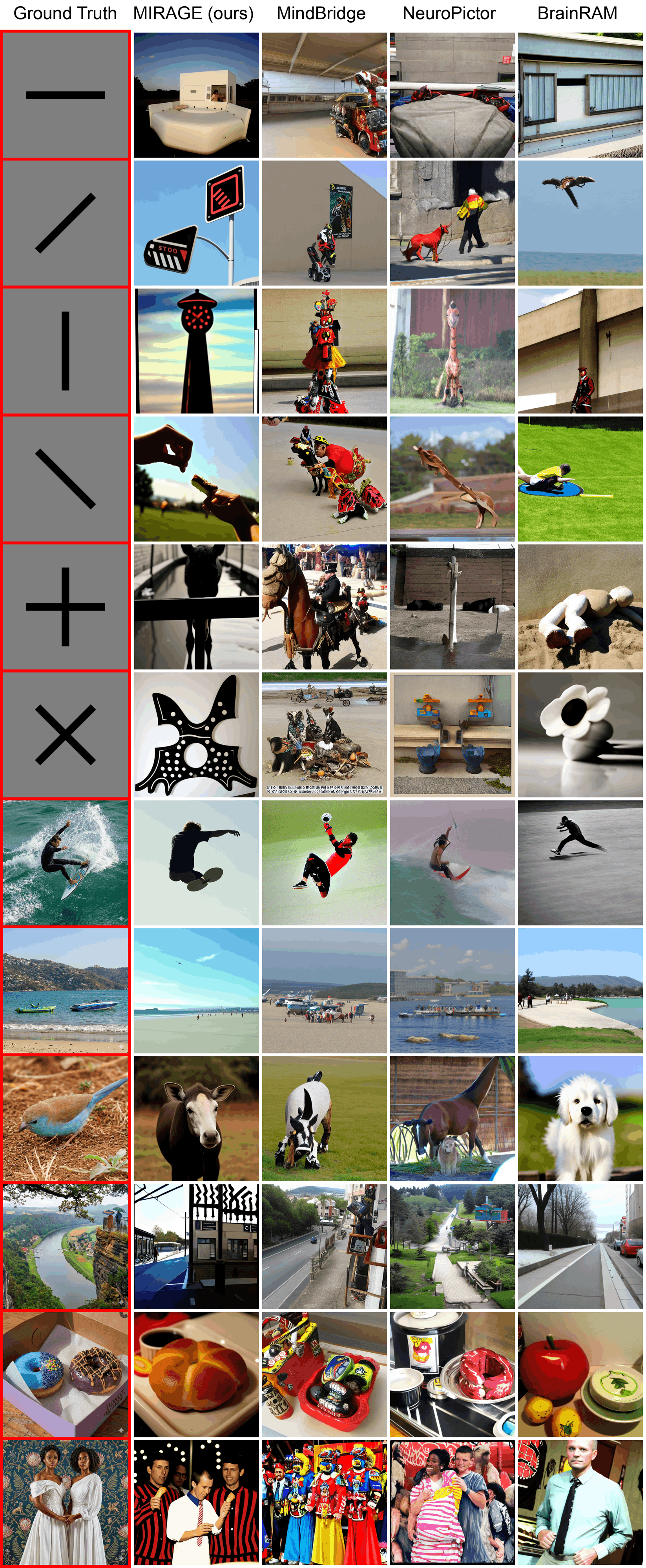}
    \caption{Median-case vision reconstructions (additional methods) from vision trials of NSD-Imagery. Samples selected as median scoring based on metrics in Table 1 of the manuscript.}
    \label{figure:vision_median_rebuttal}
\end{minipage}\hfill
\begin{minipage}[t]{0.49\textwidth}
    \vspace{0pt}
    \centering
    \includegraphics[width=0.80\textwidth]{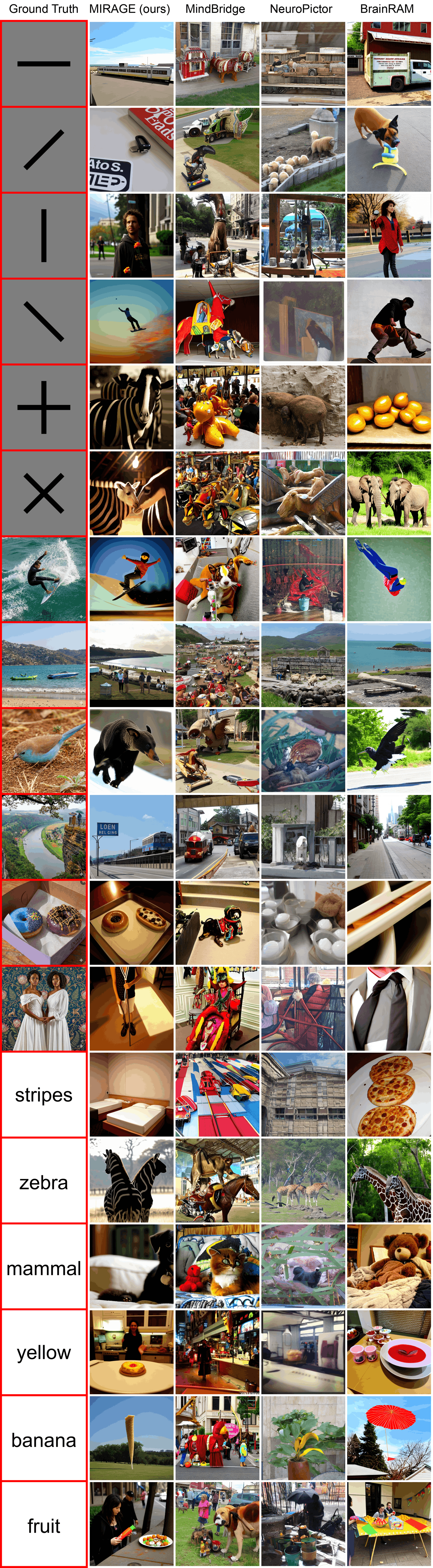}
    \caption{Median-case imagery reconstructions (additional methods) from imagery trials of NSD-Imagery. Samples selected as in Fig \ref{figure:vision_median_rebuttal}.}
    \label{figure:imagery_median_rebuttal}
\end{minipage}
\end{figure*}
\FloatBarrier
\clearpage

\begin{figure*}[!htb]
\centering
\begin{minipage}[t]{0.49\textwidth}
    \vspace{0pt}
    \centering
    \includegraphics[width=0.80\textwidth]{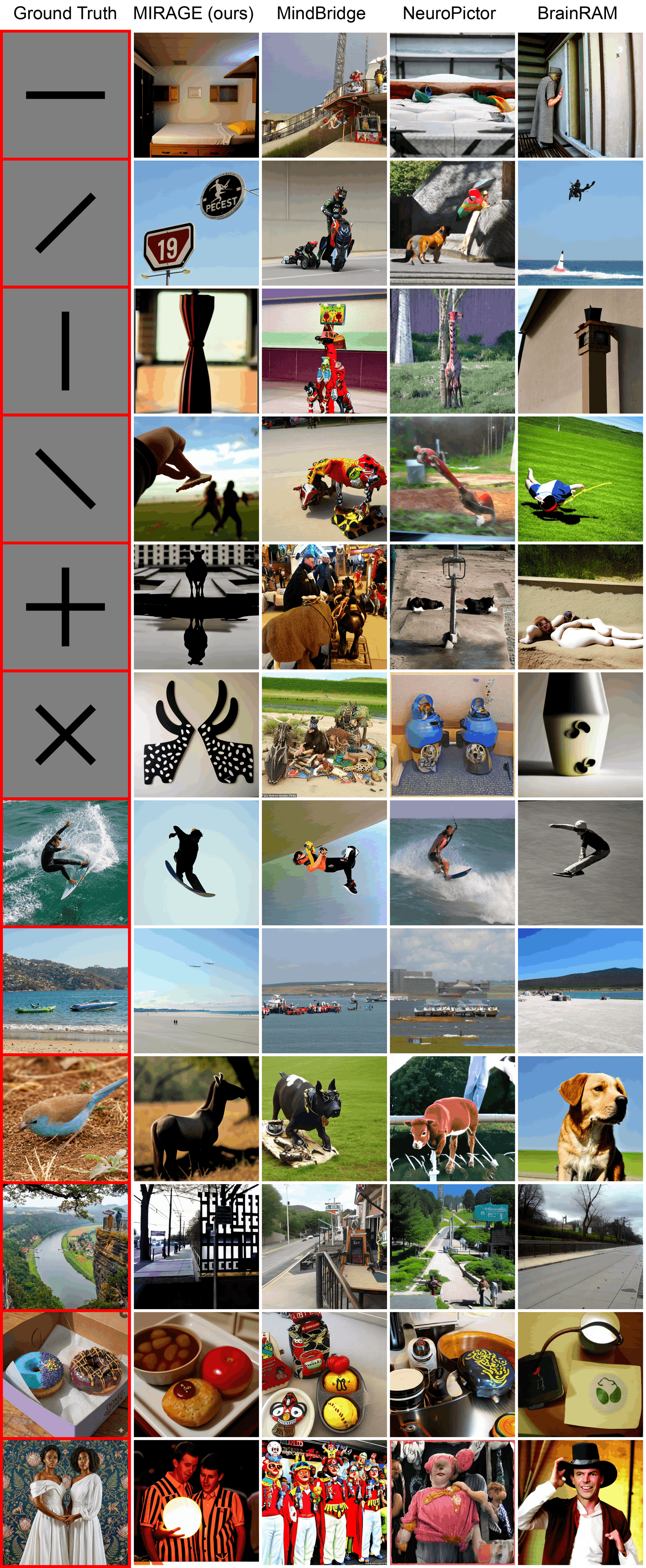}
    \caption{Worst-case vision reconstructions (additional methods) from vision trials of NSD-Imagery. Samples selected as lowest scoring based on metrics in Table 1 of the manuscript.}
    \label{figure:vision_worst_rebuttal}
\end{minipage}\hfill
\begin{minipage}[t]{0.49\textwidth}
    \vspace{0pt}
    \centering
    \includegraphics[width=0.80\textwidth]{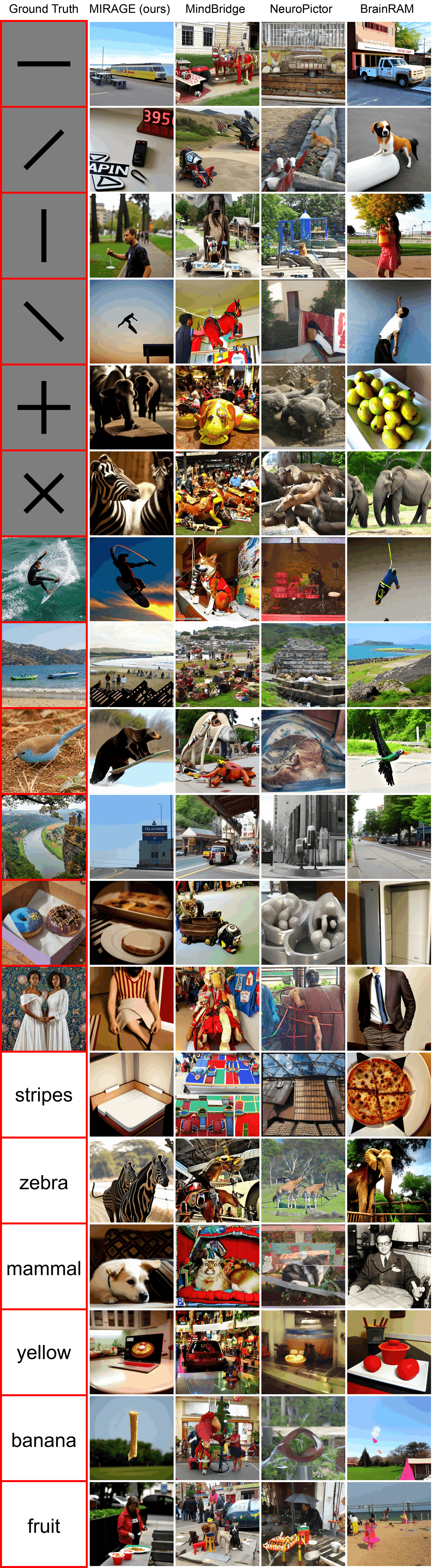}
    \caption{Worst-case imagery reconstructions (additional methods) from imagery trials of NSD-Imagery. Samples selected as in Fig \ref{figure:vision_worst_rebuttal}.}
    \label{figure:imagery_worst_rebuttal}
\end{minipage}
\end{figure*}

\FloatBarrier

\subsection{Additional evaluation metric details}
\label{app:metrics}
For the metrics in Table $1$ of the manuscript, a two-way comparison evaluates whether the feature embedding of the stimulus image is more similar to the feature embedding of the target reconstruction, or the feature embedding of a randomly selected "distractor" reconstruction. Two-way identification refers to percent correct across a set of two-way comparisons performed on a pool of distractor images. The two-way identification metrics we report, which are calculated using reconstructions of the $11$ other NSD-Imagery stimuli as distractors, are notably different from the two-way identification metrics presented in individual reconstruction papers that perform evaluations using reconstructions of the shared1000 as the pool of distractors. The pool of distractor images for NSD-Imagery is much smaller, and contains multiple distinct types of stimuli that may significantly alter the resulting identification accuracy metrics. Because of this difference, the two-way identification accuracy numbers are not directly comparable to two-way identification results evaluated on the shared1000 in our work or in other papers. Brain correlation scores are the Pearson correlation between the averaged measured brain response $\beta$ and the predicted brain response $\beta'$ produced by a brain encoding model (GNet \cite{St-Yves_heirarchy}) averaged across voxels within a respective ROI in visual cortex, including the whole visual cortex, early visual cortical regions (V1, V2, V3, and V4), and higher visual areas (set complement of visual cortex and early visual cortex). All metrics in Tables 1, Appendix Table \ref{table:simple_stimuli}, and Appendix Table\ref{table:complex_stimuli} were calculated and averaged across 10 images sampled from the output distribution of each method using a random seed. The caption metrics were computed against the ground truth image captions provided with the NSD-Imagery dataset. Metrics in Table $1$ of the manuscript are the original values reported in each of the respective papers, except for the iCNN method, which has never been benchmarked on the NSD shared1000 test set and so results reported are from our reproduction of the method utilizing the author's open source code. For the results from our method (MIRAGE) in the table, we compute values across 5 output repetitions sampled from the posterior of our method, and average those values together for the table.

\subsubsection{Normalized average of image feature metrics}
\label{app:featureaverage}
To compute the normalized average of the image feature metrics used in Fig \ref{fig:combined_exp4_ablation}B and to select the best, median, and worst reconstructions displayed in figures throughout our paper, we first standardized each metric $S_k$ to the unit interval $[0, 1]$. For metrics where lower values indicated better performance (denoted by the set $\mathcal{L}$), we applied an inverted Min-Max normalization such that the optimal raw value mapped to 1. The final score $\bar{S}_{\text{final}}$ was computed as the arithmetic mean of these normalized values, ensuring equal weighting across all image feature metrics:

\begin{equation}
\bar{S}_{\text{final}} = \frac{1}{K} \sum_{k=1}^{K} \left( 
\begin{cases} 
\frac{\max(S_k) - S_k}{\max(S_k) - \min(S_k)} & \text{if } k \in \mathcal{L} \\
\frac{S_k - \min(S_k)}{\max(S_k) - \min(S_k)} & \text{otherwise}
\end{cases} 
\right)
\label{eq:feature_aggregation}
\end{equation}

\noindent Where:
\begin{itemize}
    \item $K$ is the total number of metrics.
    \item $S_k$ denotes the raw score of the $k$-th metric.
    \item $\mathcal{L}$ is the set of metrics where a lower score indicates better performance.
    \item $\min(S_k)$ and $\max(S_k)$ are the minimum and maximum values of metric $k$ across the dataset.
\end{itemize}

\FloatBarrier
\onecolumn
\subsection{Statistical significance of metrics}
\label{app:statisticalsig}

\begin{table*}[!htb]
    \centering
    \setlength{\tabcolsep}{2pt}
    \small
    \resizebox{\textwidth}{!}{
    \begin{tabular}{lcccccccccccc}
        \toprule
        Method & \multicolumn{4}{c}{Low‑Level} & \multicolumn{4}{c}{High‑Level} & \multicolumn{3}{c}{Brain Correlation}\\
        \cmidrule(lr){2-5}\cmidrule(lr){6-9}\cmidrule(l){10-12}
        & PixCorr $\uparrow$ & SSIM $\uparrow$ & Alex(2) $\uparrow$ & Alex(5) $\uparrow$ 
        & Incep $\uparrow$ & CLIP $\uparrow$ & Eff $\downarrow$ & SwAV $\downarrow$ 
        & Early Vis. $\uparrow$ & Higher Vis. $\uparrow$ & Visual Cortex $\uparrow$\\
        \midrule
        \multicolumn{12}{c}{\textbf{Mental Imagery Reconstructions}}\\
        \midrule
        MIRAGE & $\pm$0.0061 & $\pm$0.0091 & $\pm$0.89\% & $\pm$1.06\% & $\pm$1.27\% & $\pm$1.24\% & $\pm$0.0044 & $\pm$0.0040 & $\pm$0.0075 & $\pm$0.0067 & $\pm$0.0059 \\ 
MindEye1 & $\pm$0.0082 & $\pm$0.0082 & $\pm$1.07\% & $\pm$0.72\% & $\pm$1.43\% & $\pm$1.35\% & $\pm$0.0064 & $\pm$0.0047 & $\pm$0.0073 & $\pm$0.0075 & $\pm$0.0069 \\ 
Brain Diffuser & $\pm$0.0068 & $\pm$0.0087 & $\pm$1.40\% & $\pm$1.07\% & $\pm$1.47\% & $\pm$1.46\% & $\pm$0.0053 & $\pm$0.0041 & $\pm$0.0086 & $\pm$0.0079 & $\pm$0.0070 \\ 
iCNN & $\pm$0.0081 & $\pm$0.0055 & $\pm$0.93\% & $\pm$0.54\% & $\pm$1.42\% & $\pm$1.21\% & $\pm$0.0041 & $\pm$0.0023 & $\pm$0.0055 & $\pm$0.0074 & $\pm$0.0056 \\ 
MindEye2 & $\pm$0.0085 & $\pm$0.0079 & $\pm$1.07\% & $\pm$0.87\% & $\pm$1.41\% & $\pm$1.27\% & $\pm$0.0067 & $\pm$0.0051 & $\pm$0.0073 & $\pm$0.0081 & $\pm$0.0070 \\ 
MindBridge & $\pm$0.0063 & $\pm$0.0080 & $\pm$1.38\% & $\pm$1.11\% & $\pm$1.45\% & $\pm$1.24\% & $\pm$0.0053 & $\pm$0.0045 & $\pm$0.0079 & $\pm$0.0079 & $\pm$0.0071 \\ 
NeuroPictor & $\pm$0.0062 & $\pm$0.0072 & $\pm$1.23\% & $\pm$1.22\% & $\pm$1.53\% & $\pm$1.37\% & $\pm$0.0046 & $\pm$0.0037 & $\pm$0.0087 & $\pm$0.0080 & $\pm$0.0072 \\ 
BrainRAM & $\pm$0.0074 & $\pm$0.0105 & $\pm$1.30\% & $\pm$1.21\% & $\pm$1.49\% & $\pm$1.26\% & $\pm$0.0058 & $\pm$0.0056 & $\pm$0.0088 & $\pm$0.0074 & $\pm$0.0074 \\ 
        \midrule
        \multicolumn{12}{c}{\textbf{Vision Reconstructions}}\\
        \midrule
        MIRAGE & $\pm$0.0072 & $\pm$0.0097 & $\pm$1.23\% & $\pm$1.21\% & $\pm$1.56\% & $\pm$1.41\% & $\pm$0.0040 & $\pm$0.0045 & $\pm$0.0052 & $\pm$0.0050 & $\pm$0.0044 \\ 
MindEye1 & $\pm$0.0086 & $\pm$0.0086 & $\pm$1.35\% & $\pm$1.31\% & $\pm$1.53\% & $\pm$1.53\% & $\pm$0.0035 & $\pm$0.0036 & $\pm$0.0053 & $\pm$0.0046 & $\pm$0.0042 \\ 
Brain Diffuser & $\pm$0.0052 & $\pm$0.0082 & $\pm$1.38\% & $\pm$1.35\% & $\pm$1.49\% & $\pm$1.50\% & $\pm$0.0040 & $\pm$0.0038 & $\pm$0.0055 & $\pm$0.0051 & $\pm$0.0045 \\ 
iCNN & $\pm$0.0077 & $\pm$0.0052 & $\pm$1.24\% & $\pm$1.26\% & $\pm$1.40\% & $\pm$1.40\% & $\pm$0.0021 & $\pm$0.0025 & $\pm$0.0058 & $\pm$0.0052 & $\pm$0.0044 \\ 
MindEye2 & $\pm$0.0049 & $\pm$0.0084 & $\pm$1.45\% & $\pm$1.43\% & $\pm$1.60\% & $\pm$1.51\% & $\pm$0.0034 & $\pm$0.0037 & $\pm$0.0054 & $\pm$0.0053 & $\pm$0.0048 \\ 
MindBridge & $\pm$0.0046 & $\pm$0.0041 & $\pm$1.43\% & $\pm$1.48\% & $\pm$1.54\% & $\pm$1.48\% & $\pm$0.0029 & $\pm$0.0036 & $\pm$0.0064 & $\pm$0.0048 & $\pm$0.0047 \\ 
NeuroPictor & $\pm$0.0046 & $\pm$0.0060 & $\pm$1.40\% & $\pm$1.54\% & $\pm$1.47\% & $\pm$1.44\% & $\pm$0.0021 & $\pm$0.0028 & $\pm$0.0053 & $\pm$0.0053 & $\pm$0.0048 \\ 
BrainRAM & $\pm$0.0063 & $\pm$0.0087 & $\pm$1.48\% & $\pm$1.36\% & $\pm$1.49\% & $\pm$1.43\% & $\pm$0.0044 & $\pm$0.0047 & $\pm$0.0053 & $\pm$0.0048 & $\pm$0.0044 \\ 
        \bottomrule
\end{tabular}
    }
    \caption{Standard error measurements for evaluation metrics of fMRI-to-Image reconstruction models evaluated on both the vision and mental imagery trials of NSD-Imagery. Values correspond to the standard error spread of values in Table \ref{table:combined} in the manuscript.} 
    \label{table:combined_stats} 
\end{table*}

\FloatBarrier

\subsection{NSD test set feature metric evaluations}
\label{app:shared1000}

\begin{table}[!htb]
    \centering
    \setlength{\tabcolsep}{2pt}
    \small
    \resizebox{\textwidth}{!}{
    \begin{tabular}{lccccccccccc}
        \toprule
        \textbf{NSD Shared1000 Test Set} & \multicolumn{4}{c}{Low-Level} & \multicolumn{4}{c}{High-Level} & \multicolumn{3}{c}{Brain Correlation} \\
        \cmidrule(lr){2-5} \cmidrule(lr){6-9} \cmidrule(l){10-12}
        Method & PixCorr $\uparrow$ & SSIM $\uparrow$ & Alex(2) $\uparrow$ & Alex(5) $\uparrow$ & Incep $\uparrow$ & CLIP $\uparrow$  & Eff $\downarrow$ & SwAV $\downarrow$ & Early Vis. $\uparrow$ & Higher Vis. $\uparrow$ & Visual Cortex $\uparrow$ \\
        \midrule
        \textbf{MIRAGE (ours)} & 0.285 & 0.361 & 94.30\% & 95.73\% & 91.18\% & 90.92\% & 0.732 & 0.473 & 0.337 & 0.371 & 0.372 \\
        \textbf{MindEye1 \cite{scotti_reconstructing_2023}} & 0.319 & 0.360 & 92.49\% & 96.44\% & \underline{93.55\%} & \underline{92.14\%} & \underline{0.648} & \underline{0.377} & 0.350 & \underline{0.374} & 0.378 \\
        \textbf{Brain Diffuser \cite{ozcelik2023braindiffuser}} & 0.273 & \underline{0.365} & \underline{94.39\%} & 96.64\% & 91.28\% & 90.90\% & 0.728 & 0.421 & \underline{0.353} & \textbf{0.375} & \underline{0.381} \\
        \textbf{iCNN \cite{shen_deep_2019}} & \underline{0.321} & 0.336 & 94.33\% & \underline{97.09\%} & 90.46\% & 74.47\% & 0.797 & 0.528 & \textbf{0.410} & 0.371 & \textbf{0.396} \\
        \textbf{MindEye2 \cite{Scotti2024MindEye2}} & \textbf{0.322} & \textbf{0.431} & \textbf{96.10\%} & \textbf{98.61\%} & \textbf{95.42\%} & \textbf{92.98\%} & \textbf{0.619} & \textbf{0.344} & \underline{0.360} & 0.368 & 0.373 \\
        \bottomrule
    \end{tabular}
    }
    \vspace{10pt}
    \caption{Quantitative comparison between reconstruction methods on the NSD Shared1000 Test Set. Metrics are the same as Table $1$ of the manuscript.}
    \label{table:shared1000}
\end{table}

\subsection{Comparison of image feature metrics across stimuli types}
\label{app:stimtypes}
\begin{table}[!htb]   
    \centering
    \setlength{\tabcolsep}{4pt}
    \small
    \resizebox{\textwidth}{!}{
    \begin{tabular}{lccccccccccc}
        \toprule
        Method & \multicolumn{4}{c}{Low-Level} & \multicolumn{4}{c}{High-Level} & \multicolumn{3}{c}{Brain Correlation} \\
        \cmidrule(lr){2-5} \cmidrule(lr){6-9} \cmidrule(l){10-12}
        & PixCorr $\uparrow$ & SSIM $\uparrow$ & Alex(2) $\uparrow$ & Alex(5) $\uparrow$ & Incep $\uparrow$ & CLIP $\uparrow$  & Eff $\downarrow$ & SwAV $\downarrow$ & Early Vis. $\uparrow$ & Higher Vis. $\uparrow$ & Visual Cortex $\uparrow$ \\
        \midrule
        \multicolumn{12}{c}{\textbf{Mental Imagery Reconstructions (Simple Stimuli)}} \\
        \midrule
        \textbf{MIRAGE (ours)} & 0.027 & \underline{0.511} & \textbf{53.11\%} & \textbf{67.27\%} & \underline{42.39\%} & \underline{60.30\%} & \textbf{0.939} & \underline{0.563} & \textbf{0.224} & \textbf{0.118} & \textbf{0.164} \\
        \textbf{MindEye1 \cite{scotti_reconstructing_2023}} & \underline{0.033} & 0.456 & \underline{43.71\%} & \underline{61.67\%} & 37.46\% & 58.37\% & \underline{0.974} & 0.563 & \underline{0.200} & \underline{0.107} & \underline{0.148} \\
        \textbf{Brain Diffuser \cite{ozcelik2023braindiffuser}} & 0.013 & \textbf{0.524} & 30.68\% & 50.68\% & 34.43\% & 44.51\% & 0.983 & 0.603 & 0.152 & 0.091 & 0.128 \\
        \textbf{iCNN \cite{shen_deep_2019}} & \textbf{0.063} & 0.427 & 27.42\% & 47.65\% & \textbf{45.11\%} & \textbf{67.99\%} & 1.006 & \textbf{0.546} & 0.138 & 0.045 & 0.081 \\
        \textbf{MindEye2 \cite{Scotti2024MindEye2}} & 0.011 & 0.448 & 23.37\% & 45.34\% & 31.14\% & 49.02\% & 0.987 & 0.590 & 0.074 & 0.035 & 0.051 \\
        \midrule
        \multicolumn{12}{c}{\textbf{Vision Reconstructions (Simple Stimuli)}} \\
        \midrule
        \textbf{MIRAGE (ours)} & \textbf{0.159} & \underline{0.569} & \textbf{74.24\%} & \textbf{82.77\%} & \textbf{56.78\%} & \underline{63.71\%} & \textbf{0.913} & \underline{0.537} & \underline{0.395} & \textbf{0.174} & \underline{0.279} \\
        \textbf{MindEye1 \cite{scotti_reconstructing_2023}} & 0.129 & 0.506 & \underline{62.01\%} & \underline{76.36\%} & \underline{43.33\%} &  60.64\% & \underline{0.961} & 0.549 & 0.370 & \underline{0.140} & 0.243 \\
        \textbf{Brain Diffuser \cite{ozcelik2023braindiffuser}} & 0.075 & \textbf{0.586} & 40.19\% & 66.67\% & 38.30\% & 42.20\% & 0.988 & 0.601 & 0.209 & 0.106 & 0.169 \\
        \textbf{iCNN \cite{shen_deep_2019}} & \underline{0.132} & 0.454 & 57.01\% & 74.89\% & 37.69\% & \textbf{69.02\%} & 0.992 & \textbf{0.534} & \textbf{0.447} & 0.133 & \textbf{0.278} \\
        \textbf{MindEye2 \cite{Scotti2024MindEye2}} & 0.040 & 0.487 & 50.87\% & 68.98\% & 43.52\% & 52.46\% & 0.980 & 0.577 & 0.334 & 0.108 & 0.204 \\
        \bottomrule
    \end{tabular}
    }
    \vspace{10pt}
    \caption{Quantitative comparison between reconstruction methods for both imagery and vision trials on simple stimuli. Metrics are the same as Table $1$ of the manuscript.}
    \label{table:simple_stimuli}
\end{table}

\begin{table}[!htb]
    \centering
    \setlength{\tabcolsep}{4pt}
    \small
    \resizebox{\textwidth}{!}{
    \begin{tabular}{lccccccccccc}
        \toprule
        Method & \multicolumn{4}{c}{Low-Level} & \multicolumn{4}{c}{High-Level} & \multicolumn{3}{c}{Brain Correlation} \\
        \cmidrule(lr){2-5} \cmidrule(lr){6-9} \cmidrule(l){10-12}
        & PixCorr $\uparrow$ & SSIM $\uparrow$ & Alex(2) $\uparrow$ & Alex(5) $\uparrow$ & Incep $\uparrow$ & CLIP $\uparrow$  & Eff $\downarrow$ & SwAV $\downarrow$ & Early Vis. $\uparrow$ & Higher Vis. $\uparrow$ & Visual Cortex $\uparrow$ \\
        \midrule
        \multicolumn{12}{c}{\textbf{Mental Imagery Reconstructions (Complex Stimuli)}} \\
        \midrule
        \textbf{MIRAGE (ours)} & \textbf{0.181} & \underline{0.285} & \underline{74.74\%} & 57.65\% & 62.121\% & \underline{54.62\%} & \underline{0.888} & 0.587 & \textbf{0.183} & \textbf{0.165} & \textbf{0.172} \\
        \textbf{MindEye1 \cite{scotti_reconstructing_2023}} & 0.138 & 0.243 & \textbf{75.42\%} & 60.34\% & \underline{66.591\%} & 51.06\% & 0.921 & \textbf{0.566} & \underline{0.159} & \underline{0.164} & \underline{0.161} \\
        \textbf{Brain Diffuser \cite{ozcelik2023braindiffuser}} & 0.114 & \textbf{0.278} & 73.60\% & \textbf{66.02\%} & \textbf{71.02\%} & \textbf{63.64\%} & \textbf{0.888} & 0.567 & 0.114 & 0.163 & 0.154 \\
        \textbf{iCNN \cite{shen_deep_2019}} & \underline{0.153} & 0.253 & 73.71\% & 62.84\% & 53.674\% & 15.46\% & 0.982 & \underline{0.575} & 0.089 & 0.079 & 0.081 \\
        \textbf{MindEye2 \cite{Scotti2024MindEye2}} & 0.032 & 0.231 & 70.42\% & \underline{65.11\%} & 61.97\% & 51.93\% & 0.943 & 0.601 & 0.062 & 0.074 & 0.068 \\
        \midrule
        \multicolumn{12}{c}{\textbf{Vision Reconstructions (Complex Stimuli)}} \\
        \midrule
        \textbf{MIRAGE (ours)} & \underline{0.282} & 0.315 & 83.83\% & 70.38\% & 82.727\% & 69.659\% & 0.845 & 0.555 & 0.331 & 0.350 & 0.353 \\
        \textbf{MindEye1 \cite{scotti_reconstructing_2023}} & 0.308 & \underline{0.318} & \underline{85.11\%} & \underline{85.27\%} & 81.55\% & 70.038\% & \underline{0.800} & \underline{0.471} & \underline{0.378} & \underline{0.365} & \underline{0.379} \\
        \textbf{Brain Diffuser \cite{ozcelik2023braindiffuser}} & 0.139 & \textbf{0.323} & 80.49\% & 79.02\% & \underline{83.60\%} & \underline{74.43\%} & 0.829 & 0.509 & 0.284 & 0.353 & 0.341 \\
        \textbf{iCNN \cite{shen_deep_2019}} & \textbf{0.316} & 0.316 & \textbf{86.33\%} & \textbf{87.80\%} & \textbf{84.62\%} & 29.05\% & 0.860 & 0.514 & \textbf{0.437} & 0.358 & \textbf{0.397} \\
        \textbf{MindEye2 \cite{Scotti2024MindEye2}} & 0.223 & 0.333 & 84.28\% & 85.83\% & 80.08\% & \textbf{77.46\%} & \textbf{0.794} & \textbf{0.454} & 0.378 & \textbf{0.360} & 0.376 \\
        \bottomrule
    \end{tabular}
    }
    \vspace{10pt}
    \caption{Quantitative comparison between reconstruction methods for both imagery and vision trials on complex stimuli. Metrics are the same as Table $1$ of the manuscript.}
    \label{table:complex_stimuli}
\end{table}

\FloatBarrier

\subsection{Impact of trial repetition averaging on performance}
\label{app:trial_reps}

\begin{figure}[!htb]   
\centering
\includegraphics[width=0.6\columnwidth]{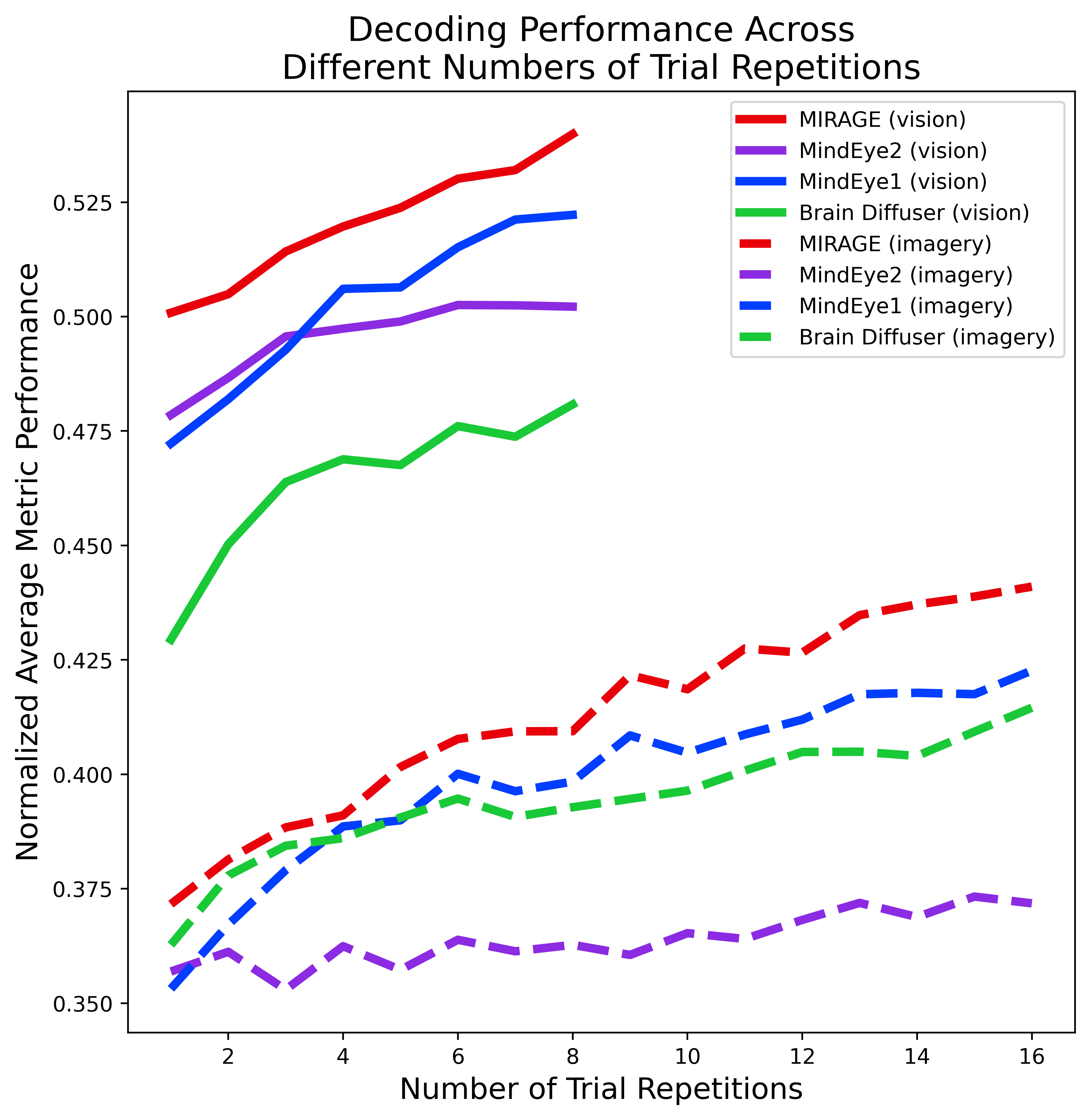}
\caption{Performance of \textbf{MIRAGE} and other methods when averaging across brain activity responses to multiple trial repetitions of the same stimulus. Y-axis is the normalized average of all metrics in Table $1$ of the manuscript, X-axis is the number of averaged trial repetitions.} 
\label{figure:trial_reps}
\end{figure}

One of the experimental details that varies between NSD \cite{allen_massive_2022} and NSD-Imagery \cite{NSDImagery} is the number of times each stimulus was presented in the experiment, also called the number of trial repetitions. NSD contained $3$ trial repetitions of each stimulus in both the training and test sets, while NSD-Imagery contains 8 trial repetitions for the vision task and $16$ trial repetitions for the imagery task. In Fig \ref{figure:trial_reps}, we plot the effect of these additional trial repetitions on the performance of \textbf{MIRAGE} relative to the other methods we compare against.

\subsection{Impact of training data scale on performance}
\label{app:scaling}

\begin{figure}[!htb]   
\centering
\includegraphics[width=0.6\columnwidth]{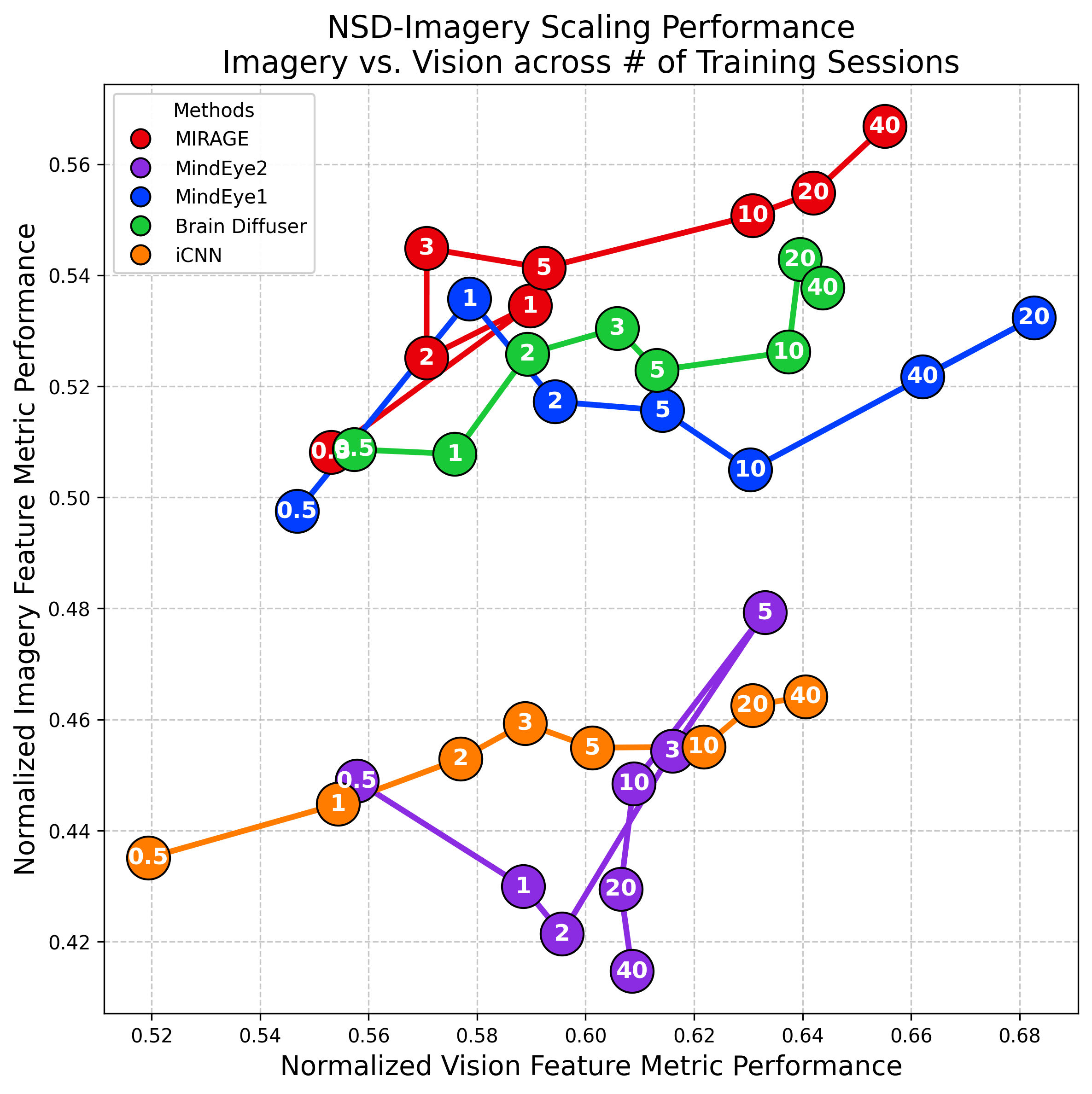}
\caption{Performance of \textbf{MIRAGE} and other methods on NSD-Imagery for Subject 1 when trained on different numbers of fMRI sessions present in NSD. Each session includes approximately one hour of fMRI data. Metrics are the normalized average of all metrics in Table $1$ of the manuscript, with imagery performance on the Y axis and vision on the X axis. Methods are indicated by color, with the number of training sessions indicated by the numbers in each dot.} 
\label{figure:scaling}
\end{figure}

An additional challenge in deploying these fMRI-to-image decoding methods lies in making them more generalizable to new subjects. 
\textbf{MIRAGE}, along with all of the other methods examined in this paper, were trained with $40$ hours of subject-specific fMRI data comprising $10,000$ unique stimuli. Collecting this much training data for new subjects in practical settings is currently impractical or impossible for certain clinical patients.
Recent work in MindEye2 \cite{Scotti2024MindEye2} has tackled this problem head-on by using a multi-subject pretraining step, however as evaluated in Fig \ref{figure:scaling}, this technique generalizes poorly to mental imagery data. By contrast, MIRAGE outperforms all other methods for 
 mental image reconstruction using only $3$ hours of fMRI training data, and continues to scale robustly up to $40$ sessions. We additionally note that the methods that used ridge regression decoding backbones (\textbf{MIRAGE}, Brain Diffuser, iCNN) all produce much more consistent scaling improvements on mental images than the models that utilize deep neural network backbones (MindEye1, MindEye2).

\subsection{Impact of diffusion strength on performance}
\label{app:priorstrength}

\begin{figure}[!htb]    
\begin{center}
\includegraphics[width=0.8\columnwidth]{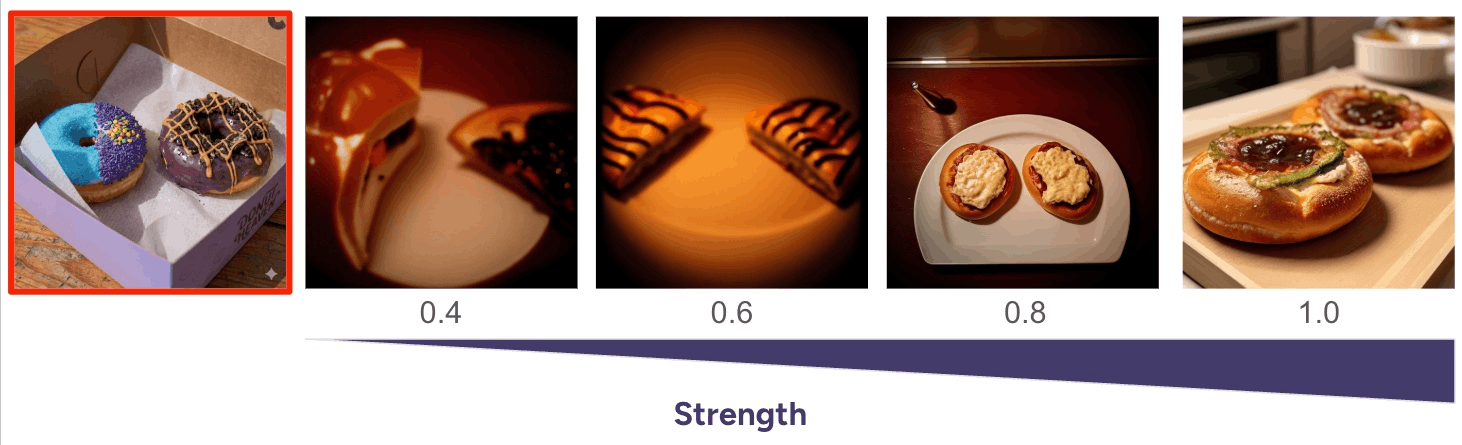}
\end{center}
\caption{Examples of reconstructions provided at different diffusion strength parameters, images are the ground truth (outlined in red) and reconstructions provided at 0.4, 0.6, 0.8, and 1.0 diffusion strength respectively. Strength values below 0.4 experience no noticeable variation due to the nonlinear dynamics of the strength parameter.} 
\label{fig:strengthrecons}
\end{figure}

\begin{figure}[!htb]    
\begin{center}
\includegraphics[width=0.6\columnwidth]{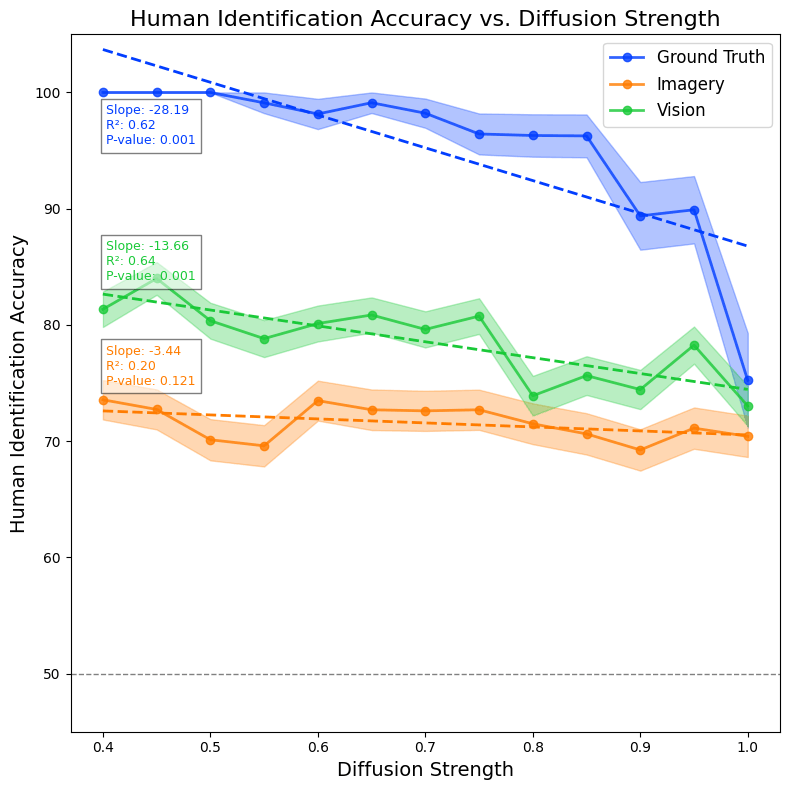}
\end{center}
\caption{Human identification accuracy of \textbf{MIRAGE} (with no CLIP-Image guidance) as a function of diffusion model strength for imagery trials (orange line), vision trials (green line), and a control experiment that used the features directly from the ground truth image and caption (blue line). A dashed line is placed at the $50\%$ chance threshold. Results are from a behavioral experiment that is identical to Experiment 1 (Fig \ref{fig:combined_2afc_exp3}A), but varied across strength parameters.} 
\label{fig:priorstrength}
\end{figure}
Recent work \cite{Shirakawa2024SpuriousRF, brainbits} has raised the question of how much of the detail in fMRI-to-image reconstructions originates in the brain and how much is simply hallucinated by the strong natural priors enforced by a diffusion model. This critical perspective makes a clear prediction: As the strength of the natural prior increases, the results should improve. One way of examining this relationship is by modulating the strength parameter in the img2img mode of the diffusion model, by which an initial image (in our case the low-level reconstruction provided by the VDVAE model discussed in Section \ref{low-level}) is partially noised and then denoised with CLIP guidance. This denoising process is where the natural priors are enforced, and the amount of denoising (and thus the amount of the final image that is guided by the natural prior) is modulated by the strength parameter. We repeated Experiment 1 presented in Fig \ref{fig:combined_2afc_exp3}A of the paper across a wide range of strength parameters to investigate the potential influence of diffusion strength on the results. We test strength parameters between $0.4$—the lowest strength value that yields meaningful variation from the input image—and $1.0$, which destroys the entire input image before denoising. In this experiment, we use only CLIP-text semantic guidance for the diffusion process to increase the contrast between the original input image and the purely semantic guidance during the denoising process, although we acknowledge that this slightly reduces the performance of \textbf{MIRAGE} during the experiment relative to the results in Fig \ref{fig:combined_2afc_exp3}A. The results demonstrate that increasing the strength parameter (and therefore increasing the influence of the prior in determining the ultimate reconstruction) induces no significant change in the rate at which humans can correctly identify mental image reconstructions as corresponding to the stimulus image, and seen image reconstructions experience a significant \textit{decrease} in identifiability as diffusion strength increases (Fig \ref{fig:priorstrength}). Although the diffusion model prior clearly plays a role in improving the quality and aesthetics of the reconstructions, our results show that there is still plenty of decoded signal present in the reconstructions to facilitate identification even with minimal guidance from the diffusion model. 

To further emphasize our improved signal recovery and decreased reliance on the priors of the image generator, MIRAGE employs the lowest diffusion strength parameter ($0.7$) among comparable “dual stream” (high-level/low-level) reconstruction methods, including as Brain Diffuser ($0.75$) and MindEye1 ($0.85$), meaning MIRAGE explicitly relies less on the diffusion model’s prior than other approaches. Additionally, MIRAGE achieves state-of-the-art (SOTA) performance on both simple and conceptual stimuli types, which are outside the natural prior of the diffusion model. These findings underscore MIRAGE’s ability to extract and utilize more imagery signal from the brain independent of the natural prior of the diffusion model, setting it apart from alternative methods.

\FloatBarrier

\subsection{Retrieval analysis}

To assess the necessity of MIRAGE's generative complexity, we compared our method against a baseline of direct image retrieval. Fundamentally, retrieval and reconstruction differ in the nature of their image priors: retrieval relies on a static image corpus, effectively acting as a prior that assigns a non-zero, uniform probability to a finite set of candidate images and zero probability to all other possible images. In contrast, reconstruction utilizes a diffusion model, which models a continuous probability distribution across the broader manifold of natural images from its training distribution. While a systematic comparison exploring the interaction between various decoded feature spaces and retrieval corpus distributions is beyond the scope of this work, we evaluated a targeted baseline to determine if the diffusion model provides distinct value over retrieval within MIRAGE's selected feature spaces and the shared1000 retrieval pool commonly used in other work \cite{Scotti2024MindEye2, scotti_reconstructing_2023}. 

We performed retrieval in the pooled ViT-L/14 image embedding space used to drive the MIRAGE generative model, and the hidden layer ViT-L/14 space utilized in the retrieval pooling step (Section \ref{sec:reconstruction}) using a candidate corpus of 1000 COCO \cite{microsoftcoco} images sourced from the NSD shared1000. As demonstrated in Table \ref{table:retrieval_appendix_split}, MIRAGE substantially outperforms all retrieval baselines on most high-level and semantic metrics, while hidden-layer retrieval remains competitive on certain low-level metrics. These results confirm that appending a diffusion model to the end of the MIRAGE pipeline yields fundamentally superior, bespoke image decoding performance compared to selecting likely approximations from a static corpus.

\begin{table*}[!htb]
    \centering
    \setlength{\tabcolsep}{2pt}
    \small
    \resizebox{\textwidth}{!}{
    \begin{tabular}{lccccccccccc}
        \toprule
        Method & \multicolumn{4}{c}{Low‑Level} & \multicolumn{4}{c}{High‑Level} & \multicolumn{3}{c}{Brain Correlation} \\
        \cmidrule(lr){2-5}\cmidrule(lr){6-9}\cmidrule(l){10-12}
        & PixCorr $\uparrow$ & SSIM $\uparrow$ & Alex(2) $\uparrow$ & Alex(5) $\uparrow$ 
        & Incep $\uparrow$ & CLIP $\uparrow$ & Eff $\downarrow$ & SwAV $\downarrow$ 
        & Early Vis. $\uparrow$ & Higher Vis. $\uparrow$ & Visual Cortex $\uparrow$ \\
        \midrule
        \multicolumn{12}{c}{\textbf{Mental Imagery Reconstructions (Simple Stimuli)}}\\
        \midrule
        \textbf{MIRAGE (ours)}                  & \textbf{0.027} & \underline{0.511} & \textbf{53.11\%} & \textbf{67.27\%} & \textbf{42.39\%} & \textbf{60.30\%} & \underline{0.939} & \textbf{0.563} & \textbf{0.224} & \textbf{0.118} & \textbf{0.164} \\
        Top-1 Retrieval                         & -0.022 & 0.446 & 28.79\% & 42.42\% & \underline{36.36\%} & \underline{53.03\%} & 1.007 & 0.603 & 0.053 & 0.099 & 0.105 \\
        Top-1 Retrieval (hidden layer)          & \underline{0.015} & \textbf{0.531} & \underline{33.33\%} & \underline{51.52\%} & 31.82\% & 48.48\% & \textbf{0.880} & \underline{0.596} & \underline{0.136} & \underline{0.116} & \underline{0.145} \\
        \midrule
        \multicolumn{12}{c}{\textbf{Vision Reconstructions (Simple Stimuli)}}\\
        \midrule
        \textbf{MIRAGE (ours)}                  & \textbf{0.159} & \underline{0.569} & \textbf{74.24\%} & \textbf{82.77\%} & \textbf{56.78\%} & \textbf{63.71\%} & \underline{0.913} & \underline{0.537} & \textbf{0.395} & \textbf{0.174} & \textbf{0.279} \\
        Top-1 Retrieval                         & -0.062 & 0.500 & 12.12\% & 46.97\% & 37.88\% & 48.48\% & 0.971 & 0.602 & 0.080 & 0.067 & 0.111 \\
        Top-1 Retrieval (hidden layer)          & \underline{0.034} & \textbf{0.640} & \underline{42.42\%} & \underline{56.06\%} & \underline{54.18\%} & \underline{62.18\%} & \textbf{0.730} & \textbf{0.430} & \underline{0.134} & \underline{0.127} & \underline{0.170} \\
        \midrule
        \multicolumn{12}{c}{\textbf{Mental Imagery Reconstructions (Complex Stimuli)}}\\
        \midrule
        \textbf{MIRAGE (ours)}                  & \textbf{0.181} & \underline{0.285} & \textbf{74.74\%} & \textbf{57.65\%} & \textbf{62.12\%} & \underline{54.62\%} & \underline{0.888} & \underline{0.587} & \textbf{0.183} & \textbf{0.165} & \textbf{0.172} \\
        Top-1 Retrieval                         & 0.090 & 0.229 & \underline{63.64\%} & \underline{56.09\%} & \underline{45.45\%} & \textbf{63.64\%} & 0.952 & 0.601 & \underline{0.181} & 0.065 & 0.137 \\
        Top-1 Retrieval (hidden layer)          & \underline{0.153} & \textbf{0.352} & 43.94\% & 34.85\% & 37.88\% & 48.48\% & \textbf{0.756} & \textbf{0.480} & 0.164 & \underline{0.111} & \underline{0.139} \\
        \midrule
        \multicolumn{12}{c}{\textbf{Vision Reconstructions (Complex Stimuli)}}\\
        \midrule
        \textbf{MIRAGE (ours)}                  & \textbf{0.282} & \underline{0.315} & \textbf{83.83\%} & \textbf{70.38\%} & \textbf{82.73\%} & \textbf{69.66\%} & \underline{0.845} & 0.555 & \textbf{0.331} & \textbf{0.350} & \textbf{0.353} \\
        Top-1 Retrieval                         & \underline{0.196} & 0.294 & \underline{78.79\%} & \underline{68.18\%} & 53.03\% & \underline{56.06\%} & 0.866 & \underline{0.541} & -0.016 & \underline{0.238} & 0.150 \\
        Top-1 Retrieval (hidden layer)          & 0.179 & \textbf{0.382} & 69.70\% & 54.55\% & \underline{81.33\%} & 34.85\% & \textbf{0.716} & \textbf{0.447} & \underline{0.187} & 0.235 & \underline{0.231} \\
        \bottomrule
\end{tabular}
    }
    \caption{Quantitative comparison between MIRAGE and two Top-1 Retrieval baselines (pooled and hidden layer CLIP ViT-L/14 embeddings), separated by simple and complex stimuli and averaged across all subjects. Metrics are the same as Table \ref{table:combined}. Bold indicates the best performance between MIRAGE and the retrieval baselines within each stimulus category, and underlines indicate second-best.}

    \label{table:retrieval_appendix_split}
\end{table*}

\begin{figure}[!htb]
  \centering
  \begin{minipage}[t]{0.48\textwidth}
    \centering
    \textbf{A}\\[4pt]
    \includegraphics[width=\linewidth]{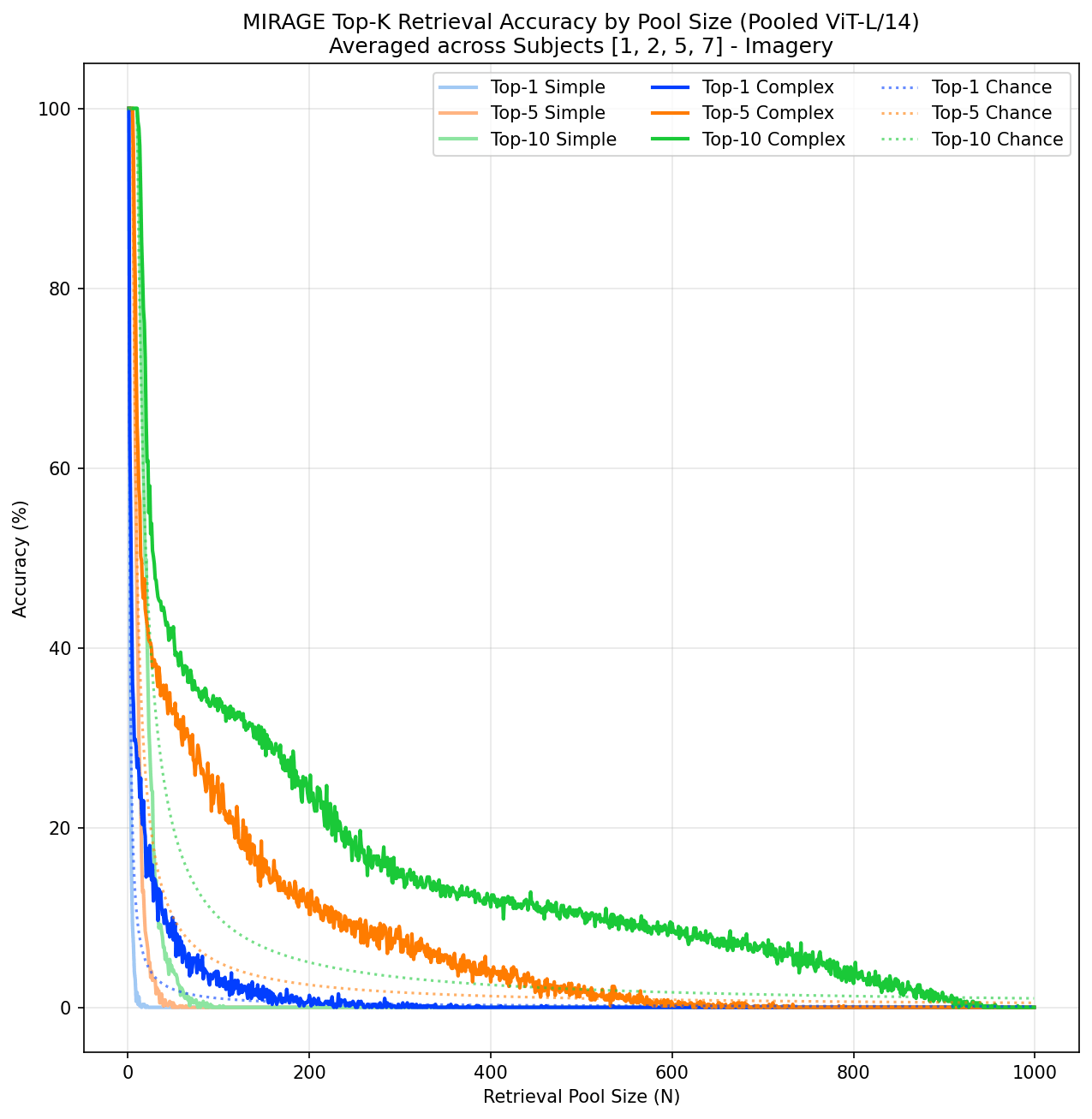}
    \textbf{C}\\[4pt]
    \includegraphics[width=\linewidth]{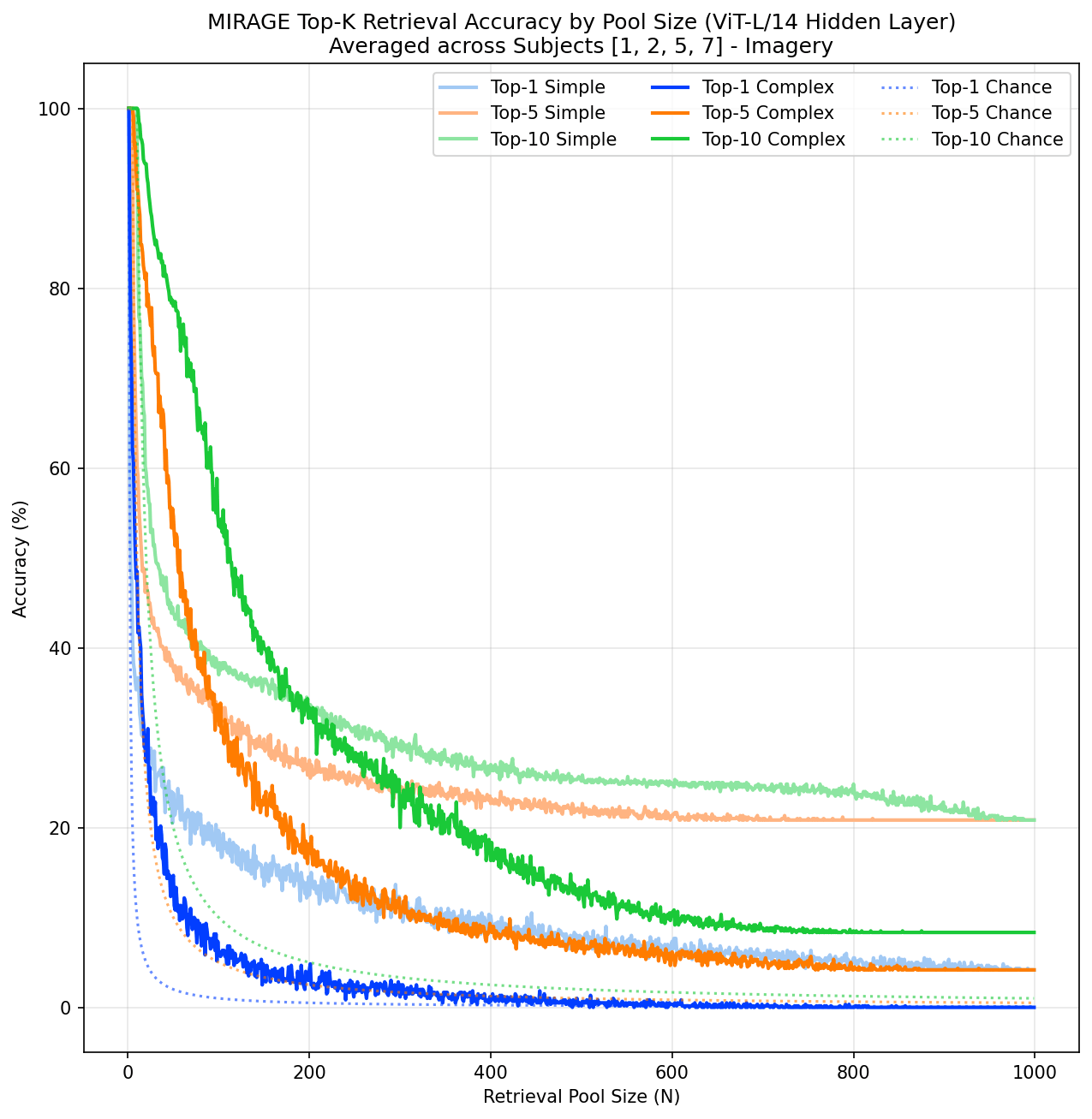}
  \end{minipage}\hfill
  \begin{minipage}[t]{0.48\textwidth}
    \centering
    \textbf{B}\\[4pt]
    \includegraphics[width=\linewidth]{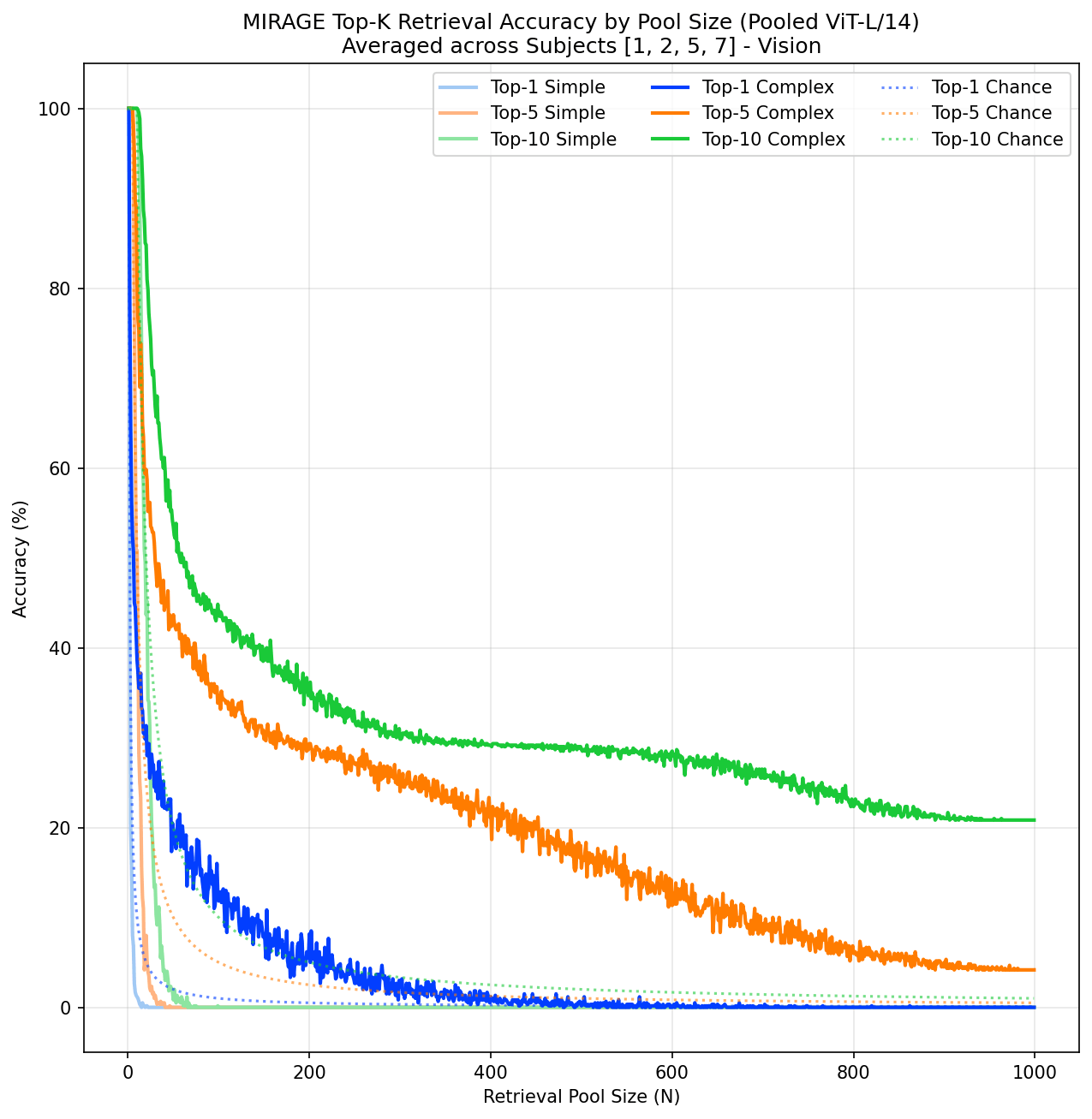}
    \textbf{D}\\[4pt]
    \includegraphics[width=\linewidth]{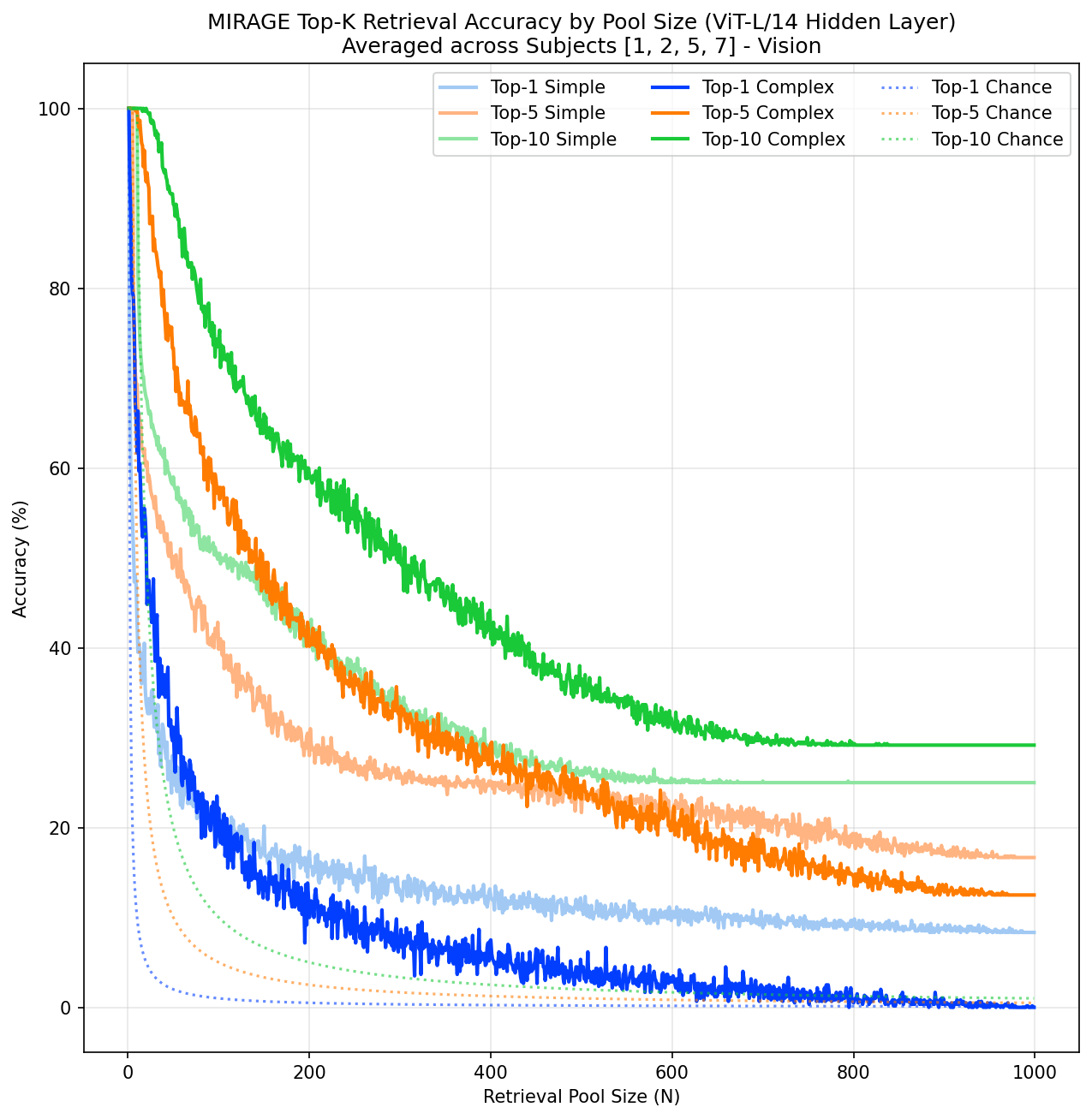}
  \end{minipage}

  \vspace{-6pt}
  \caption{\textbf{Top-K retrieval performance vs. pool size for Subjects 1, 2, 5, and 7.} Accuracy (y-axis) is evaluated across varying distractor pool sizes (x-axis) for both mental imagery (left: \textbf{A, C}) and vision trials (right: \textbf{B, D}). The top row (\textbf{A, B}) evaluates retrieval in the pooled ViT-L/14 image embedding space used to drive the MIRAGE generative model, while the bottom row (\textbf{C, D}) uses the hidden layer ViT-L/14 space utilized in the retrieval pooling step (Section \ref{sec:reconstruction}). Curves denote top-1, top-5, and top-10 performance for simple (light lines) and complex (dark lines) stimuli, with chance levels indicated by corresponding dotted lines. To calculate accuracy, the ground-truth NSD-Imagery stimulus is shuffled with $N$ random distractor images from the NSD shared1000 pool; a success is recorded when the target image ranks within the top $K$ closest matches to the subject's brain-predicted embedding. All curves are bootstrapped across 100 randomly sampled distractor pools for each value of $N$.
  }
  \vspace{-8pt}
  \label{fig:retrieval_curves}
\end{figure}

We also evaluated the dynamics of a set of common "Top-K" retrieval metrics (top-1, top-5, and top-10) as a function of the size of the retrieval pool being used (Fig \ref{fig:retrieval_curves}). Since these metrics measure the reliability of being able to extract the exact stimulus image from a pool, we added the set of NSD-Imagery stimulus images to our retrieval pool for this analysis, while noting that these stimuli are often out of distribution for this particular retrieval pool. Predictably, retrieval accuracy when using the pooled CLIP ViT-L/14 embeddings fails for simple stimuli, as these features do not capture the low-level features (e.g., orientation) that define these images. Conversely, retrieval accuracy is well above chance for complex stimuli, with vision retrieval predictably outperforming imagery retrieval. For the hidden layer ViT-L/14 embeddings, we see improvements across the board for all stimulus types relative to the pooled embeddings, reinforcing our conclusion from Section \ref{sec:reconstruction} that these sparser image embeddings provide a lot of utility in a retrieval context.

\FloatBarrier
\subsection{Behavioral experiment}
\label{app:behavioral}

\subsubsection{Experiment protocols}
\label{app:experimentprotocols}
We conducted a set of behavioral experiments on $500$ human raters online. For our experiment, we identified no risks to the human participants, and our institution's IRB approved our experiment. We probed $3$ experiments intermixed into two discrete sections within the same behavioral tasks, with each experiment consisting of trials sampled evenly from the 18 different stimuli, 3 stimulus types, and the 4 NSD subjects who completed all 40 scanning sessions (subjects 1, 2, 5, 7). After sampling the target reconstruction, the distractor reconstruction was sampled from a pool of reconstructions from the same subject, stimulus type, and whether it was a vision/imagery reconstruction, but a different specific stimulus. The experimental trials within each task were shuffled and $36$ trials were presented to each subject. Our subjects were recruited through the \href{http://www.prolific.ac.uk)}{Prolific platform}, with our experimental tasks hosted on \href{http://meadows-research.com}{Meadows}. Each human rater was paid $\$1.50$ for the completion of the experiment, and the median completion time was $6$ minutes and $17$ seconds, resulting in an average payment rate of $\$14.32$/hour. Each human rater was presented with $6$ attention check trials during the experiment. An attention check is a trial in which the ground truth image is presented as a candidate image during the trial. Because the ground truth image will always be the image that is most similar to itself, these trials were used to identify whether subjects were paying attention to the task and the instructions. We identified $5$ human raters who failed at least 2 attention checks and removed those raters from our data before conducting our analysis. Raters were blind to the conditions of the experiment. The primary way we ensured consistency across trials was to keep the experiment very short (~5m) and to sample a very large number of subjects (>500). Our goal with this design was to minimize any effects produced by the subject’s rating profile changing over time as they saw more and more trials, and to minimize the biases of any one particular subject on the result of the experiment. Code to reproduce our experiment can be found in \href{https://anonymous.4open.science/r/mental_imagery_behavioral_analysis-2214/}{our anonymized GitHub repository}. All human subjects provided written informed consent. All procedures were approved the Institutional Review Board at the University of Minnesota.

\subsubsection{2AFC identification task}
\begin{figure}[!htb]    
\begin{center}
\includegraphics[width=0.8\columnwidth]{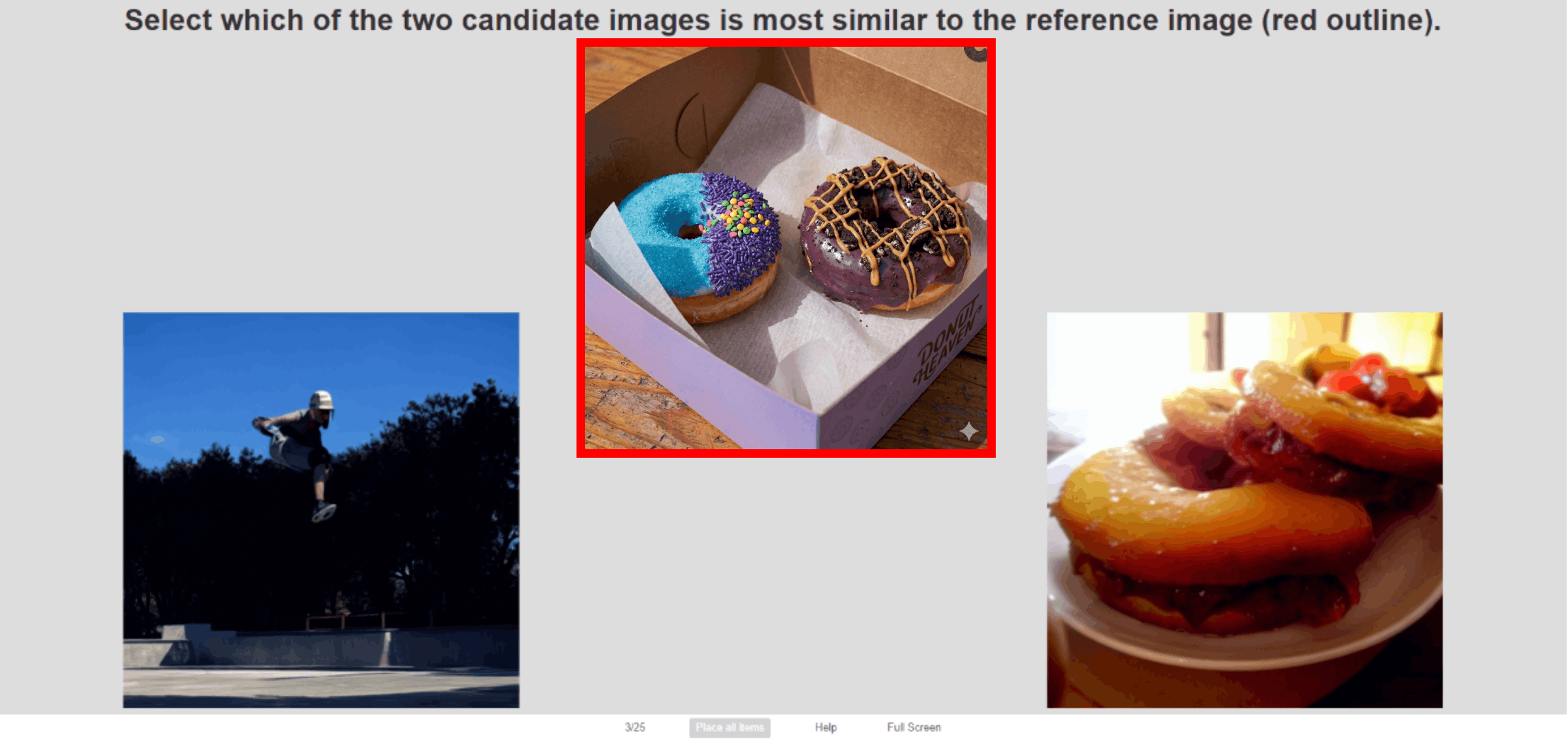}
\end{center}
\caption{An example of the 2 alternative forced choice task used in the first behavioral experiment performed by human raters.} 
\label{figure:task1}
\end{figure}

Our first experiment, which made up the entirety of the first task, was a $2$ alternative forced choice task (2AFC) facilitated by the "Match-To-Sample" task on the Meadows platform. An example of the first experiment can be seen in Fig \ref{figure:task1}. In this experiment, human raters were asked to select which of two candidate images was more similar to a reference image. The reference image provided is the ground truth image the NSD-Imagery subject either saw or imagined, and the $2$ candidate images were the target reconstruction of the reference image, or a randomly selected reconstruction from an fMRI scan corresponding to a different stimulus of the same stimulus type. The two candidate images were always sampled from the same reconstruction method and NSD-Imagery subject. This experiment was repeated for all reconstruction methods, visual modalities, NSD subjects, and across 10 reconstructions sampled from the output distribution of each reconstruction method. With the results presented in Section \ref{humanratings}, we establish a baseline for human-rated image identification accuracy of mental image reconstructions, as no other paper has conducted behavioral evaluations of mental image reconstructions.

\subsubsection{Continuous similarity rating task}
\begin{figure}[!htb]    
\begin{center}
\includegraphics[width=0.8\columnwidth]{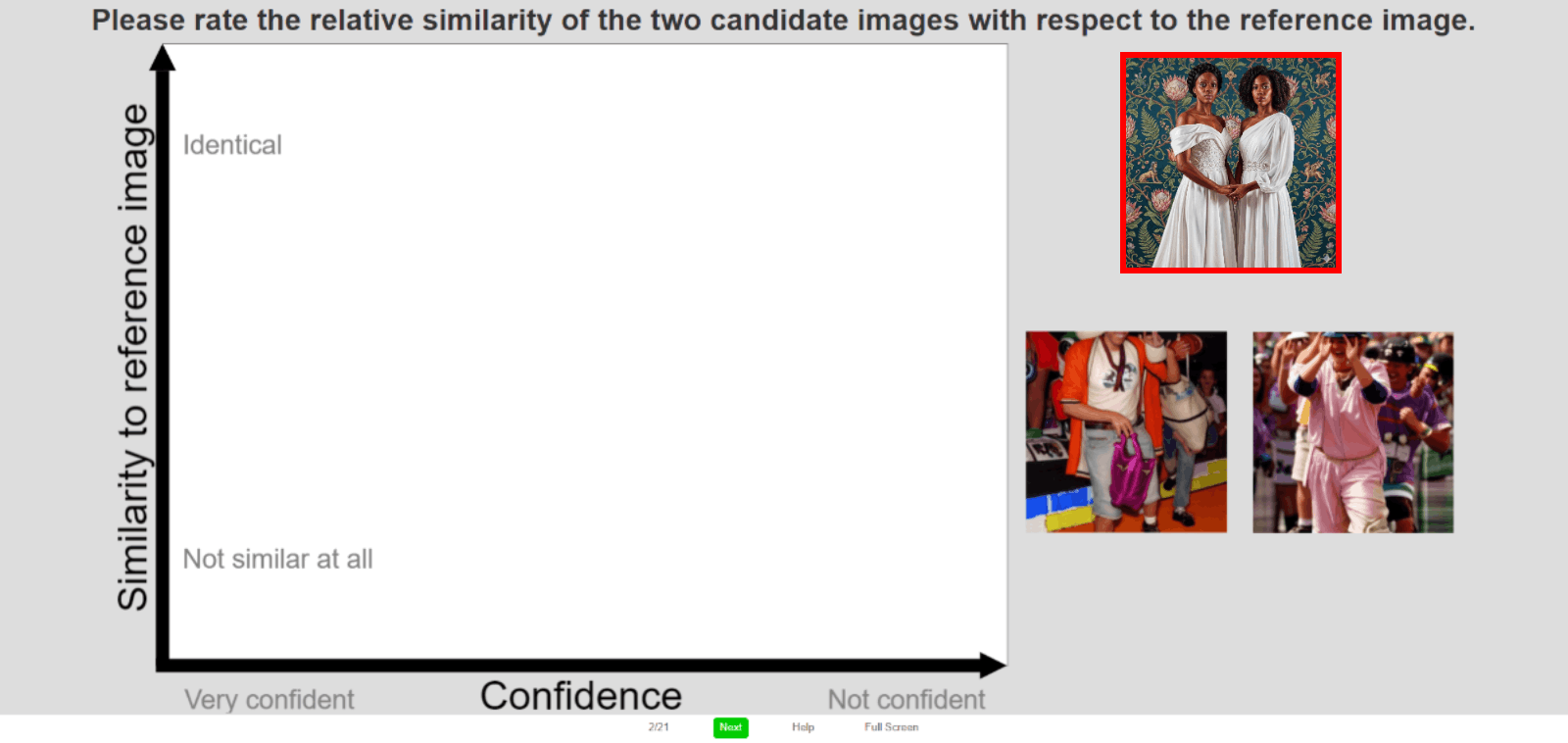}
\end{center}
\caption{An example of similarity score task used in experiments $2$ and $3$ of the behavioral experiment performed by human raters.} 
\label{figure:task23}
\end{figure}

The second and third experiments we conducted were shuffled together inside the second task of the experiment, which was facilitated by the "Drag-Rate" task on the Meadows platform. An example of the task used in experiments $2$ and $3$ can be seen in Fig \ref{figure:task23}. In this task, human raters were presented with a reference image, two candidate images, and a continuous two-dimensional plot that they could drag the candidate images onto, where the Y-axis represented "similarity to the reference image" and the X-axis represented the rater's confidence. The reference image provided was always the ground truth image the NSD-Imagery subject either saw or imagined. For experiment $2$, the $2$ candidate images were reconstructions of the reference image from the imagery and vision trials of the NSD-Imagery trials. Experiment $2$ was repeated for the simple and complex stimuli (as conceptual stimuli do not have meaningful vision reconstructions), all reconstruction methods, NSD subjects, and across 10 reconstructions sampled from the output distribution of each reconstruction method. For experiment $3$, we designed a head-to-head experiment for conceptual reconstructions, where the candidate images were reconstructions of the imagery trials for the conceptual stimuli produced by different reconstruction methods. Experiment $3$ contained trials for all NSD subjects, 10 reconstructions sampled from the output distribution of each reconstruction method, and 3 unique combinations of reconstruction methods for each sample. One-dimensional similarity ratings—like the ones used in this section of the experiment—can be extremely sensitive to the context of the alternative samples being compared against, and so are primarily useful for comparing the relative similarity of the candidate stimuli presented during each individual trial. The two comparison tasks evaluated within this task of the experiment were designed with this in mind, each configured to more directly compare the difference in quality between reconstructions of vision and imagery, as well as to compare the differences in quality between reconstruction methods on the conceptual stimuli. Our analysis of these results in Section \ref{humanratings} provides a detailed analysis of how reconstruction performance scales across vision and imagery, and of how each method performs on the conceptual stimuli.

\subsection{iCCN implementation}
\label{app:icnnchanges}
Originally introduced in \citet{shen_deep_2019}, and first trained on NSD in  \citet{Shirakawa2024SpuriousRF}, we adapt the author's open source implementation to try and faithfully replicate their results, making the following changes to the implementation:
\begin{enumerate}
    \item \textbf{Normalization of images:} We disabled normalization of images when computing VGG19 features. During our initial trials, normalization led to unexpected color distortions in the reconstructed images. Removing normalization allowed the reconstructions to maintain their original color integrity, which is particularly crucial for visual comparisons in tasks requiring precise color representation.
    \item \textbf{Feature decoding with Ridge Regression:} Instead of the \texttt{fastl2lir} library, we employed the Ridge Regression implementation from the \texttt{sklearn} library. This change enhanced compatibility with the rest of our workflow and provided better support for managing memory-intensive computations. For VGG19 layers with a large feature space, feature decoding was performed in chunks. This approach enabled the simultaneous calculation of features and fitting of the Ridge Regression model without requiring intermediate results to be saved to disk, thereby optimizing both time and memory usage.
\end{enumerate}

\subsection{Public code release}
We provide a \href{https://github.com/MedARC-AI/MIRAGE}{public GitHub repository} with the code to reproduce our method.

\subsection{AI-Generated Images and Copyright Compliance}
Several figures in this manuscript contain synthetic images generated by artificial intelligence. Specifically, the AI-generated panels within Fig 1, 2, 3, 4, 5, and B, C, D, E, F, G, H, I, J, K, L, O, R, and S in S1 Text were created using the Stable Cascade model developed by Stability AI.

In accordance with requirements for publishing under the Creative Commons Attribution 4.0 International (CC BY 4.0) license, we confirm that the use of these images complies with the software's terms of use. The Stability AI Non-Commercial Research Community License Agreement explicitly states under its definition of Derivative Works: "For clarity, Derivative Works do not include the output of any Model." Because the model creators do not claim copyright over the generated output, we, as the creators of these specific image outputs, license them under CC BY 4.0 for this publication.

Terms of Service Link: \href{https://huggingface.co/stabilityai/stable-cascade/blob/main/LICENSE}{https://huggingface.co/stabilityai/stable-cascade/blob/main/LICENSE}

In addition to the model outputs generated by the MIRAGE architecture, the "ground-truth" stimuli images for the complex stimuli containing natural scenes have also been replaced with AI-generated proxy images using Stable Cascade. The original experimental methodology utilized specific images from the Microsoft Common Objects in Context (MS COCO) dataset as visual stimuli during fMRI acquisition. While the annotations within the MS COCO dataset are open access, the underlying images retain the copyright of their original Flickr authors. Because several of the specific images used in our experimental subset carried restrictive licenses (e.g., NonCommercial, NoDerivatives, or All Rights Reserved) that are incompatible with open-access republication, we generated visually and semantically similar proxy images to serve as illustrative substitutes in the figures. We emphasize that these AI-generated proxy images shown in the manuscript are for illustrative purposes in the manuscript only. They are not the exact images shown to the human subjects during the fMRI data collection phase. All quantitative evaluations, fMRI-to-image model training, and performance metrics discussed in this paper were conducted using the original, exact MS COCO images. The substitution of these images in the published figures does not alter the underlying data, the MIRAGE model's architecture, or the reported quantitative results.

\end{document}